\documentclass{acmtrans2m}

\usepackage{amssymb,latexsym,epsfig,graphicx,psfrag}

\usepackage{comment}

\makeatletter
\def\@begintheorem#1#2{\rm \trivlist \item[\hskip \labelsep{\bf #1\ #2}]}
\def\@opargbegintheorem#1#2#3{\rm \trivlist
      \item[\hskip \labelsep{\bf #1\ #2\ (#3)}]}
\makeatother

\newcommand{\mycolon}{\,\colon}
\mathcode`\:="8000%
{\catcode`\:=\active \gdef:{\mycolon}}

\newcommand{\epunc}{\enspace} 
\newcommand{\code}[1]{\textsf{#1}} 
\newcommand{\dt}[1]{\emph{#1}} 
\newcommand{\ol}{\overline}
\newcommand{\Rule}[2]{          
  \begin{array}[b]{c}
  #1 \\\hline
  #2
  \end{array}}

\newcommand{\dom}{{\mathit dom}}
\newcommand{\ldb}{[\![}
\newcommand{\rdb}{]\!]}
\newcommand{\means}[1]{\ldb {#1}\rdb}
\newcommand{\union}{\cup}
\newcommand{\intersect}{\cap}
\newcommand{\proves}{\vdash}
\newcommand{\ext}[3]{[#1\mid#2\!\mapsto\!#3]}
\newcommand{\lam}[2]{\lambda #1 \bullet #2} 
\newcommand{\all}[2]{\forall #1 \bullet #2}
\newcommand{\some}[2]{\exists #1 \bullet #2}
\newcommand{\Empty}{\varnothing}

\newcommand{\mletml}[3]{\MLET\;#1=#2\;\MIN\;#3}
\newcommand{\mifthenelse}[3]{\MIF\;#1\;\MTHEN\;#2\;\MELSE\;#3}
\newcommand{\MIN}{\mbox{\textsf{in}}}
\newcommand{\MLET}{\mbox{\textsf{let}}} 
\newcommand{\MTHEN}{{\textsf{then}}}
\newcommand{\MELSE}{{\textsf{else}}}
\newcommand{\MIF}{{\textsf{if}}}

\newcommand{\aproves}{\rhd} 
\newcommand{\ncomp}{\mathrel{ >\!\! \!\!\! \not\leq }} 
\newcommand{\etaconf}[3]{\mathit{conf}\,#1\,(#3, #2)}
\newcommand{\conf}[1]{\mathit{conf}\,#1}
\newcommand{\REL}{\mathcal{R}} 
\newcommand{\rel}[3]{\REL\;#1\;#2\;#3}
\newcommand{\SREL}{\mathcal{G}} 
\newcommand{\srel}[4]{\SREL\;#1\;#2\;#3\;#4}
\newcommand{\srell}[3]{\SREL\;#1\;#2\;#3}
\newcommand{\veq}[3]{#2\mathrel{\sim_{#1}}#3}
\newcommand{\nats}{{\mathbb{N}}}

\newcommand{\keyword}[1]{\mbox{\textrm{\textbf{#1}}}}

\newcommand{\CLASS}{\keyword{class}}
\newcommand{\SUPER}{\keyword{super}}
\newcommand{\UNIT}{\keyword{unit}}
\newcommand{\ITT}{\keyword{it}} 
\newcommand{\EXT}{\keyword{extends}}
\newcommand{\NEW}{\keyword{new}}
\newcommand{\CON}{\keyword{con}}
\newcommand{\IS}{\keyword{is}}
\newcommand{\NULL}{\keyword{null}}
\newcommand{\BOOL}{\keyword{bool}}
\newcommand{\OBJECT}{\keyword{Object}}
\newcommand{\ELSE}{\keyword{else}}
\newcommand{\THEN}{\keyword{then}}
\newcommand{\FALSE}{\keyword{false}}
\newcommand{\TRUE}{\keyword{true}}
\newcommand{\IF}{\keyword{if}}
\newcommand{\FI}{\keyword{fi}}
\newcommand{\DO}{\keyword{do}}
\newcommand{\OD}{\keyword{od}}
\newcommand{\IN}{\keyword{in}}
\newcommand{\SKIP}{\keyword{skip}}
\newcommand{\ABORT}{\keyword{abort}}

\newcommand{\TO}{\mathord{\rightarrow}}
\newcommand{\construct}[1]{\CON\{\, #1 \,\}}
\newcommand{\class}[3]{\CLASS\;{#1}\;\EXT\;{#2}\; \{ \;#3\; \} }
\newcommand{\ifelse}[3]{\IF\;#1\;\THEN\;#2\;\ELSE\;#3\;\FI}
\newcommand{\assym}{\mbox{$\mathord{\: \!:\!= \:}$}} 
\newcommand{\gassym}{\mathord{ :\!\!:\!= }} 

\newcommand{\assign}[2]{#1 \assym #2}
\newcommand{\fassign}[3]{#1.#2 \assym #3}
\newcommand{\faccess}[2]{#1.#2}
\newcommand{\eqtest}[2]{#1 = #2}
\newcommand{\new}[2]{\NEW\;#1} 
\newcommand{\mcall}[3]{#1.#2(#3)}
\newcommand{\cast}[2]{(#1)\;#2}
\newcommand{\var}[3]{#1 \assym #2\;\IN\;#3}
\newcommand{\seq}[2]{#1;\;#2}
\newcommand{\is}[2]{#1\;\IS\;#2} 

\newcommand{\mylet}[1]{\MLET\;#1\;\MIN\;}
\newcommand{\semtrue}{{\mathit{true}}}
\newcommand{\semfalse}{{\mathit{false}}}
\newcommand{\NIL}{\mathit{nil}}
\newcommand{\IT}{\mathit{it}}

\newcommand{\Case}{\textsc{Case}} 
\newcommand{\self}{\textsf{self}}
\newcommand{\result}{\textsf{result}}

\newcommand{\T}{T} 
\renewcommand{\S}{S} 
\newcommand{\CL}{CL} 
\newcommand{\M}{M} 

\newcommand{\npoints}{\mathrel{\not\leadsto}}
\newcommand{\npointsx}[1]{\npoints^{#1}}
\newcommand{\hext}{\mathrel{\unlhd}} 
\newcommand{\subclasses}{\mathord{\downarrow}} 
\newcommand{\cdep}{\mathrel{\sqsubset}}
\newcommand{\cdeps}{\mathrel{\prec}}
\newcommand{\cdepseq}{\mathrel{\precsim}}
\newcommand{\well}{\mathrel{\ll}}

\newcommand{\implies}{\mathbin{\:\Rightarrow\:}}
\renewcommand{\iff}{\mathbin{\:\Leftrightarrow\:}}

\newcommand{\dfields}{\textit{dfields}} 
\newcommand{\fields}{\textit{fields}}
\newcommand{\super}{\textit{super}}
\newcommand{\constr}{\textit{constr}}
\newcommand{\mtype}{\textit{mtype}}
\newcommand{\pars}{\textit{pars}}
\newcommand{\mscope}{\textit{mscope}}
\newcommand{\protected}{\textit{prot}}
\newcommand{\loctype}{\mathit{loctype}}
\newcommand{\cnames}{\mathit{ClassNames}}
\newcommand{\menv}{\mathit{MEnv}}
\newcommand{\type}{\mathit{type}}
\newcommand{\state}[1]{\mathit{state}\,#1}
\newcommand{\heap}{\mathit{Heap}}
\newcommand{\Loc}{\mathit{Loc}}
\newcommand{\fresh}{\mathit{fresh}}
\newcommand{\restr}{\mathit{restr}}
\newcommand{\Oh}{\mathord{\mathit{Oh}}}
\newcommand{\Rh}{\mathord{\mathit{Rh}}}
\newcommand{\Ch}{\mathord{\mathit{Ch}}}

\newcommand{\locsOR}{locs(Own\subclasses,Rep\subclasses)}

\newtheorem{theorem}{Theorem}[section]

\newtheorem{lemma}[theorem]{Lemma}
\newtheorem{assumption}[theorem]{Assumption}
\newdef{definition}[theorem]{Definition}
\newdef{remark}[theorem]{Remark}
\newdef{example}[theorem]{Example}

\newenvironment{infig}[1]{%
\medskip\noindent\textbf{\textit{#1}}\\ \hbra\begin{center}}{%
\vspace{-1ex}\end{center}\hket\medskip}

\newenvironment{infign}[1]{
\medskip\noindent\textbf{\textit{#1}}\\ \hbra}{%
\vspace{-1ex}\hket\medskip}

\newcommand{\hbra}{
\hbox to \linewidth{\vrule width0.3mm height 1.8mm depth-0.3mm
                    \leaders\hrule height1.8mm depth-1.5mm\hfill
                    \vrule width0.3mm height 1.8mm depth-0.3mm}}
\newcommand{\hket}{
\hbox to \linewidth{\vrule width0.3mm height1.5mm
                    \leaders\hrule height0.3mm\hfill
                    \vrule width0.3mm height1.5mm}}

\title{Ownership Confinement Ensures Representation Independence for
  Object-Oriented Programs
}

\author{
ANINDYA BANERJEE\\
Department of Computing and Information Sciences\\
Kansas State University\\
Manhattan KS 66506 USA
\and DAVID A.\ NAUMANN\\
Department of Computer Science\\
Stevens Institute of Technology\\
Hoboken NJ 07030 USA
}

\begin{abstract}
Dedicated to the memory of Edsger W.\ Dijkstra.

\vspace{\baselineskip}

\noindent Representation independence or relational parametricity formally
characterizes the encapsulation provided by language constructs for
data abstraction and justifies reasoning by simulation.
Representation independence has been shown for a variety of languages
and constructs but not for shared references to mutable state; indeed
it fails in general for such languages.  This paper formulates
representation independence for classes, in an imperative,
object-oriented language with pointers, subclassing and dynamic
dispatch, class oriented visibility control, recursive types and
methods, and a simple form of module.  An instance of a class is
considered to implement an abstraction using private fields and
so-called representation objects.  Encapsulation of representation
objects is expressed by a restriction, called confinement, on
aliasing.  Representation independence is proved for programs
satisfying the confinement condition.  A static analysis is given for
confinement that accepts common designs such as the observer and
factory patterns.  The formalization takes into account not only the
usual interface between a client and a class that provides an
abstraction but also the interface (often called ``protected'')
between the class and its subclasses.
\end{abstract}


\begin{document}
\begin{bottomstuff}
This is an expanded and revised version of a paper originally
appearing in \emph{ACM Symposium on Principles of Programming Languages}, 2002.\\
Banerjee was supported by NSF grants EIA-9806835, CCR-0209205 and NSF
Career award CCR-0093080/CCR-0296182.\\
Naumann was supported by NSF grants INT-9813854 and CCR-0208984 and 
a grant from the New Jersey Commission on Science and Technology.\\
Banerjee and Naumann were also supported by EPSRC grant GR/S03539,
``Abstraction, Confinement and Heap Storage''.
\end{bottomstuff}
\maketitle

{\small
\tableofcontents 
}

\section{Introduction}

You have implemented a class~\cite{DhalNygaard,JavaLang}, \code{FIFO},
whose instances are FIFO queues with public methods \code{enqueue} and
\code{dequeue} as well as method \code{size} that reports the
number of elements in the queue. The class, implemented in some Java-like
object-oriented language,  is part of a library and is used by many
programs, most unknown to you. The queue is \emph{represented} using 
a singly linked chain of nodes that point to elements of the queue.
There is also a sentinel node \cite{Cormen}. 
Each instance of \code{FIFO} has a field \code{num} with the number
of nodes and a field \code{snt} that references the sentinel.
You realize that a simpler, more efficient implementation can be
provided without the sentinel, using two fields, \code{head} and
\code{tail}, pointing to the end nodes in the chain. 
You revise method \code{size} to return \code{num} instead of
\code{num}$-1$ and revise the other methods suitably.  You are guided
to the necessary revisions by thinking about the correspondence,
sometimes called a \dt{simulation relation}, between the representations
for the two versions.   

Can the revisions affect the behavior of clients, that is, programs that
use class \code{FIFO} in some way or other? The answer would be yes, if some client
determined the number of nodes by reading field \code{num} directly.
A client that refers to field name \code{snt} would no longer compile.
But you have taken care to \emph{encapsulate} the queue's
representation: the fields are declared to be private.  
By using programming language constructs like private fields you aim
to ensure that client programs depend only on the \emph{abstraction}
provided by the class, not on its representation.   
If client behavior is independent from the representation of
\code{FIFO}, it is enough for you to ensure equivalent visible
behavior of the revised methods.

For scalable systems, scalable system-building tools, and scalable development methods,
abstraction is essential.  
For reasoning about a single component, e.g., a class, module, or local block, 
abstraction makes it possible to consider other components in terms of
their behavioral interface rather than their internal representation.\footnote{  
Even a primitive type like \textbf{int} is an abstraction from the
machine representation.}
Abstraction is needed for the automated reasoning embodied in static
analysis tools~\cite{CousotC77} and it is needed for formal and
informal reasoning about functional correctness during development and
evolution~\cite{Milner71,Hoare:data:rep}.   
Modular reasoning has always been a central issue in software engineering and in
static analysis. With the ascendancy of mobile code it has become
absolutely essential.  For example, it is possible for clients of
\code{FIFO} 
to be linked to it only at runtime, so it is impossible  to check all
uses to determine whether the revisions affect them.  

The need for flexible but robust encapsulation mechanisms to support
data abstraction has been one of the driving forces in the evolution of
programming language design, from type safety and scoped local
variables to module and abstract data type constructs \cite{LiskovGuttag}.
There is a rich theoretical literature on the subject (e.g., 
\cite{Plotkin73,Reynolds74,Donahue79,Haynes84,Reynolds84,refine:refine,Mitchell96,Lynch-Vaandrager,deRdataref}). 
Many different language constructs have been studied. There is
considerable variation in the details of these theories, partly
because the intended applications vary from justifying general tools
for program analysis and transformation to justifying proof rules to
be applied to specific programs as in the \code{FIFO} example.  The 
common thread is that two implementations of a component are linked by a 
simulation relation between the two representations. 

Unfortunately, these theories are inadequate for object-oriented
programs.  They deal well with the encapsulation of data structures
that correspond directly to some language construct, such as modules,
local variables, or private fields.  But the \code{FIFO} example also involves
encapsulation of a data structure composed of heap cells and
pointers, including aliasing with the \code{tail} field as depicted in
Fig.~\ref{fig:confFIFO}.

The problem is that encapsulation provided by language constructs
often runs afoul of aliasing.   
For variables and parameters, aliasing can be prevented through
syntactic restrictions that are tolerable in practice (and often 
assumed in formal logics and theories).  Aliasing via pointers is an
unavoidable problem in object oriented programming where shared
mutable objects are pervasive.  
Yet unintended aliasing can be catastrophic.  A version of the Java
access control system was rendered insecure because a leaked reference
to an internal data structure made it possible to forge crytographic
authentication~\cite{Vitek00}. 
In simply typed languages, types offer limited help: variables $x,y$ are not
aliased if they have different types.  Even this help is undercut by subclass
polymorphism: in Java, a variable $x$ of type \OBJECT\ can alias $y$
of any type.

The ubiquity and practical significance of the issue is articulated well in the
manifesto of~\citeN{hogg92geneva}. A number of subsequent papers in
the object-oriented programming literature propose disciplines to
control aliasing.
Of particular relevance are disciplines that impose some form of
\emph{ownership confinement} that restricts access to designated
``representation objects'' except via their  ``owners'', to 
prevent  \emph{representation exposure} \cite{SRC160toplas}. 
A good survey on confinement, especially ownership, can be found in the
dissertation of~\citeN{ClarkeDiss}; see also
\citeN{JavaConcur},\citeN{Vitek00}, \citeN{NoblePotter},
\citeN{FTfJPmuller}, \citeN{BoylandBury},
\citeN{Aldrich02}, and the related work section of this paper.

In Figure~\ref{fig:confFIFO}, an instance of class \code{FIFO} (the
owner) uses private fields to point to objects intended to be part of
its encapsulated representation, as indicated by the dashed rectangle. 
\begin{narrowfig}{18em} 
  \begin{center}
\psfrag{xxFIFO}{\code{FIFO}}
\psfrag{xxUser}{\code{User}}
\psfrag{xxElement}{\code{Element}}
\psfrag{xxNum}{\code{num 4}}
\psfrag{xxTail}{\code{tail}}
\psfrag{xxHead}{\code{head}}
\includegraphics{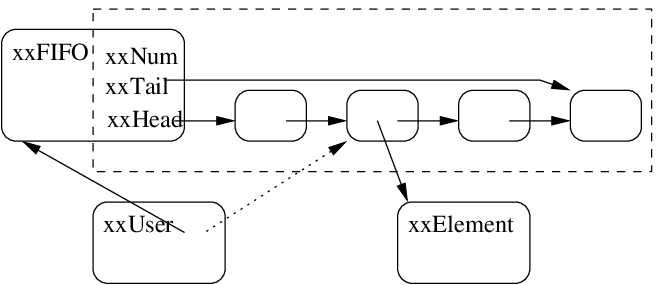}     
  \end{center}
    \caption{A \code{FIFO} object with its encapsulated
      representation: private fields and nodes of a list (within the
      dashed rectangle).  One element of the queue is shown as well as
      a user of the queue, but other objects and references are
      omitted.
      The dotted reference is an example of representation exposure.
}
    \label{fig:confFIFO}
\end{narrowfig}   
The contribution of this paper is a theory of representation independence
for encapsulation of data in the heap, using ownership confinement. 
We follow~\citeN{Reynolds84} in calling our main result an
\emph{abstraction theorem}.  Some readers may prefer the term
\dt{relational parametricity}.

The literature on confinement is largely concerned with static or
dynamic checks to ensure invariance of various confinement properties.
One of our contributions is to show how established semantic
techniques can be used to evaluate confinement disciplines.
To prove our abstraction theorem, we use a semantic formulation of confinement.
Separately, we give a modular, syntax-directed static analysis for
confinement and show that it accepts some interesting example programs that
embody important object-oriented design patterns. 

There are a number of ways in which abstractions can be expressed
using constructs of contemporary object-oriented languages,
including  modules, classes, local variables, object instances, not to
mention heap structures such as object groups.
We treat the most common situation: an instance of some class
is viewed as representing an abstraction, possibly using some other
objects as part of its representation.    

We are aware of no previous results on representation independence
that address encapsulation of objects in the heap.  Thus it is
tempting to present the ideas in the setting of a 
simple idealized language, say a simple imperative language with
pointers to mutable heap cells.  But this would leave open some
challenging issues, such as how class-based scoping rules fit with
instance-based abstraction.
We have chosen to consider a rich imperative object-oriented
language with class-based visibility, inheritance and dynamic binding,
type casts and tests, recursive types, and other features sufficient
for programs that fit common 
design patterns such as observer and factory \cite{DesPat}.

Previous work on representation independence has been concerned with
relating two versions of a component with respect to programs that use
the component.  But the designer of a class needs to consider not only
users (the client interface) but also subclasses (the \emph{protected
  interface}).  This is a source of complication in our treatment
of confinement and, to a lesser extent, in our treatment of
representation independence.  
Our results consider replacement of one version of a class by another
with the same public interface, in the context of arbitrary classes
that use it or are subclasses of it.

\paragraph*{Overview and readmap}

Sect.~\ref{sect:rif} introduces the language for which our results
are proved and  describes a simple example with which we review
the formalization of representation independence using simulation
relations.  The example is extended to one showing how
representation independence can be invalidated 
by leaked references to representation objects. 
The section concludes with an informal statement of our abstraction theorem.

Sect.~\ref{sect:rifa} discusses more elaborate examples that typify
object-oriented programs.  A version of a
Meyer-Sieber~\cite{Meyer:Sieber} example shows how higher order
programs can be expressed.  
Versions of the observer pattern \cite{DesPat} illustrate  challenges in formulating 
robust but practical notions of confinement.  The section concludes
with an informal description of our notion of ownership confinement.

Sect.~\ref{sect:syn} formalizes the syntax and typing rules.
Sect.~\ref{sect:sem} gives a surprisingly simple denotational semantics
in the manner of~\citeN{Strachey}.   
Confinement, the semantic notion, is defined formally in Sect.~\ref{sect:conf}.
Sect.~\ref{sect:abs} gives the first main result, an abstraction
theorem for confined programs.
Sect.~\ref{sect:app} shows in detail how the theorem applies to the examples in
Sect.~\ref{sect:rifa} and to further variations on the observer pattern.
Sect.~\ref{sect:osub} considers examples of the interface between
an owner class and its subclasses.  To achieve a sufficiently flexible form
of confinement for subclasses of the owner class, we add a simple
module construct to the language. 
Sect.~\ref{sect:sabs} proves a second abstraction theorem, for this extended
language and for a generalized notion of simulation needed for owner subclasses.
Sect.~\ref{sect:static} wraps up the technical development by
defining a static analysis for confinement that accepts 
the examples of Sections~\ref{sect:rif}, \ref{sect:rifa},
\ref{sect:app}, and \ref{sect:osub};   
soundness with respect to (semantic) confinement is shown.  
Sect.~\ref{sect:disc} discusses related work and open challenges.

Detailed proofs are given, as the complexity of similar
languages has led to errors in published proofs, e.g., of type soundness.
Appendices [to be put on line but not in print] give some additional proofs
and executable Java code for all the examples.

The organization of the paper is intended to make it possible for the
casual reader to skip some technical material and still get the gist of the
results.  
Readers who wish to study the details may still prefer to
skip, on a first reading, material concerning object constructors and
proofs that involve fixpoints and inheritance.

\paragraph*{Differences from the preliminary version}

Outgoing references, from representation objects to client objects,
were disallowed in the preliminary version of this paper
\cite{popl02}. We conjectured that they could be allowed if
restricted to read-only access as in~\cite{FTfJPmuller,SRC160toplas}.   
Here we allow them without restriction, as is needed to handle
examples such as the observer pattern where observers may well change
state in response to events.
We have also added constructors to the language, at the cost
of some complexity in proofs due to the interdependence of semantics
for commands and for constructors.  The benefit is succinct
formulation of an abstraction theorem sufficient for transparent
application to realistic examples.  
The other major additions are as follows:
module-scoped methods, the generalized abstraction theorem, substantial worked
examples, and the static analysis for confinement.  

In~\cite{popl02} we discuss simulation proofs of the equivalence of
``security passing style''~\cite{WallachAF00}  with the lazy ``stack
inspection'' implementation of Java's privilege-based access control
mechanism~\cite{Gong99}, and then extend our language to include access
control.  We give an abstraction theorem for this extended language. 
It was this study that led us to the main results but in
retrospect it seems tangential and is omitted.
  

\section{Representation independence}\label{sect:rif}

We begin this expository section with a very simple example of representation independence,
contrived mainly to introduce the Java-like language that we will use.
Building on this example we show how pointer aliasing can invalidate
representation-independence.  We conclude with an 
informal statement of the main results.    
Sect.~\ref{sect:rifa} deals with more challenging
examples including the observer pattern~\cite{DesPat} and gives a more
precise description of ownership confinement.

\subsection{A first  example}\label{sect:bool}

The concrete syntax for classes is based on that of Java
\cite{JavaLang} but using more conventional notation for simple
imperative constructs.  Keywords are 
typeset in bold font and comments are preceded by double slash.
A program consists of a collection of class declarations like the following
one.  

\pagebreak 

\begin{quote}\sf
\begin{tabbing}
\CLASS\ Bool \EXT\ \OBJECT\ \{  \\
\hspace*{1em}\= \BOOL\ f;  \hspace{10em}                  \= // \textrm{private field} \+ \\
                \CON\{ \SKIP\  \}                         \> // \textrm{public constructor} \\
                \UNIT\ set(\BOOL\ x)\{ \self.f\assym x \} \> // \textrm{public method}   \\
                \BOOL\ get()\{ \result\assym \self.f\ \}  \>// \textrm{public method} \-\\
\}
\end{tabbing}
\end{quote}
There are two associated methods:  \code{set} takes a boolean
parameter and returns nothing; 
\code{get} takes no parameter and returns a boolean value.
Methods are considered to be public, that is, visible to methods in all classes.
(Module-scoped methods are added in Sect.~\ref{sect:sabs}.)  
Every method has a return type; the primitive type \UNIT, with only a
single value ($\ITT$),  corresponds to Java's ``void'' and is used for methods
like \code{set} that are called only for their effect on state.    

Instances of class \code{Bool} have a field \code{f} of (primitive)
type \code{bool}. 
A field \code{f} is accessed in an expression of the form \code{e.f}, and
in particular  \code{\self.f} is used for fields of the current object; a
bare identifier like \code{x} is either a parameter or a local
variable.   
The distinguished variable $\result$ provides the return value; it is
initialized with the default for its type ($\semfalse$ for
$\BOOL$ and $\NIL$ for class types). 
Fields are considered to be private, that is,
visible only to the methods declared in the class. 
Visibility is class-based, as in many mainstream
object-oriented languages: an object can directly access the private fields of
another object of the same class.

When a new object is constructed, each field is initialized with the default value for
its type.  Then the constructor commands are executed: the
constructors declared in superclasses are executed before the declared one
which is designated by keyword \CON.
We refrain from considering constructors with parameters. 
In subsequent examples we omit the constructor if it is \SKIP. 

The observable behavior of a \code{Bool} object can be achieved using
an alternate implementation in which the complement is stored in a field:
\begin{quote}\sf
\begin{tabbing}
\CLASS\ Bool \EXT\ \OBJECT\  \{ \\
\hspace*{1em}\= \BOOL\ f; \+ \\
                \CON\{ \self.f\assym \TRUE\ \} \\
                \UNIT\ set(\BOOL\ x)\{ \self.f\assym $\neg$x \}    \\
                \BOOL\ get()\{ \result\assym $\neg$(\self.f)\ \} \}
\end{tabbing}
\end{quote}
We do not formalize class types (``interfaces'' in Java) separately
from class declarations.  Class names are used as types
and we use the term \dt{class} loosely to mean  
the name of a declared class.
But we are concerned with relating comparable versions of a class: 
as in the example above, a comparable version has the same name and
methods with the same names and signatures.   

We claim that no client program using \code{Bool} can distinguish one
implementation from the other; thus we are free to replace one by the other.   
Of course this is not the case if we consider aspects of client
behavior such as real time or the size of object code ---but these are
not at the level of abstraction of source code. 
Moreover, input and output for end users is of some limited type like
\textbf{int} or \code{String}. If a \code{Bool} could be output
directly, say displayed in binary on the screen, then an end user
could distinguish between the implementations.
So we consider only clients that use \code{Bool} objects in temporary
data structures and not as input or output data.  

An example of such a client is method \code{main} in the
following class.
It declares a local variable \code{b} of type \code{Bool}, with scope 
beginning at the keyword \IN.  In the absence of explicit braces,
the scope of a local variable extends to the end of the method body.
\begin{quote}\sf
\begin{tabbing}
\CLASS\ Main\ \EXT\ Object \{ \\
\hspace*{1em}\=  String inout; 
\+ \\
\UNIT\ main()\{ \= 
Bool b\assym \NEW\ Bool \IN\   \+ \\ 
\hspace*{1em}\=  
  \IF\ \ldots\self.inout\ldots \=\THEN\
      b.set(true) 
      \ELSE\ b.set(false)  \FI; \+ \\
      \self.inout\assym convertToString(b.get()) \} \}
\end{tabbing}
\end{quote}
We may consider method \code{main} as a main program for which the observable state
consists of field \code{inout}.  Its final value depends on some
condition ``\code{\ldots\self.inout\ldots}'' on its initial value.
No object of type \code{Bool} is reachable in the state of a \code{Main}
object after invocation of \code{main}, so there is no observable
difference between its behavior using one implementation of
\code{Bool} and its behavior using the other.

The claim is that we need not consider specific clients; there is no
use of \code{Bool} that can distinguish between the two implementations.
The standard reasoning goes as follows.
\begin{enumerate}
\item\label{rulea} Suppose \code{o} is an object of type \code{Bool} for the first
  implementation and $\code{o}'$ an object for the second.  The
  correspondence between their states is described by the
  \dt{basic coupling relation} 
\[ \code{o.f} = \neg(\code{o}'.\code{f}) \epunc . \]

\item
This relation has the \dt{simulation property}: 
  \begin{itemize}
  \item it holds   initially (once the constructor has been executed),
    and 
  \item if the two versions of \code{set} (respectively, \code{get}) are
  executed from related states  then the outcomes are related.  
(As we consider sequential programs, the outcome is the updated heap
and the return value if any.)
  \end{itemize} 
In short, the relation is established by the constructor and
preserved by the methods of \code{Bool}.

\item\label{rulec} To consider client programs we must consider program states
  consisting of local variables (and parameters) along with the heap,
  which may contain many instances of \code{Bool} as well as other objects.
  For states, we define the \dt{induced coupling relation}. Primitive values
  and locations are related by equality.\footnote{Later we refine this point.}
  A pair of heaps are related if
  there is a one-to-one correspondence between \code{Bool} objects such that they
  are pairwise related by the basic coupling of (1), and everything else is
  related by equality.   

The induced coupling relation is preserved by all commands in methods of all
classes.  This is the \dt{abstraction theorem}. 

\item\label{ruled} For a pair of states related by the (induced) coupling, if no \code{Bool}
  objects are reachable then the states are equal.   This fact, known
  as the \dt{identity extension lemma}, holds by definition of the
  induced coupling.
\end{enumerate}
It is a consequence of (\ref{rulec}) and (\ref{ruled}) that the two implementations cannot
be distinguished by a client that does not input or output \code{Bool}
objects. Any initial state for such a client is related to itself, by
(\ref{ruled}).  We can consider an execution of the client using either of the
two implementations of \code{Bool}; the final states are related,
according to (\ref{rulec}).  And thus they are equal, by (\ref{ruled}).  

Identity extension confirms that the chosen notion of
coupling relation is suited to the chosen form of encapsulation.
(Here, encapsulation means private fields and objects not input or output.) 
It is typically a straightforward consequence of the definitions.

For program refinement, identity can be replaced by inequality in step
(\ref{ruled}).  In this paper we do not emphasize  refinement, but the
requisite adaptation of our results is straightforward. 
For applications in program analysis, other relations are used in step
(\ref{ruled}), e.g., for secure information flow the relation
expresses equivalence from the point of low-security observers
\cite{VolpanoSI96}.\footnote{Our formulation of the abstraction
  theorem can be applied directly to prove  command and class
  equivalences for a specific program.  For applications of simulation
  in static analysis, the problem is usually to show that a syntax
  directed system of types and effects approximates some property like
  secure information flow, for all programs in a language.   We have
  not attempted to formulate an abstraction theorem general enough to
  apply directly in such analyses; they use analysis-specific typing
  systems rather than the language's own types and syntax.  But the
  essence of our result is that the language is relationally
  parametric, given suitable confinement conditions.   Indeed, in work
  subsequent to this paper, \citeN{csfw02} use the same language and
  semantic model for a relational analysis of secure information
  flow.} 

The abstraction theorem is a non-trivial property of the language.
It would fail, for example, if the language had constructs that
allowed client programs to read the private fields of
\code{Bool} ---or to enumerate the names of the private fields, or to
query the number of boolean fields that are currently true.
Such operations would be considered strange indeed.  

Familiar operations on pointers, however, can also violate
abstraction.  For example, with pointer arithmetic one can distinguish
between two representations that differ only in the size of storage
used (e.g., representing a boolean value using one bit of an integer
versus one bit of a character).  Even in the absence of pointer
arithmetic, shared references lead to the following problem.

\subsection{Representation exposure}\label{sect:repe}

Consider the following class \code{OBool} which provides functionality
similar to that of \code{Bool}, in fact using \code{Bool}.
For clarity we have chosen different method names, to emphasize that
we are not comparing this class with \code{Bool}.
\begin{quote}\sf
\begin{tabbing}
\CLASS\ OBool \EXT\ \OBJECT\  \{ \\
\hspace*{1em}\= Bool g; \+ \\
\UNIT\ init()\{ \self.g\assym \NEW\ Bool; \self.g.set(true) \}    \\
\UNIT\ setg(\BOOL\ x)\{ \self.g.set(x)\ \}    \\
\BOOL\ getg()\{ \result\assym \self.g.get()\ \} \}
\end{tabbing}
\end{quote}
To simplify the formal development,  we sidestep the complicated
interactions between subclassing and method calls in constructors 
by confining attention to constructors without parameters or method
calls.
In cases where this is inadequate, an ordinary method can be used
(like \code{init} in this example). 


Here is an alternate implementation of \code{OBool}.
\begin{quote}\sf
\begin{tabbing}
\CLASS\ OBool \EXT\ \OBJECT\  \{ \\
\hspace*{1em}\= Bool g; \+ \\
                \UNIT\ init()\{ \self.g\assym \NEW\ Bool; \self.g.set(false) \}    \\
                \UNIT\ setg(\BOOL\ x)\{ \self.g.set($\neg$ x)\ \}    \\
                \BOOL\ getg()\{ \result\assym $\neg$(\self.g.get())\ \} \}
\end{tabbing}
\end{quote}
To describe the connection between the two implementations 
a suitable basic coupling (recall (\ref{rulea}) in Sect.~\ref{sect:bool}) 
is the following relation between an object state \code{o} for the first
implementation of \code{OBool} and $\code{o}'$ for the alternate one:
\begin{trivlist}
\item \hfill 
$(\code{o.g} = \NIL = \code{o}'.\code{g}) \lor (\code{o.g} \neq \NIL \neq
\code{o}'.\code{g} \land \code{o.g.f} = \neg(\code{o}'.\code{g.f})) $
\epunc.
\hfill $(*)$ 
\end{trivlist}
If $\code{o}$ and $\code{o}'$ are newly constructed, the first disjunct
holds; method 
\code{init} establishes the second disjunct.
Invocations of \code{setg} and \code{getg}  maintain the
relation: From related initial states, either both abort (due
to dereferencing $\NIL$ because \code{init} has not been called) or both
terminate in related states.  

For these implementations, it is not just a private field that is to be
encapsulated, but also the object referenced by that field.  This is
apparent in the coupling $(*)$ which involves both. 
To describe the roles of the objects involved, we call class
\code{OBool} an \dt{owner} class. Its instances ``own'' objects of class
\code{Bool}, their representation objects, which are called \dt{reps} 
for short.  Together, an owner and its reps constitute what we call an
\dt{island} (cf.\ Fig.~\ref{fig:confFIFO}), following~\citeN{Hogg}.

Here is a suitable client for \code{OBool}.
\begin{quote}\sf
\begin{tabbing}
\CLASS\ Main\ \EXT\ Object \{ \\
\hspace*{1em}\=  String inout; 
\+ \\
\UNIT\ main()\{ \= 
OBool z\assym \NEW\ OBool \IN\  z.init();  \+ \\ 
  \IF\ \ldots\self.inout\ldots \=\THEN\
      z.setg(true) 
      \ELSE\ z.setg(false)  \FI; \\
      \self.inout\assym convertToString(z.getg()) \} \}
\end{tabbing}
\end{quote}
This does not distinguish between the two implementations of
\code{OBool} nor does it violate the intended encapsulation boundary. 

Suppose we add to both versions of \code{OBool} the following method
which ``leaks'' a reference to the rep object.
\begin{quote}\sf
  Bool bad()\{ \result\assym \self.g \}       
\end{quote}
The method gives its caller an alias to the object pointed to by the
private field \code{g}.  This makes the location of the encapsulated
object visible to clients.    
In and of itself, access to this location is not harmful.\footnote{
To make this clear, one could  assume that, for both versions of
\code{OBool}, the \code{Bool} object is allocated at the same location.
The assumption can be formalized by adding a conjunct 
$ \code{o.g} = \code{o}'.\code{g} $ to coupling $(*)$ and assuming
that method \code{init} preserves this equality.  It is then preserved
by all the methods of \code{OBool} including \code{bad}.
Another justification is given in Sect.~\ref{sect:sabs} where
we show formally how the language is ``parametric in locations''.}   
Like the other methods, method \code{bad} preserves $(*)$.
But a client class \code{C} can exploit the leak as in the following command. 
\begin{quote}\sf
\begin{tabbing}
OBool z\assym \NEW\ OBool \IN\ z.init(); \\ 
Bool\ w\assym z.bad() \IN\ 
\IF\ w.get() \THEN\ \SKIP\ \ELSE\ \ABORT\  \FI 
\end{tabbing}
\end{quote}
The command aborts if the new \code{OBool} is an object $\code{o}'$ for the
second implementation of \code{OBool}, 
but it does not abort for an object $\code{o}$ for the first implementation.  
The client command preserves the relation $(*)$, indeed  it does not
alter the state of the objects it accesses.  But the relation is not the
identity for the rep object states: we have 
$\code{o.g}=\code{o}'.\code{g}$ but 
$\code{o.g.f}$ is not equal to $\code{o}'.\code{g.f}$.
So the relation is not the identity for the client to which the reps
are visible.
An attempt to argue using the steps in Sect.~\ref{sect:bool} breaks
down because identity extension (\ref{ruled}) fails.

The abstraction theorem, step (\ref{rulec}), can also fail. Consider the following client
command.
\begin{quote}\sf
\begin{tabbing}
OBool z\assym \NEW\ OBool \IN\ z.init(); Bool\ w\assym z.bad() \IN\ w.set(true)
\end{tabbing}
\end{quote}
This does not preserve relation $(*)$. To see why, suppose 
$\code{o},\code{o}'$ are a related pair of \code{OBool} objects assigned
to \code{z} and satisfying $(*)$.  
After the assignment to \code{w}, the effect of
\code{w.set(true)} is to make $\code{o.g.f}=\code{o}'.\code{g.f}$, contrary to
the relation $(*)$.  This is very different from the effect of \code{z.setg(true)}. 

The examples show that both ingredients of representation independence
---identity extension and preservation--- can fail if a rep is leaked.
The challenge is to confine pointers in a way that
disallows harmful leaks and thus admits a robust representation
independence property ---without imposing 
impractical restrictions.  The challenge is made more difficult by
various features of Java-like languages, for example, type casts.  We
consider casts now; other challenges are deferred to Sect.~\ref{sect:rifa}.

Suppose we change the return type for method \code{bad}, attempting to hide
the type of the rep object.
\begin{quote}\sf
  \OBJECT\ bad()\{ \result\assym \self.g \}       
\end{quote}
Class \OBJECT\ is the root of the subclassing
hierarchy so by subsumption it allows references to objects of any class. 
The client can use a \code{(Bool)} cast to assert that the result  of
\code{z.bad()} has type \code{Bool}. (In a state where the assertion is
false, the cast would cause abortion.) 
\begin{quote}\sf
\begin{tabbing}
OBool z\assym \NEW\ OBool \IN\ z.init(); \\ 
Bool\ w\assym (Bool)(z.bad()) \IN\ 
\IF\ w.get() \THEN\ \SKIP\ \ELSE\ \ABORT\  \FI 
\end{tabbing}
\end{quote}
Again, the client is dependent on representation.

Note that the cast could not be used if the scope of class name \code{Bool}
did not include the client.  This suggests a focus on modules (``packages''
in Java) for confinement of pointers, as has been studied by
\citeN{Vitek00} among others (see Sect.~\ref{sect:disc}).  But in our
example the field has private 
scope, each rep is associated with a single owner, and the coupling relation
is expressed in terms of a single owner. 
Our results  account for this sort of instance-based encapsulation, which
is common in practice and which is similar to the value-oriented
notions used for representation independence in functional languages
\cite{Reynolds84,Mitchell86,Mitchell91}.

\subsection{Overview of results}\label{sect:ov}

In the examples above, class \code{OBool} is viewed as providing an
abstraction. It is just as sensible to consider \code{Bool} as providing an
abstraction for which \code{OBool} is a client. 
We do not annotate programs with a fixed designation of owners and reps.
Rather, we study how to reason about a class, say $Own$, one has
chosen to view as an abstraction with encapsulated representation.
Objects of any subclass of $Own$ are also considered to be owners.  
A second class, say $Rep$, is designated as the type of reps for
$Own$.  (In practice, $Rep$ could be an interface or class type; this
generalization  is straightforward but would complicate the
formalization.)

A complete program is a closed collection of class declarations,
called a \dt{class table}.
We consider an idealized Java-like language similar to the sequential
fragment of C++ (without pointer arithmetic), Modula-3, Oberon, C\#,
Eiffel, and other class-based languages.
It includes subclassing and  dynamic dispatch, class oriented visibility control,
recursive types and methods, type casts and tests (Java's
\texttt{instanceof}), and a simple form of module.  

Roughly speaking, a class table $CT$ is \dt{confined}, for $Own$ and $Rep$,
if all of its methods preserve confinement.  A confined heap
is one where the objects can be partitioned into some owner islands (recall
Fig.~\ref{fig:confFIFO}) along with a block of client objects as in
Fig.~\ref{fig:confineObs}.   Furthermore, there are no references from
clients to reps.   (We use the term \dt{client} for all objects except
owners and reps.)

Sect.~\ref{sect:rifa} discusses confinement in more detail and 
the formal definitions are the subject of Section~\ref{sect:conf}.  
The full significance of the definitions does not become clear until
Sect.~\ref{sect:osub} where we study subclasses of $Own$: an object of
such a type inherits the methods and private fields of $Own$, which
manipulate reps.  To be useful, owner subclasses must have some 
access to reps. On the other hand, full access cannot be granted; to 
do so would be to study not the class as unit of encapsulation but a
class together with its subclasses, which would be revised in concert.

Our objective is to compare versions of $Own$ that may use different
reps.  We say $CT$ and $CT'$ are \dt{comparable} if they
are identical except for having different versions of class $Own$, and
those two versions declare the same public methods.  
The two versions of $Own$ may well use different rep
classes, say $Rep$ and $Rep'$.  Without loss of generality, our
formalization has $Rep$ and $Rep'$ both present in $CT$ and
$CT'$.

An interesting question is how to formalize basic couplings, step
(\ref{rulea}) of the proof method outlined in Sect.~\ref{sect:bool}.
To allow useful data structures, we need to allow representations to
include pointers to client objects (e.g., elements of the queue in
Fig.~\ref{fig:confFIFO}).  But if the programmer is required 
to define a relation involving the state of objects outside the encapsulated data,
how can this be done in a modular way?  We have chosen to use relations on the
encapsulated state only. Put differently: those things on which a
coupling depends are considered as part of the island.  
Although other alternatives merit study, this one makes for 
transparent application of the formal results to interesting
examples (this is done in Sects.~\ref{sect:app} and \ref{sect:osub}).
Moreover, it is straightforward to define the induced coupling.

A \dt{basic coupling} is a relation between a pair of owner islands for
comparable $CT$ and $CT'$.  A simple example is given by  $(*)$ above
in Sect.~\ref{sect:repe}.
More interesting is the observer example, discussed in
Sect.~\ref{sect:rifa}, which uses a linked list of client objects (the
observers). In Fig.~\ref{fig:basicSimObs} on
page~\pageref{fig:basicSimObs}, a basic coupling is depicted in which
the observer objects occur as dangling pointers from the corresponding
islands.  The point is that both versions are manipulating the same
observer objects in the same way, including the invocation of
methods on those objects.  So the state of the observer objects is not
relevant in the basic coupling ---nor could it be, if the argument is
to be carried out in a modular way independent of the particular
clients.  

In a related pair of islands, both owners have the same class, which
may well be a proper subclass of $Own$. 

The induced \dt{coupling relation} for heaps relates $h$ to $h'$ just
if there are confining partitions for which corresponding islands are
pairwise related by the basic coupling.   Moreover, there is an exact 
correspondence between client objects in $h$ and $h'$.  Primitive
values are related by equality. Locations are related by an
arbitrary bijection.  

The induced relation is a \dt{simulation} if it is preserved by the
methods of class $Own$ in $CT$ and in $CT'$.  A method declared in one
version of $Own$ may be inherited in the other version; it is the
behavior of those methods that matters.

The abstraction theorem says that a simulation is preserved by all
methods of all classes, provided that both class tables are confined.
The identity extension lemma says that the induced relation is the
identity, after garbage collection, for client states in which no
owners are reachable.

Sect.~\ref{sect:abs} gives the formal definitions for couplings and
simulation in the special case where locations of objects other than
reps are related by equality.  The abstraction and identity extension
results are proved there
in detail.  Sect.~\ref{sect:sabs} generalizes the definitions to allow
an arbitrary bijection on locations; abstraction and
identity extension are proved for the general case.
The special case is of interest because it is adequate for some 
applications in program analysis (e.g., \cite{csfw02}) and for
non-trivial examples like those of Sect.~\ref{sect:rifa} (as shown
in Sect.~\ref{sect:app}).  Examples that require the general case are
given in Sect.~\ref{sect:osub}; they are subclasses of $Own$ that 
construct reps and pass them to methods of $Own$ as in the factory
pattern \cite{DesPat}.  
Notation is more complicated for the general case but the proofs are not
very different from the special case.

These results are proved in terms of a semantic formulation of
confinement; indeed, the details of this formulation come directly
from what is needed in the proofs.
Sect.~\ref{sect:static} gives a syntax-directed static analysis: typing 
rules that characterize \dt{safe} programs and a proof that safety
implies confinement (soundness).  Our objective is to round out the story 
by showing how confinement can be achieved in practice, not to give a
definitive treatment of static analyses.  But our analysis accepts many
natural examples and the constraints are clearly motivated in the proof of soundness.
The analysis is modular: It does not require code annotations and the only constraint 
it imposes on client programs is that they cannot manufacture representation objects.


\section{Ownership confinement}\label{sect:rifa}

This section considers two substantial examples of
representation-independence.  The first is an 
object-oriented version of an example given by \citeN{Meyer:Sieber}
as a challenge for semantics of Algol. 
It illustrates the expressiveness of object-oriented constructs,
specifically the use of \emph{callbacks} which go against the
hierarchical calling structure which typifies the simplest forms of
procedural and data abstraction.

The second example is an instance of the observer pattern
\cite{DesPat} which is widely used in object-oriented programs.  In
addition to callbacks it involves a non-trivial data structure and
outgoing references from representation objects to clients.  Note that
we use the term \emph{client} not just for objects that use an 
abstraction (by instantiating it or calling its methods) but for any
objects except instances of the abstraction of interest or its
encapsulated representation.

The section concludes with an overview of our semantic notion of confinement.

\subsection{Callbacks}\label{sect:mse}

\citeN{Meyer:Sieber} consider the following pair of Algol commands:

\medskip

\begin{sf} 
\noindent\hspace*{2em}
\keyword{var} n\assym 0; P(n\assym n+2); \IF\ n \keyword{mod} 2 = 0 \THEN\
\ABORT\ \ELSE\ \SKIP\ \FI \hfill $(*)$\medskip \\
\medskip
\hspace*{2em}
\keyword{var} n\assym 0; P(n\assym n+2); \ABORT\ \hfill ($\dagger$)\\
\end{sf}
Both invoke some procedure \code{P}, passing to it the command 
\code{n\assym  n+2} that acts on local variable \code{n}.  
(That is, \code{P} is passed a parameterless procedure whose calls have
the effect \code{n\assym  n+2}.)
For any \code{P}, the commands are equivalent.
The reason is that in the first example \code{n} is invariably even:
\code{P} is declared somewhere not in the scope of \code{n} so the variable
can only be affected by (possibly repeated) executions of \code{n\assym n+2} and
this maintains the invariant.  

The difficulty in formalizing this argument is due to the difficulty of
capturing the semantics of lexically scoped local variables and procedures
in a language where local variables can be free in procedures 
that can be passed as arguments to other procedures.
(It appears even more difficult, and remains an open problem, to cope with
assignment of such procedures to variables~\cite{OTbook}.)

Now we consider a Java-like adaptation of the example, due to Peter
O'Hearn.
In place of local variable \code{n}  
it uses a private field \code{g} in a class \code{A}. Instead of
passing the command \code{n\assym n+2} as argument, an \code{A}-object passes a reference to
itself; this  gives access to a public method \code{inc} that adds 2
to the field. 
\begin{quote}\sf
\begin{tabbing}
\CLASS\ A \EXT\ Object \{ \\
\hspace*{1em}\= \keyword{int} g; // (the default integer value is 0) \+ \\
  \UNIT\ callP(C y)\{ y.P(\self); 
                      \IF\ \self.g \keyword{mod} 2 = 0 
 \THEN\ \ABORT\  \ELSE\ \SKIP\  \FI\ \} \\
  \UNIT\ inc()\{ \self.g\assym \self.g + 2 \} \}
\end{tabbing}
\end{quote}
In the context of this class and some declaration of class \code{C} with
method \code{P}, the Algol command $(*)$ corresponds to the
command
\medskip

\begin{sf}
\noindent
\medskip
\hspace*{2em}
C y\assym \NEW\ C \IN\ A x\assym \NEW\ A \IN\ x.callP(y)  \hfill ($\ddagger$)\\
\end{sf}
This aborts because after calling \code{y.P}, method \code{callP} aborts.
The command ($\dagger$) also corresponds to ($\ddagger$) but in the
context of an alternative implementation of class \code{A}:
\begin{quote}\sf
\begin{tabbing}
\CLASS\ A \EXT\ Object \{ \\
\hspace*{1em}\= \keyword{int} g;  \+ \\
  \UNIT\ callP(C y)\{ y.P(\self); \ABORT\ \} \\
  \UNIT\ inc()\{ \self.g\assym \self.g + 2 \} \}
\end{tabbing}
\end{quote}
In Example~\ref{ex:ms},
we use the abstraction theorem to  prove equivalence of the two versions
using coupling relation
\[ \code{o.g} = \code{o}'.\code{g} \land \code{o.g}\mathbin{\keyword{mod}} 2 = 0 
\epunc .\]
This relation is preserved by arbitrary \code{P} because \code{P} can
affect the private field \code{g} only by calls to \code{inc}.

As~\citeN{REYNOLDS78B} shows (see also~\cite{Reddy-classes}),
instance-based object-oriented constructs can be 
expressed in Algol-like languages, but the latter are in some ways
significantly more powerful.  The Java version of the example can be seen
as giving an explicit closure to represent the command \code{n\assym
  n+2} in the form of method \code{inc}. 
Indeed the simplicity of the semantic model for our language can be
explained by saying the language is defunctionalized 
\cite{REYNOLDS72A,BanerjeeHR01}  and lacks true higher order constructs.
If the example is written in such a language, \code{P} ranges over more
limited procedures than in Algol. 
The root problem for Algol 
semantics~\cite{ReynoldsEssence,OHearn:Tennant} and proof rules~\cite{OlderogAchieve,Lfour}
is the interaction between arbitrary nesting of variable and procedure
declarations and possibility of passing procedures as arguments.  
In imperative languages like C and Modula-3, procedures can be passed as
arguments and even stored in variables, but only if their free variables
are in outermost scope.  
This restriction greatly simplifies implementation
of the language, and it suffices to admit simple but adequate semantic
models.\footnote{\citeN{sdr} uses such a model to prove an abstraction theorem and
apply it to Meyer-Sieber examples.  The simpler of their examples can
be proved directly in the model without use of simulations
\cite{semho}.}
The constructs of a Java-like language offer similar expressive power
and also admit simple models.  

The example also illustrates what are known as \dt{callbacks} in object-oriented programs.
When an \code{A}-object invokes \code{y.P(\self)} it passes a reference to itself, by
which \code{y} may invoke a method on the \code{A}-object which is in the middle of
executing method \code{callP} ---a callback to \code{A}.
If in $(\ddagger)$ we replace \code{x.callP(y)} by \code{x.callP(\self)},
and assume that $(\ddagger)$ is a constituent of a method of class \code{C},
then we get a callback to \code{C}.

The point of the Algol example is modular reasoning about $(*)$ and
$(\dagger)$ independent from the definition of \code{P}.  For the
object-oriented version we can also consider reasoning independent from
subclasses of \code{A}.  If instead of $(\ddagger)$ we consider a method 
\begin{quote}\sf 
\UNIT\ m(C y, A x)\{ x.callP(y) \}
\end{quote}
then there is the possibility that \code{m} is passed an argument \code{x}
of some subtype of \code{A} that overrides \code{inc}.  
By dynamic binding, the overriding implementation would be invoked by
\code{callP} and our 
reasoning above would no longer be sound. For modular reasoning, we could
require that any overriding declaration of \code{inc} must preserve the
intended invariant that \code{g} is even.  
To impose such a requirement ---and a corresponding one for \code{callP}---
is to require behavioral subclassing~\cite{LiskovWing,Dhara-Leavens96}.
One important application of simulations is in the formalization 
of behavioral subclassing but that is beyond the scope of this paper.

Unlike much work on reasoning about object-oriented programs, our results
do not depend on behavioral subclassing.  Representation independence holds
for clients and abstractions that do not exhibit behavioral subclassing
(see Sect.~\ref{ssect:beh}).  

\subsection{The observer pattern}\label{sect:sconf}

In this subsection we consider variations on an often-used design
known as the observer  pattern~\cite{DesPat} which involves a non-trivial
recursive data structure using multiple rep objects and outgoing
references to client objects.  Further variations are given in
Sect.~\ref{sect:app}.  

We focus attention on  the abstraction provided by an
\code{Observable} object (sometimes called the ``subject'').  It
maintains a list of so-called observers to be notified when some event occurs.  Its public method
\code{add} allows the addition of an observer object to the list.  The public method
\code{notifyAll} represents the event of interest; its effect is to invoke
method \code{notify} on each observer in the list. 
What \code{notify} does is not relevant, so long as it is
confined.\footnote{In Java, class \texttt{Object} declares methods \code{notify} and
\code{notifyAll}.  Here we assume that no superclass of
\code{Observer} declares \code{notify} and no superclass of
\code{Observable} declares \code{notifyAll}.  In the Java versions of
our examples we use different names.}

The abstraction involves a collection of objects, a well-worn example for data
representations.  Simple collections are essentially mutable sets of 
pointers to client objects. Testing whether a reference is in the set 
requires only pointer equality.  To facilitate lookup by key, and to
facilitate implementations like binary search trees, it may be necessary
for the abstraction to invoke a comparison method on the client objects in
the collection.  This is similar to the call to \code{notify} in the observer pattern.  

In the first version of the observer example, Fig.~\ref{fig:obs},
most of the work is done by the owner class \code{Observable}, which uses
rep class \code{Node} to store observers in a singly linked list.    
A more object-oriented version appears in Fig.~\ref{fig:obsc} of
Sect.~\ref{sect:app}; it exemplifies the use of class-based
visibility.    

\begin{figure}[t] 
  \begin{center}
\sf  \begin{tabbing}
\CLASS\ Observer \EXT\ Object \{   // ``abstract class'' to be overridden
in clients\\
\hspace*{1em}\= \UNIT\  notify()\{ \ABORT\  \} \}  \\
\\
\CLASS\ Node \EXT\ Object \{ // rep for Observable
\+ \\
  Observer ob; \\
  Node nxt; // next node in list\\
  \UNIT\  setOb(Observer o)\{ \self.ob\assym o \} \\
  \UNIT\  setNext(Node n)\{ \self.nxt\assym n \} \\
  Observer getOb()\{ \result\assym \self.ob \} \\
  Node getNext()\{ \result\assym \self.nxt \} \}  
\- \\
\\
\CLASS\ Observable \EXT\ Object \{ // owner 
\+ \\
  Node fst; // first node in list\\ 
  \UNIT\  add(Observer ob)\{ 
Node n\assym \NEW\  Node; n.setOb(ob); n.setNext(\self.fst); \self.fst\assym n \}
\\
  \UNIT\  notifyAll()\{ 
    Node n\assym \self.fst; \keyword{while} n $\neq$ \NULL\  \DO\
    n.getOb().notify(); n\assym n.getNext() \OD\ \} \}   
  \end{tabbing}
  \end{center}
\caption{First version of observer pattern, in procedural style.}
\label{fig:obs}
\end{figure}

Fig.~\ref{fig:obsCli} gives example client classes \code{AnObserver}
and \code{Main}.   Class \code{AnObserver} records notifications in its state.    
Method \code{main} constructs and initializes an \code{Observable},
installs an observer, and invokes \code{notifyAll};  
upon termination, $\code{ob.count} = 1$ and no \code{Observable} is reachable.  
\begin{figure}[t]
\sf\begin{tabbing}
\CLASS\ AnObserver \EXT\ Observer \{ \\
\hspace*{1em} \= int count;  \+ \\
     \UNIT\  notify()\{ \self.count\assym \self.count+1 \} \} 
\- \\
\\
\CLASS\ Main \EXT\ Object \{ \+ \\
   AnObserver ob; \\ 
  \UNIT\  main()\{  \\  
\hspace*{1em} ob\assym \NEW\  AnObserver; Observable obl\assym \NEW\  Observable; 
   obl.add(ob); obl.notifyAll() \} \} 
\end{tabbing}
  \caption{Example client for \code{Observable}.}
  \label{fig:obsCli}
\end{figure}

Fig.~\ref{fig:obsaa} gives another version of \code{Observable}, using
a sentinel node \cite{Cormen}, for the sake of an example.  A more
compelling use of sentinels is the version of Fig.~\ref{fig:obsca} (in
Sect.~\ref{sect:app}), which also uses subclassing and dynamic dispatch.

\begin{figure}[t] 
\sf\begin{tabbing}
\CLASS\ Node2 \EXT\ Object \{ // rep for Observable \\
\hspace*{1em} \= Observer ob; \+ \\
  Node2 nxt; \\
  \UNIT\ setOb(Observer o)\{ \self.ob\assym o \} \\
  \UNIT\  setNext(Node2 n)\{ \self.nxt\assym n \} \\
  Observer getOb()\{ \result\assym \self.ob \} \\
  Node2 getNext()\{ \result\assym \self.nxt \} \}  
\- \\

\CLASS\ Observable \EXT\ Object // owner \{ 
\+ \\
  Node2 snt; // sentinel node pointing to list \\
  \CON\{ \self.snt\assym \NEW\ Node2 \} \\
  \UNIT\ add(Observer ob)\{ \\
\hspace*{1em} \=    
Node2 n\assym new Node2; n.setOb(ob); n.setNext(\self.snt.getNext()); \self.snt.setNext(n); 
\}
 \\
  \UNIT\ notifyAll()\{  \+ \\
    Node2 n\assym \self.snt.getNext(); \keyword{while} n $\neq$ \NULL\  \DO\
    n.getOb().notify(); n\assym n.getNext() \OD\ \} \}   
\end{tabbing}
\caption{Version of observable that uses sentinel node, in procedural style}
\label{fig:obsaa}
\end{figure}

In Sect.~\ref{sect:app} we show equivalence of the versions of Figs.~\ref{fig:obs} 
and~\ref{fig:obsaa} as an application of the abstraction theorem and
identity extension. 
The coupling relation describes the correspondence between a pair of
lists, one with and one without a 
sentinel node (see Fig.~\ref{fig:basicSimObs}).
It is enough to say that the same \code{Observer} locations are stored in
the lists, in the same order.  The state of the \code{Observer} is not
relevant ---nor could it be in a modular treatment, as class
\code{Observer} has no fields.  To reason about outgoing calls, namely to
\code{notify}, it is enough to show that the two implementations make the
same calls.  Those calls may lead to calls back to the \code{Observable}, but
encapsulation ensures that those calls are the only way the behavior of
\code{notify} can depend on, or affect, the \code{Observable}.

Except for the \code{bad} method of Sect.~\ref{sect:bool}, all of
the examples discussed so far satisfy the confinement conditions discussed next.

\subsection{Confinement}\label{ssect:conf}

We need a notion of confinement to prevent representation exposures
that invalidate simulation-based reasoning, as discussed in
Sect.~\ref{sect:bool}.    A related issue is how to 
formulate simulation.  In all the examples, our discussion centered on a
corresponding pair of instances for two implementations of the owner class.
In particular, the coupling relations are described for a pair of
instances as discussed in Sect.~\ref{sect:ov}.   
A class- or module-based notion of confinement might rule out
leaks, but we aim for an instance-based notion of simulation suited to the
kind of examples we have discussed.  These involve an abstraction provided by a
single instance (the owner object) using a representation accessed via its
private fields.
So we need to prevent problematic sharing not only between client and
owner but also between different instances of the owner class.

Fig.~\ref{fig:confineObs} illustrates instance-based owner confinement;
in this case \code{Node}s are confined to their owning \code{Observable}.  
\begin{narrowfig}{23em} 
  \begin{center}
\psfrag{xxMain}{\code{Main}}
\psfrag{xxObserver}{\code{Observer}}
\psfrag{xxObservable}{\code{Observable}}
\psfrag{xxNode}{\code{Node}}
\includegraphics{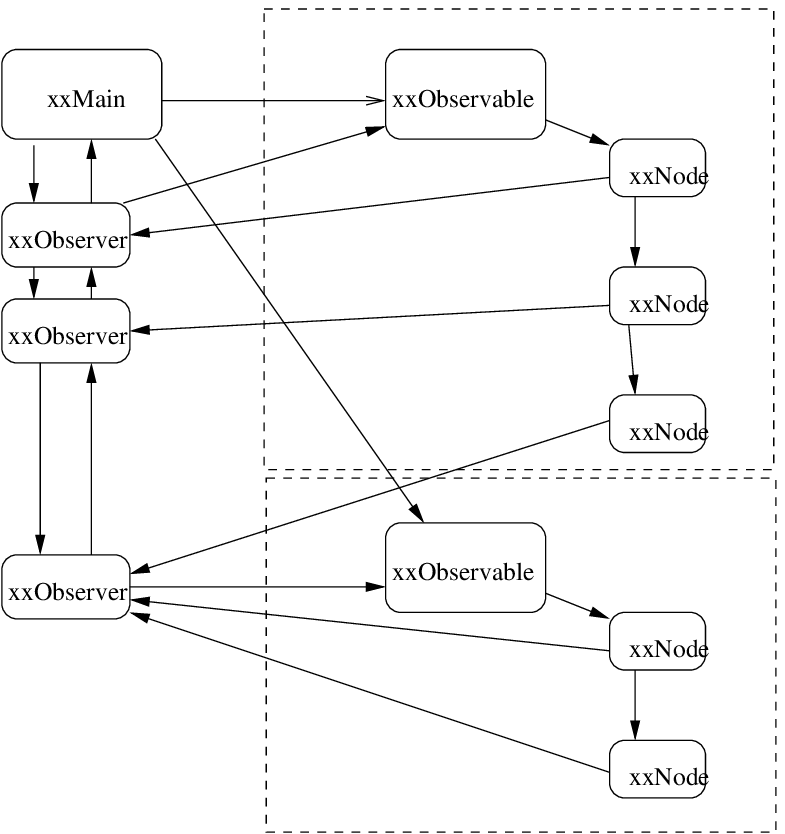}     
  \end{center}
    \caption{Confinement example.  Rounded boxes are instances of the
      indicated class.  Solid arrows represent allowed pointers.  Dashed boxes
      indicate owner islands, each consisting of one owner and its reps.}
    \label{fig:confineObs}
\end{narrowfig}   
Following~\citeN{Hogg}, we use the term \dt{island} for the sub-heap
consisting of an owner and its reps.  Dashed lines in the Figure depict 
two islands.  
Our notion of owner confinement imposes four conditions on
islands; here are the first three:
\begin{enumerate}
\item there are no references from a client object to a rep;
\item there are no references from an owner to reps in a different island;
\item there are no references from a rep into a different island.
\end{enumerate}
The Figure exhibits most allowed references, but we also allow an owner to
reference another owner (see Fig.~\ref{fig:part} on page
\pageref{fig:part}).
An example is given in Sect.~\ref{ssect:osm}.
Note that heap confinement is a state predicate.  
The full definition, formalized in Sect.~\ref{sect:conf}, 
deals with preservation of this predicate by commands and also with leaks
via parameter passing in outgoing method calls from island to client.

In class-based languages with inheritance, there is a subclass (or ``protected'')
interface in addition to the public one. This raises the possibility of
expressing encapsulation of reps for not only (instances of) the owner
class but also its subclasses.  We have chosen the alternative that
subclasses are like clients in that  fields they declare may \emph{not} point to
reps. To the list of conditions above we add: 
\begin{itemize}
\item[(4)]   references from an owner's fields to its reps are only in
  the private fields of the owner class.
\end{itemize}
In order not to abandon the expressiveness of subclassing, however, we
allow subclass methods to manipulate reps: they may be constructed, stored
in local variables, and passed to the owner.
This fits well with the factory
pattern~\cite{DesPat} which allows owner behavior to be adapted in owner subclasses
without violating encapsulation.  To balance the paper, we have deferred 
the relevant examples to Sect.~\ref{sect:app}.

Confinement is formulated using class names.  Two incomparable class names,
$Own$ and $Rep$, are designated.  An object is considered to be an owner
(respectively, a rep) if its type is $Own$ (resp.\ $Rep$) or a subtype
thereof.
Incomparability is a mild restriction that enforces a widely-followed 
discipline of distinguishing between rep objects (e.g., nodes in a linked
list) and objects representing abstractions (e.g., a list).
The technical benefit of incomparability is that if $C$ and $D$ are
incomparable, which we write $C\ncomp D$, then an expression of type
$C$ never has a value of type $D$.

We aim for a perspicuous separation between the semantic property
needed for the abstraction theorem and the syntactic conditions used
for static analysis.  The ``semantic'' property in fact includes
conditions on method signatures.
For example, we impose the restriction that the return type of a public owner method
is incomparable to $Rep$; this disallows method \code{bad} of
Sect.~\ref{sect:bool}.

Our use of types to formulate alias restrictions allows heterogeneous data
structures, but is slightly restrictive in that there is a single common
superclass for all reps.   For more flexibility in practical applications,
our theory could be adapted by taking $Own$ and $Rep$ to be ``class types''
(``interfaces'' in Java),  rather than class implementations.
The generalization is straightforward and not illuminating.

The more substantial restriction is due to the fact that class \OBJECT\ is
comparable to all classes.  Because Java lacks parametric polymorphism,
\OBJECT\ is often used to express generics, e.g., a list containing elements
of arbitrary type. 
A method to enumerate the list would have return type \OBJECT, which violates our
restriction on owner methods.  
This restriction could be dropped in favor of more sophisticated conditions to ensure
that no rep is returned (see  Sect.~\ref{sect:disc}). But in practice many
generics have some sort of constraint expressed by a class or interface type ---like
\code{Observer} in our examples, or \code{Comparable} for data structures that depend on an
ordering.  These do not run afoul of our restriction. 
In any case, the use of \OBJECT\ for generics   
is widely deplored because it undercuts the benefits of typing; parametric types
are clearly preferable.   

Some works on confinement have considered all the confinement
properties intended to be satisfied by a program, using hierarchical
notions of ownership \cite{NoblePotter,Mueller01}.  
For example, a \code{Set} could own the 
header of a list which in turn owns the nodes of the list.  
This is not necessary for our purposes (see Sect.~\ref{sect:disc}).  
To analyse the abstraction provided by the set, we would consider both
the header and nodes to be reps, with a common superclass $Rep$.   
On the other hand, to replace one header implementation by another, 
\code{Set} is irrelevant; we choose $Own$ to be the header and $Rep$
for the nodes.


\section{Syntax}\label{sect:syn}

This section formalizes the language, for which purpose we adapt some
notations from Featherweight Java~\cite{Featherweight}.\footnote{But the languages
  differ, e.g., ours has imperative features and private fields.}
To avoid burdening the reader with straightforward technicalities 
we deliberately confuse surface syntax with abstract syntax. 
We do not distinguish between classes and class types.
We confuse syntactic categories with names of their typical elements.
Barred identifiers like $\ol{\T}$ indicate finite lists, e.g., $\ol{\T}\:\ol{f}$
stands for a list $\ol{f}$ of field names with corresponding types $\ol{\T}$.  
The bar has no semantic import; $\ol{\T}$ has nothing to do with $\T$.

The grammar is based on given sets of class names (with typical element
$C$), field names ($f$), method names ($m$), and names ($x$) for parameters 
and local variables.  In most respects $\self$ and $\result$ are like
any other variables  but $\self$ cannot be the target of assignment.  

\begin{samepage}
\begin{infig}{Grammar}
$\begin{array}{l@{\hspace{-.07em}}cl@{\hspace{-.3em}}l}
\T &\gassym & \BOOL \mid \UNIT \mid  C 
&\mbox{data type} \\[.7ex]
\CL&\gassym& \class{C}{C}{\ol{\T}\;\ol{f};\;\construct{\S}\;\ol{\M}}
\quad
&\mbox{class declaration}\\[.7ex]
\M &\gassym& \T\; m(\ol{\T}\;\ol{x})\;\{\S \}
&\mbox{method declaration}\\[.7ex]
\S &\gassym& \assign{x}{e}\mid \fassign{e}{f}{e} 
&\mbox{assign to variable, to field}\\[.7ex]
& \mid & \assign{x}{\new{C}{}} 
&\mbox{object construction}\\[.7ex]
& \mid & \assign{x}{\mcall{e}{m}{\ol{e}}} \mid 
\assign{x}{\mcall{\SUPER}{m}{\ol{e}}} &\mbox{method calls}\\[.7ex]
& \mid & \var{\T\; x}{e}{\S} &\mbox{local variable block}\\[.7ex]
& \mid &  \ifelse{e}{\S}{\S} \mid \seq{\S}{\S} 
&\mbox{conditional, sequence}\\[.7ex]
e  &\gassym& x \mid \NULL \mid \TRUE \mid \FALSE \mid \ITT & \mbox{variable,
  constant}\\[.7ex]
&\mid& \faccess{e}{f} \mid \eqtest{e}{e} 
&\mbox{field access, equality test}\\[.7ex]
&\mid& \is{e}{C} \mid \cast{C}{e} & \mbox{type test, cast} 
\end{array}$
\end{infig}
\end{samepage}

Without formalizing it precisely, we assume there is a class $\OBJECT$
with no fields or methods which can be used as a superclass. 
Additional primitive types, such as integers, can be treated in the same way as
$\BOOL$ and $\UNIT$ (integers can also be represented, e.g., in unary
using linked lists).

In the formal language, expressions do not have side effects. 
Object construction, \NEW, occurs only as a command $\assign{x}{\new{C}{}}$ 
that assigns to a local variable.
Method calls are not expressions but rather occur in special assignments
$\assign{x}{\mcall{e}{m}{\ol{e}}}$ to allow both heap effects and a return
value.

\begin{remark}[syntactic sugar]\label{rem:sugar}
In examples we use several abbreviations:
  \begin{itemize}
  \item A method call command $\mcall{e}{m}{\ol{e}}$, e.g.,
    $\code{\self.g.set(true)}$, abbreviates a call assigning to an
    otherwise unused local variable.
\item Assignment of a new object to a field abbreviates a local block 
assigning the new object to a variable that is then assigned to the field.
\item Methods that return values but do not mutate state are used in
  expressions, e.g., the argument in $\code{\self.inout\assym
    convertToString(z.getg())}$ and the target object in 
  $\code{n.getOb().notify()}$.  These are easily desugared using fresh
  variables and suitable assignments.
 \end{itemize}
As the language has general recursion, we omit loops.  
For desugaring loops it would be  convenient to have local or
private method declarations, but the module-scoped methods added in
Sect.~\ref{sect:sabs} suffice.
The issue is discussed in Sect.~\ref{ssect:peq}.\qed
\end{remark}

A program is given as a \dt{class table} $CT$, a finite partial function sending
class name $C$ to its declaration $CT(C)$ which may make mutually recursive
references to other classes.  Well formed class tables are characterized using
typing rules which are expressed using some auxiliary
functions that in turn depend on the class table, as is needed to allow
mutual recursion.  Consider a declaration 
\[ CT(C) =  \class{C}{D}{\ol{\T}_1\;\ol{f};\;\construct{\S_1}\;\ol{\M}}  
\epunc . \]
To refer to the constructor, we define $\constr\:C = \S_1$.
For the direct superclass of $C$, we define $\super\:C = D$.
Let $\M$ be in the list $\ol{\M}$ of method declarations, with 
\[ \M = \T\; m(\ol{\T}_2\;\ol{x})\;\{\S_2 \} 
\epunc .\]
We record the
typing information by defining $\mtype(m,C) = \ol{\T}_2 \TO\T$. 
(Note that $\ol{\T}_2 \TO\T$ is not a data type in the language.)
The parameter names are given by $\pars(m,C) = \ol{x}$.  
If $m$ has no declaration in $CT(C)$ but $\mtype(m,D)$ is defined
then $m$ is an inherited method: we define $\mtype(m,C)=\mtype(m,D)$ and
$\pars(m,C)=\pars(m,D)$. 
For the declared fields, we define 
$\type(\ol{f},C) =\ol{\T}_1$ and   
$\dfields\, C = (\ol{f}:\ol{\T}_1)$.
Here $\ol{f}:\ol{\T}_1$ denotes a finite mapping of field names to types.
To include inherited 
fields, we define $\fields\,C = \dfields\,C \union \fields\,D$ and
assume $\ol{f}$ is disjoint from the names in $\fields\,D$.  The
built-in class \OBJECT\ has no methods and
$\fields(\OBJECT)$ is the empty list.  

A \dt{typing context} $\Gamma$ is a finite mapping from variable and
parameter names to data types, such that $\self\in\dom\,\Gamma$.
Whereas the Java format \mbox{\code{T x}} is used in code to give 
\code{x} type \code{T}, it is written \code{x:T} in typing contexts.
Typing of commands for methods declared in class $C$ is expressed using
judgements $\Gamma\proves S $ where $\Gamma\,\self= C$.
Moreover, if $\mtype(m,C)=\ol{\T}\TO \T$ then $\Gamma\,\ol{x} = \ol{T}$ and 
$\Gamma\,\result = \T$.\footnote{In~\cite{popl02} we make $C$ an explicit,
  and redundant, part of the judgement, and we use separate return
  statements rather than variable $\result$.}
For brevity, we sometimes say ``command'' to refer to a derivable judgement
$\Gamma  \proves S $. 
The judgement $\Gamma\proves e:\T$ says that expression $e$ has type $\T$.
The constructor is typed using a judgement
$\self:C\proves S:\CON $ which is distinguished from the typing of $\S$ as
a command, as the former is used to define the semantics of $\S$ as a
constructor, which in turn is used in the semantics of object construction
(\NEW). 

\begin{definition}[{\bf subtyping , $\leq$}]\label{def:subty}
The class table determines a subtyping relation $\leq$ as follows. 
If $T$ or $U$ is $\BOOL$ or $\UNIT$ then define $T\leq U$ iff
$T=U$.  For class types $C$ and $D$, define $C\leq D$ iff either $C=D$ or
$\super\, C\leq D$.\qed
\end{definition}
Subsumption is built into the rules for specific constructs. 
For example, the assignment rule allows 
$x:D,y:E,\self:C \proves \assign{x}{y}$ 
provided that $E\leq D$.

The constructor for one class may construct objects of other classes
(Fig.~\ref{fig:obsaa} is an example).
But we prefer not to model divergence due to cyclic constructor
dependencies as in the following.
\[
\begin{array}{l}
\class{B}{\OBJECT}{B\;f;\;\construct{\assign{\faccess{\self}{f}}{\new{C}}}} \\
\class{C}{B}{\construct{\SKIP\ }} 
\end{array}
\]
(Recall that to initialize a $C$ object both the $B$- and $C$-constructor are applied.)

\begin{definition}[{\bf constructor dependence, $\cdep$}]\label{def:cdep}
For $B,C$ ranging over declared classes,
we say that $C$ has constructor dependence on $B$, written 
$B\cdep C$, iff $B\cdep (\super\,C)$ or
$\assign{x}{\new{B}{}}$ occurs in $\constr\:C$, for some $x$.\qed
\end{definition}
Note that $B\cdep C$ just if the constructor of $C$ or one of its ancestor
classes contains $\new{B}{}$ (by which we mean $\assign{x}{\new{B}{}}$ for
some $x$). 
The transitive closure $\cdep^+$ has
$B\cdep^+ C$ just if construction of a $C$-object entails construction
of a $B$-object.
For the example above we have $C\cdep B$ and $C\cdep C$.

\begin{definition}[{\bf well formed class table}]\label{def:wfct}
A class table is well formed provided it satisfies the following
conditions.  
\begin{itemize}
\item 
Each class declaration 
$\class{C}{D}{\ol{\T}\;\ol{f};\;\construct{\S}\;\ol{\M}}$
is well formed, that is, each method declaration $\M$ in $\ol{M}$ is
well formed, and  $\self:C\proves \S:\CON$, according to the rules to follow.
\item If $C$ occurs as the type of a field or parameter in some class then $CT(C)$ is
  defined.  No field or method has multiple declarations in a class. 
\item The subclass relation $\leq$ is antisymmetric.
\item Transitive constructor dependence, $\cdep^+$, is antisymmetric and irreflexive.  
\qed 
\end{itemize}
\end{definition}
The rules are straightforward renderings of the typing rules for Java, 
for private fields, public methods and public classes~\cite{JavaLang}.  


\begin{infig}{Typing of constructors}
$\Rule{
S = \constr\:C \quad 
\self:C\proves S \quad
\mbox{no method calls occur in $S$}
}{
\self:C\proves S:\CON
}$
\end{infig}

\begin{infig}{Typing of method declarations}
$\Rule{
\begin{array}{c}
\ol{x}:\ol{\T},\self:C,\result:T  \proves S \\
\mtype(m,\super\, C) \mbox{ is undefined or equals } \ol{\T} \TO \T \\
\pars(m,\super\, C) \mbox{ is undefined or equals } \ol{x}
\end{array}}
{ \rule{0pt}{2.5ex} 
C \proves \T\; m(\ol{\T}\;\ol{x})\{\S\}}
$
\end{infig}

In this method rule, the condition on $\mtype$ is the
standard invariance restriction on method types, as in Java
\cite{JavaLang,AbadiCardelli}.  
The last antecedent in the rule, concerning $\pars(m,D)$, ensures that all 
declarations of a method use the same parameter names.  This loses no
generality and slightly streamlines the formalization of the semantic
domains in the sequel.

\pagebreak

\begin{infig}{Typing of expressions}
$
\begin{array}{c}
\Gamma \proves x:\Gamma x 
\quad 
\Gamma \proves \NULL:B
\quad 
\Gamma \proves \ITT:\UNIT
\quad 
\Gamma \proves \TRUE:\BOOL
\quad 
\Gamma \proves \FALSE:\BOOL
\\[2.5ex]

\Rule{
\begin{array}{c}
\Gamma \proves e_1:\T_1 \quad
\Gamma \proves e_2:\T_2 
\end{array}}
{\Gamma \proves \eqtest{e_1}{e_2}:\BOOL}
\qquad 
\Rule{
\begin{array}{c}
\Gamma \proves e:(\Gamma\,\self)\quad 
(f:\T) \in \dfields(\Gamma\,\self)
\end{array}} 
{\Gamma\proves \faccess{e}{f}:\T}

\\[2.5ex]

\Rule{\Gamma \proves e:D \quad B\leq D}
     {\Gamma \proves\cast{B}{e}:B}
\qquad 
\Rule{\Gamma \proves e:D \quad B\leq D}
     {\Gamma \proves\is{e}{B}:\BOOL}
\end{array}
$
\end{infig}

The rule for equality test allows comparison of arbitrary data types,
and is reference equality in the case of class types.  But
if $e_1$ and $e_2$ have types not related by $\leq$, the test 
$\eqtest{e_1}{e_2}$ is false except when both are nil.
The rule for field access enforces private visibility: only a
method declaration in class $C$ can access fields declared in $CT(C)$.
It can access those fields on any object of its type; to access
its own fields the expression is $\self.f$.  
The rule for cast is standard.\footnote{It is not adequate for
  expressions that arise through substitutions used in program logic
  (see~\citeN{FM99}) and in small-step semantics (see
  \citeN{Featherweight}); the latter source uses the term ``stupid
  cast'' for the typing rule that allows $\cast{B}{e}$ when $B$ is not
  a subclass of the static type of $e$.}

\begin{infig}{Typing of commands}
$
\begin{array}{c}

\Rule{
\begin{array}{c}
\Gamma \proves e:\T \quad \T\leq \Gamma\, x
\quad x\neq\self 
\end{array}}
{\Gamma \proves\assign{x}{e} } 
\qquad 
\Rule{
\begin{array}{c}
\Gamma \proves e_1:(\Gamma\,\self) \quad (f:\T) \in \dfields(\Gamma\,\self) \\
\Gamma \proves e_2:U \quad U\leq T
\end{array}}
{\Gamma\proves \assign{\faccess{e_1}{f}}{e_2} } 
\\[2.5ex]

\Rule{
\begin{array}{c}
\Gamma \proves e:D \quad
\mtype(m,D) = \ol{\T}\TO \T \\
\Gamma  \proves\ol{e}:\ol{U} \quad \ol{U}\leq\ol{T} \quad x\neq \self \quad T\leq \Gamma\,x 
\end{array}}
{\Gamma \proves \assign{x}{\mcall{e}{m}{\ol{e}}} }
\quad 
\Rule{
\begin{array}{c}
\mtype(m,\super(\Gamma\,\self)) = \ol{\T}\TO \T \\
\Gamma \proves\ol{e}:\ol{U} \quad \ol{U}\leq\ol{T}
\quad x\neq\self
\quad T\leq \Gamma\,x
\end{array}}
{\Gamma\proves \assign{x}{\mcall{\SUPER}{m}{\ol{e}}} }
\\[2.5ex]

\Rule{B\leq\Gamma x \quad x\neq \self \quad B\neq\OBJECT
}{\Gamma \proves\assign{x}{\new{B}{\ }} }
\qquad 
\Rule{
\begin{array}{c}
\Gamma \proves \S_1 
\quad
\Gamma \proves \S_2 
\end{array}}
{\Gamma \proves \seq{\S_1}{\S_2} }
\\[2.5ex]

\Rule{
\begin{array}{c}
\Gamma \proves e:\BOOL\quad
\Gamma \proves \S_1  \quad
\Gamma \proves \S_2  
\end{array}}
{\Gamma \proves \ifelse{e}{\S_1}{\S_2} }
\qquad 
\Rule{
\begin{array}{c}
\Gamma \proves e:U \quad U\leq\T \quad x\neq\self \quad 
(\Gamma,x:\T) \proves \S  
\end{array}}
{\Gamma \proves\var{\T\; x}{e}{S} }
\end{array}
$
\end{infig}

The command rules have hypotheses involving partial functions  which
must be defined for the hypothesis to be satisfied.  For example, in the
rule for super calls, $\mtype(m,\super\, C)$ must be defined and equal
to $\ol{T}\TO\T$.

Each expression and command construct is the conclusion of exactly one
typing rule, and there are no other rules.  Thus we have the following.

\begin{lemma}\label{lem:tyun}
A typing $\Gamma  \proves S $ or $\Gamma  \proves e:\T$ 
has at most one derivation.\qed
\end{lemma}

\begin{definition}[{\bf inheritance}]\label{def:inh}
Method $m$ is \dt{inherited in $C$ from $B$} if 
$C\leq B$, there is a declaration for $m$ in $B$,
and there is no declaration for $m$ in any $D$ such that $C\leq D<B$.
To make the class table explicit, we also say $m$ is inherited from
$B$ in $CT(C)$.\qed
\end{definition}
Because the language has single inheritance, the subtyping relation $\leq$
is a tree: if $D\leq B$ and $D\leq C$ then $B\leq C$ or $C\leq B$.  
If $\mtype(m,C)$ is defined for some $C$ then it is defined for all
subclasses of $C$. For a given method name $m$ and class $C$,
there is a unique ancestor class declaring $m$ that is least with respect
to $\leq$.   

Lemma~\ref{lem:tyun} allows proofs by structural induction on
typings.  The following notion facilitates induction on inheritance chains.

\begin{definition}[{\bf method depth}]\label{def:dep}
For any $m$ and $C$ such that $\mtype(m,C)$ is defined,  
the \dt{method depth of $C$ for $m$ in $CT$} is defined by 
$depth(m,C)  = 1 + depth(m,\super\, C) $ if $\mtype(m,\super\, C)$ is
defined; otherwise,  $depth(m,C)=0$.\qed
\end{definition}
An immediate consequence is that 
if $\mtype(m,C)$ is defined and  $depth(m,C) = 0$ then $CT(C)$ has a
declaration for $m$. 

Finally, we consider ramifications of constructor dependence.
Note that $\OBJECT \not\cdep C$ for all $C$, by the typing rule for $\NEW$.

\begin{definition}[{\bf semantic dependence, $\well$}]\label{def:semdep}
As an auxiliary notation, we define $ B \cdepseq C$ iff $\{D \mid  D\cdep^+ B \}
\subseteq \{D \mid  D\cdep^+ C \}$ and write $B\cdeps C$ if this inclusion is proper.
For classes $B,C$ declared in the class table, define
$B\well C$ iff $B\cdeps C$ or both $B\cdepseq C$ and $B>C$.\qed
\end{definition}

\begin{lemma}\label{lem:cdep} For a well formed class table we have the following.
  \begin{enumerate}
  \item \label{cdepc}
$\well$ is well founded.
  \item \label{cdepa}
$\super\,C\well C$ for all $C$.
  \item \label{cdepb}
$B\cdep C$ implies $B\well C$ for all $B$ and $C$.
  \end{enumerate}
  \begin{proof}
Note that $\cdepseq$ is a preorder but not antisymmetric, so $\well$ is not
a lexicographic order per se.  
To prove (\ref{cdepc}), define $deps\,C=\{D\mid D \cdep^+ C\}$ for any $C$.
Then we have $B\well C$ iff $(deps\,B, B)\lessdot (deps\,C, C)$, where 
$\lessdot$ is defined by $(X,B)\lessdot (Y,C)$ iff
$X\subsetneq Y$ or $X\subseteq Y$ and $B>C$  
(where $\subsetneq$ means proper subset).
This is logically equivalent to: $X\subsetneq Y$ or $X=Y$ and $B>C$, which
shows that the definition is the lexicographic coupling of $\subsetneq$ and
$>$.  As $\subsetneq$ here is for finite subsets of declared class names,
both $\subsetneq$ and $>$ are well founded, hence so is their lexicographic coupling.

For (\ref{cdepa}), if $D\cdep^+ \super\,C$ then
$D\cdep^+ C$ by definition of $\cdep$; hence $\super\,C\cdepseq C$.  Also,
$\super\,C > C$, so (\ref{cdepa}) holds by definition of $\well$.

For (\ref{cdepb}), suppose $B\cdep C$. Then, by transitivity,
$\{D \mid  D\cdep^+ B \} \subseteq \{D \mid  D\cdep^+ C \}$.
Also, we have $B\cdep^+ C$ but $B\not\cdep^+ B$,  
by well formedness of the class table,
so the inclusion is proper.  That is, $B\cdeps C$, whence
$B\well C$ by definition of $\well$. 
  \end{proof}

\end{lemma}


\section{Semantics}\label{sect:sem}

This section defines the semantic domains, then the semantics of
expressions and commands, and finally the semantics of well formed class tables.

Because methods are associated with classes rather than with instances, the
semantic domains are rather simple.  There are no recursive domain
equations to be solved: subclassing ($\leq$) is acyclic and the cycle of
recursive references via class fields is broken via the heap.
Mutually recursive method invocations can arise through direct calls on a single
object and also through callbacks between reachable objects, as for example 
in the observer pattern.  We impose no restrictions on
such calls. A fixpoint construction is used for the method environment
which comprises the semantics of the class table. 

The interdependence between constructors and object construction commands
($\NEW$) is a bit complex; things pertaining to constructors may be skipped
on first reading.  As a way of explaining the fine points, we prove
in some detail that the semantics is well defined (Lemma~\ref{lem:semty}).

Often we write $=$ between expressions involving partial functions
such as those used in typing.  Unless otherwise indicated, it means  strong equality:
both sides are defined and equal.

\subsection{Semantic domains}

The state of a method in execution is comprised
of a \dt{heap} $h$, which is a finite\footnote{The preliminary version
\cite{popl02} of this paper has a bug: infinite heaps are allowed, and
it is not required that there be unallocated locations at every type.}  
partial function from locations to
object states, and a \dt{store} $\eta$, which assigns
locations and primitive values to the local variables and
parameters given by a typing context $\Gamma$.\footnote{In~\cite{popl02} we
  use the term ``environment'' for 
  $\eta$, wishing to avoid the irrelevant connotations of ``stack'';
here we use ``store'', following~\citeN{ReynoldsPtrs}.}
An \dt{object state} is a mapping from field names to values.   
Function application associates to the left, so $h\,\ell\,f$ is the value
of field $f$ of the object $h\,\ell$ at location $\ell$.

A command denotes a function mapping each initial state $(h,\eta)$ either
to a final state $(h_0,\eta_0)$ or to the distinguished value $\bot$.  
We use the term \dt{global state} for $(h,\eta)$, to distinguish it from
object states.
The improper value $\bot$ represents non-termination as well as runtime
errors: attempts to dereference $\NIL$ or cast a location to a type it does not have.  

In some languages it is a runtime error to dereference a dangling pointer,
i.e., one not in the domain of the heap.  In Java dangling 
pointers cannot arise: there is no command for deallocation and a correct
garbage collector never deallocates reachable objects. 
For our purposes, garbage collection need not be modelled.
Commands act on heaps and stores that are closed in the sense that all
locations that occur are in the domain of the heap. 
The following paragraphs formalize our assumptions about
locations and then define the semantic domains.

For locations, we assume that a countable set $\Loc$ is given, along
with a distinguished value $\NIL$ not in $\Loc$.  
To track each object's class we assume given a function
$\loctype:\Loc\to\cnames$ such that for each $C$ there are infinitely many
locations $\ell$ with $\loctype\;\ell=C$.   
We use the term \dt{heap} for any partial function $h$ such that $\dom\,h
\subseteq_{\mathit{fin}}\Loc$ and each $h\,\ell$ is an object state of type
$\loctype\;\ell$.  Object states are formalized later.
Because the domain of a heap is finite, the assumption about $\loctype$
ensures an adequate supply of fresh locations. 

\begin{definition}[{\bf allocator, parametric}]\label{def:param}
An \dt{allocator} is a location-valued function $\fresh$ such
that $\loctype(\fresh(C,h))= C$ and $\fresh(C,h) \not\in dom\, h$, for
all $C,h$.  An allocator is \dt{parametric} if $dom\,h_1 \intersect
locs\,C = dom\,h_2 \intersect locs\,C$ implies $\fresh(C,h_1) =
\fresh(C,h_2) $.\qed
\end{definition}
For example, taking $\Loc=\nats$,  a parametric allocator is given by the
function $\fresh(C,h) = min\{ \ell \mid \loctype\, \ell=C \land \ell\not\in dom\,h\}$.

Typical implementations encode the object class as part of its state.  
One could uncurry this representation of heaps and take 
$\Loc$ to be $\nats\times ClassNames$.
Then  $\fresh(C,h)$ could return $(n,C)$ where
$n$ is the least address of an unused memory segment of sufficient size for the state of $C$.  
This is an allocator but not parametric because the presence of objects of
one class affect the availability of memory for objects of other
classes.

We define the semantics in terms of an arbitrary allocator $\fresh$.
The assumption of parametricity is stated explicitly where it
is needed, namely for the first abstraction theorem (Sect.~\ref{sect:abs}) but
not the second (Sect.~\ref{sect:sabs}).
Parametricity of the allocator is a reasonable assumption for some
applications but not all.  The assumption streamlines the proof of the
abstraction theorem, allowing us to highlight other issues. 

In addition to heaps, it is convenient to name a number of other semantic categories
that are explained in due course.  

\begin{infig}{Semantic categories}
$ \theta \; \gassym \; T \mid \Gamma \mid  \state{C} \mid \heap
\mid \heap\otimes \Gamma \mid \heap \otimes T
\mid \theta_{\bot}
\mid C,\,\ol{x},\,\ol{\T}\TO\T \mid \menv 
$
\end{infig}

In order to define the more complicated semantic domains, we need to define
closed stores.
Stores are among the simpler semantic domains, which are defined as follows.

\begin{infig}{Semantics of types, object states, and stores}
$\begin{array}{lcl}
\means{\BOOL} &=&\{\semtrue, \semfalse\}\\[.5ex]
 \means{\UNIT}&=&\{\IT\} \\[.5ex]
 \means{C}    &=& \{\NIL\} \union locs(C\subclasses) \\[.5ex]
\means{\state{C}}&= &
\{ s \mid \dom\,s = \dom(\fields\,C) \land 
\all{(f:T)\in\fields\,C}{s f \in \means{\T}} \} \\[.5ex]
\means{\Gamma}&= & \{ \eta \mid \dom\,\eta=\dom\, \Gamma \land
 \eta\,\self\neq \NIL 
\land \all{x\in\dom\,\eta}{\eta\, x\in\means{\Gamma\, x}} \} 
\end{array}
$
\end{infig}
We write $locs\;C$ for $\{\ell \in Loc \mid \loctype\,\ell = C\}$, and 
$locs(C\subclasses)$ for $\{\ell \mid \loctype\,\ell \leq C\}$.
There is no independent meaning for $C\subclasses$.
As small dot has another use, we use the fat dot $\bullet$ to separate a
bound variable from its scope.
Note that $\means{\Gamma}$ is defined for $\Gamma$ both with and without
$\result$ in its domain.

\begin{definition}[{\bf closed heap and store}]
A heap $h$ is \dt{closed}, written $closed\:h$, iff
$rng(h\,\ell)\intersect Loc\subseteq \dom\,h$, for all $\ell\in\dom\,h$.     
A store $\eta\in\means{\Gamma}$ is \dt{closed in heap} $h$,
written $closed(h,\eta)$, iff $rng\,\eta\intersect Loc\subseteq \dom\,h$.\qed
\end{definition}
Recall that fresh locations should occur nowhere in the
global state.  For a closed store and heap, this follows from the
requirement that $\fresh(C,h) \not\in dom\, h$.\footnote{If dangling pointers were
allowed, the definition of freshness would need to be with respect to both
the store and all object states in the heap. The issue becomes
apparent in the proof of Lemma~\ref{lem:c} in the sequel, which uses
closure.
Most of the other definitions and results can be formulated without
restricting heaps to be closed, so we mistakenly neglected closure in 
\cite{popl02}.  
}

\begin{infig}{Semantics of global states and methods}
$\begin{array}{lcl}
\means{\heap}&=&
\{ h \mid \dom\,h\subseteq_{\mathit{fin}}\Loc \land
closed\,h \land 
\all{\ell\in\dom\,h}{h\ell\in\means{\state{(\loctype\;\ell)}}} \}
\\[.5ex]
\means{\heap\otimes\Gamma}&=&
\{ (h,\eta) \mid h\in\means{\heap} \land \eta\in\means{\Gamma} \land closed(h,\eta) \}  \\[.5ex]
\means{\heap\otimes\T}&=&
\{ (h,d) \mid h\in\means{\heap} \land d\in\means{T} \land (d\in Loc\implies
 d\in\dom\,h )\}  \\[.5ex]
\means{\theta_\bot} & = & \means{\theta}\union \bot
\quad\mbox{(where $\bot$ is some fresh value not in $\means{\theta}$)}
\\[.5ex]
\means{C,\,\ol{x},\,\ol{\T}\TO\T} &=&
\means{\heap \otimes (\ol{x}:\ol{\T}, \self:C)}\to
\means{(\heap\otimes\T)_\bot} \\[.5ex]

\means{\menv}  &=& \{ \mu \mid 
\begin{array}[t]{l}
\all{C,m}{} \mu C m \mbox{ is defined iff $\mtype(m,C)$ is
  defined,}  \\
\mbox{and }\mu C m \in \means{C,\pars(m,C),\mtype(m,C)} \mbox{ if $\mu C m$ defined} 
\; \}
\end{array}
\end{array}
$
\end{infig}

Just as a class declaration $CT(C)$ gives a collection of method
declarations, the semantics of a class table is a \dt{method environment}
that assigns to each class $C$ a method meaning $\mu\,C\,m$ for each $m$
declared or inherited in $C$.  

For the fixpoint construction of 
the method environment denoted by a class table, we
need to impose order on the semantic domains.
We use the term \dt{complete partial order} for a poset with least upper bounds of
countable ascending chains~\cite{Davey:P}.  
The degenerate case is ordering by equality, which is the order we use for
the semantics of $T$, $\Gamma$, $\state{C}$, $\heap$, $(\heap\otimes \Gamma)$, and 
$(\heap \otimes T)$.
Then $\means{(\heap\otimes\Gamma)_\bot}$ and
$\means{(\heap\otimes\T)_\bot}$ are complete partial orders with the
``flat'' order: $\bot$ is below anything and other comparable elements
are equal.
The set $\means{C,\,\ol{x},\,\ol{\T}\TO\T}$ is defined to be the
space of total functions
$\means{\heap \otimes (\ol{x}:\ol{\T}, \self:C)}\to\means{(\heap\otimes\T)_\bot}$, 
all of which are continuous
because $\heap \otimes (\ol{x}:\ol{\T},\self:C)$ is ordered by equality.  The
function space itself is ordered
pointwise, making it a complete partial order with minimum element $\lam{(h,\eta)}{\bot}$. 
Finally, we order $\means{\menv}$ pointwise.  All method environments $\mu$
in $\means{\menv}$ have the same domain, determined by $CT$, so this is
also a complete partial order, taken pointwise.  It has a minimum element, namely
$\lam{C}{\lam{m}{\lam{(h,\eta)}{\bot}}}$.

Whereas $\means{\state\,C}$ consists of the states for objects of exactly
class $C$, the set $\means{C}$ is downward closed.  
For data types $T_1,T_2$ we have $T_1\leq T_2 \implies \means{T_1}\subseteq 
\means{T_2}$.  

\begin{definition}[{\bf incomparable, $\ncomp$}]\label{def:ncomp}
We write $C\ncomp B$ for $C\nleq B\land C\not\geq B$.  
For a list $\ol{C}$, $\ol{C}\ncomp B$ means $C\ncomp B$ for all $C$ in $\ol{C}$.\qed
\end{definition}

\begin{lemma}\label{lem:ncomp}
For classes $C,B$, if $C\ncomp B$ then  $\means{C}\intersect \means{B} = \{\NIL\}$.
For primitive $T$ we have $\means{T}\intersect\means{B} = \Empty$.\qed
\end{lemma}
The result is a direct consequence of the definitions.  
We often use the contrapositive: if there is a non-$\NIL$
location in both $\means{B}$ and $\means{C}$ then $B\leq C$ or $C\leq B$.

\subsection{Semantics of expressions, commands, constructors and methods}

For expressions and commands, the semantics is defined by induction on typing derivations.  
As a consequence of uniqueness of typing derivations,
Lemma~\ref{lem:tyun}, the semantics is a function of typings.
The meaning of a command $\Gamma\proves S$ will be defined to be a function 
\[ \means{\Gamma\proves S } \in \means{\menv}\to\means{\heap\otimes\Gamma}\to
\means{(\heap\otimes\Gamma)_\bot}
\epunc. \]
The meaning of an expression $\Gamma\proves e:T$ will be defined to be a function
\[  \means{\Gamma\proves e:\T} \in
\means{\heap\otimes\Gamma}\to\means{\T_\bot}\]
such that the result value is always in the domain of the heap if it is a
location.\footnote{We have
  chosen a simple but slightly inelegant formulation. We express 
closure of the result for commands in the semantic domain whereas for
expressions there is no returned heap and we express closure as a property
of the semantic function.
The presentation could be made more elegant by introducing categories
$\mathbf{exp}(\Gamma,T)$ and $\mathbf{com}(\Gamma)$ with 
$ \means{\mathbf{com}(\Gamma)} =
\means{\menv}\to\means{\heap\otimes\Gamma}\to\means{(\heap\otimes\Gamma)_\bot}
$ and imposing the restriction on return values in the definition of 
$ \means{\mathbf{exp}(\Gamma,T)}$ as a subset of 
$\means{\heap\otimes\Gamma}\to\means{T_\bot}$.  
We could even restrict the meanings to those that are confined, 
but the gain in elegance would come at the expense of complexity that not
all readers would find illuminating.
We have chosen to treat confinement and parametricity as properties to be
proved after the semantics is defined, downplaying the model as an
independent structure.  Thus little would be gained by naming categories
$\mathbf{exp}(\Gamma,T)$ and $\mathbf{com}(\Gamma)$.}
This is part of Lemma~\ref{lem:semty}, the proof of which serves as an 
exposition for some details of the semantic definitions.

The command and expression constructs are strict in $\bot$,
except, as usual, for the then- and else-commands in $\IF-\FI$.
To streamline the treatment of $\bot$ in the semantic 
definitions we use a metalanguage construct which some readers will recognize 
as the bind operation of the lifting monad~\cite{MoggiMonad}.  
The construct  $\mletml{d}{E_1}{E_2}$ has the
following meaning: If the value of $E_1$ is $\bot$ then that is the
value of the entire let expression; otherwise, its value is the value
of $E_2$ with $d$ bound to the value of $E_1$.  

We let  $(h,\eta)\in\means{\heap\otimes\Gamma}$ in the following
definitions.  Identifiers are as in the corresponding typing rules.  
For semantic values we use the identifier $d$, but sometimes 
$\ell$ for elements of the sets $\means{C}$.

For expressions the semantics is straightforward; we choose the Java
semantics for casts and tests.  

\begin{infig}{Semantics of expressions}
$
\begin{array}{lcl}
\means{\Gamma \proves x:\T}(h,\eta) &=& \eta x\\[.5ex]
\means{\Gamma \proves \NULL: B}(h,\eta) &=& \NIL\\[.5ex]
\means{\Gamma \proves \ITT: \UNIT}(h,\eta) &=& \IT \\[.5ex]
\means{\Gamma \proves \TRUE: \BOOL}(h,\eta) &=& \semtrue \\[.5ex]
\means{\Gamma \proves \FALSE: \BOOL}(h,\eta) &=& \semfalse \\[.5ex]
\means{\Gamma \proves\eqtest{e_1}{e_2}:\BOOL}(h,\eta) 
&=& 
\begin{array}[t]{l}
\mylet{d_1 = \means{\Gamma \proves e_1:\T_1}(h,\eta)}\\[.5ex]
\mylet{d_2 = \means{\Gamma \proves e_2:\T_2}(h,\eta)}\\[.5ex]
\mifthenelse{d_1=d_2}{\semtrue}{\semfalse}  \\[.5ex]
\end{array}
\\[.5ex]
\means{\Gamma \proves\faccess{e}{f}:\T}(h,\eta) & =&
\begin{array}[t]{l}
\mylet{\ell  = \means{\Gamma \proves e:(\Gamma\,\self)}(h,\eta)}\\[.5ex]
\mifthenelse{\ell=\NIL}{\bot}{h\, \ell\, f }
\end{array}
\\[.5ex]
\means{\Gamma \proves \cast{B}{e}: B}(h,\eta) 
&=&\begin{array}[t]{l}
\mylet{\ell  = \means{\Gamma \proves e:D}(h,\eta)}\\[.5ex]
\mifthenelse{\ell=\NIL\lor\loctype\;\ell\leq B }{\ell}{\bot} \\[.5ex]
\end{array}
\\[.5ex]
\means{\Gamma \proves\is{e}{B}:\BOOL}(h,\eta) 
&=&\begin{array}[t]{l}
\mylet{\ell  = \means{\Gamma \proves e:D}(h,\eta)}\\[.5ex] 
\mifthenelse{\ell\neq\NIL\land\loctype\;\ell\leq B}{\semtrue}{\semfalse}  
\end{array}
\end{array}
$
\end{infig}

The semantics of commands is defined by structural induction on the
command, except for object construction $\assign{x}{\new{C}{}}$ which also
depends on the constructor semantics of the constructor, $\constr\:C$, of $C$. That in turn
depends on the constructor of $\super\,C$, and on the command semantics of
$\constr\:C$.  
Well foundedness of this dependence is part of the proof of
Lemma~\ref{lem:semty}.   

In the semantics of commands, we write $\fields\,B\mapsto \mathit{defaults}$ as an
abbreviation for the function sending each $f\in \dom(\fields\,B)$ to the
default value for  $\type(f,B)$. The defaults are $\semfalse$ for $\BOOL$, $\IT$
for $\UNIT$, and $\NIL$ for classes.
Function update or extension is written, e.g., $\ext{\eta}{x}{d}$.  
We write $\downharpoonright$ for domain
restriction: if $x$ is in the domain of $\eta$ then
$\eta\downharpoonright x$ is the function like $\eta$ but without $x$
in its domain.

\begin{infig}{Semantics of commands}
$
\begin{array}{lcl}
\means{\Gamma \proves \assign{x}{e} %
}\mu(h,\eta) &=&
\begin{array}[t]{l} 
\mylet{d = \means{\Gamma \proves e:\T}(h,\eta)} 
(h,\ext{\eta}{x}{d})\\[.5ex]
\end{array}
\\[.5ex]
\means{\Gamma \proves \assign{\faccess{e_1}{f}}{e_2} %
}\mu(h,\eta) 
&=&
\begin{array}[t]{l}
\mylet{\ell = \means{\Gamma\proves e_1:(\Gamma\,\self)}(h,\eta) }\\[.5ex]
\mifthenelse{\ell=\NIL}{\bot}{} \\[.5ex]
\mylet{d = \means{\Gamma \proves e_2:U}(h,\eta)}\\[.5ex]
  (\ext{h}{\ell}{\ext{h\ell}{f}{d}},\eta)\\[.5ex]
\end{array}
\\[.5ex]
\means{\Gamma \proves\assign{x}{\new{B}{\ }} %
}\mu(h,\eta)
&=&
\begin{array}[t]{l}
\mylet{\ell = \fresh(B,h)} \\[.5ex]
\mylet{h_1 = \ext{h}{\ell}{[\fields\,B\mapsto \mathit{defaults}]}}\\[.5ex]
\mylet{\eta_1 = [\self\mapsto  \ell]} \\[.5ex]
\mylet{h_0 =  \means{\self:B\proves \constr\:B:\CON}\mu(h_1,\eta_1)}\\[.5ex]
(h_0, \ext{\eta}{x}{\ell}) 
\end{array}
\\[.5ex]
\means{\Gamma \proves \assign{x}{\mcall{e}{m}{\ol{e}}} %
}\mu(h,\eta) 
&=&
\begin{array}[t]{l}
\mylet{\ell = \means{\Gamma \proves e:D}(h,\eta) }\\[.5ex]
\mifthenelse{\ell=\NIL}{\bot}{} \\[.5ex]
\mylet{\ol{x} = \pars(m, D)}\\[.5ex]
\mylet{\ol{d} = \means{\Gamma \proves\ol{e}:\ol{U}}(h,\eta)}\\[.5ex]
\mylet{\eta_1 = [\ol{x}\mapsto\ol{d}, \self\mapsto\ell]} \\[.5ex]
\mylet{(h_1, d_1) = \mu(\loctype\,\ell)m(h,\eta_1)}  \\[.5ex]
(h_1,\ext{\eta}{x}{d_1})\\[.5ex]
\end{array}
\\[.5ex]
\means{\Gamma \proves \assign{x}{\mcall{\SUPER}{m}{\ol{e}}} %
}\mu(h,\eta) 
&=&
\begin{array}[t]{l}
\mylet{\ell = \eta\,\self } \\[.5ex]
\mylet{\ol{x} = \pars(m, \Gamma\,\self)}\\[.5ex]
\mylet{\ol{d} = \means{\Gamma \proves\ol{e}:\ol{U}}(h,\eta)}\\[.5ex]
\mylet{\eta_1 = [\ol{x}\mapsto\ol{d}, \self\mapsto\ell]} \\[.5ex]
\mylet{(h_1, d_1) = \mu(\super(\Gamma\,\self) )m(h,\eta_1)}  \\[.5ex]
(h_1,\ext{\eta}{x}{d_1})\\[.5ex]
\end{array}
\\[.5ex]
\means{\Gamma \proves \seq{\S_1}{\S_2} %
}\mu(h,\eta)
&=&
\begin{array}[t]{l}
\mylet{(h_1,\eta_1) = \means{\Gamma \proves \S_1 %
}\mu(h,\eta)}\\[.5ex]
\means{\Gamma \proves \S_2 %
}\mu(h_1,\eta_1) \\[.5ex]
\end{array}
\\[.5ex]
\means{\Gamma \proves \ifelse{e}{\S_1}{\S_2} %
}\mu(h,\eta)
&=& \begin{array}[t]{l}
\mylet{b =\means{\Gamma \proves e:\BOOL}(h,\eta)}\\[.5ex]
\mifthenelse{b}{\means{\Gamma \proves \S_1 %
}\mu(h,\eta)}{}  
\means{\Gamma \proves \S_2 %
}\mu(h,\eta) \\[.5ex]
\end{array}
\\[.5ex]
\means{\Gamma \proves\var{\T\; x}{e}{S} %
}\mu(h,\eta) 
&
=&
\begin{array}[t]{l}
\mylet{d = \means{\Gamma \proves e:U}(h,\eta)}\\[.5ex]
\mylet{\eta_1 = \ext{\eta}{x}{d} } \\[.5ex]
\mylet{(h_1,\eta_2) = \means{(\Gamma, x:T) \proves \S}\mu(h,\eta_1)} \\[.5ex]
(h_1,(\eta_2 \mathord{\downharpoonright} x)) \\[.5ex]
\end{array}
\end{array}
$
\end{infig}

Method calls of the form $\assign{x}{\mcall{e}{m}{\ol{e}}} $ are
dynamically bound: the method meaning is determined by $\loctype\,\ell$ in
the semantic definition, where $\ell$ is the value of $e$.
By typing, $\loctype\,\ell\leq D$ and $\pars(m, \loctype\,\ell) = \pars(m,
D)$.  Super-calls are statically bound: the method meaning used, 
$\mu(\super\,C)m$, is determined by the static class $C$.
Note that if $\mtype(m,\super\,C)$ is defined, as required by the typing
rule, then $\pars(m, C)=\pars(m, \super\,C)$. 

The meaning of a command $\S$ as a constructor is a function 
\[ \means{\self:C\proves S:\CON}\in\means{\menv}\to\means{\heap\otimes \self:C}
\to\means{\heap_\bot} 
\epunc .\]
Dependence on $\means{\menv}$ is a formal technicality:
the semantic definition uses the command semantics of $\S$, but the typing rule
disallows method calls in $\S$.  

\begin{infig}{Semantics of constructor}
\(\begin{array}{lcl}
\means{\self:C\proves S:\CON}\mu(h,\eta) &=&
\mylet{B = \super\,C} \\[.5ex]
&&\mylet{S_0 = \constr\:B} \\[.5ex]
&& \MLET\ h_1 =   
  \begin{array}[t]{l}
\MIF\ B\neq\OBJECT \\[.5ex]
\MTHEN\ \means{\self:B\proves S_0:\CON}\mu(h,\eta)\ \MELSE\ h\ \MIN 
\end{array}
\\[.5ex]
&& \mylet{(h_0,-) = \means{\self:C\proves S}\mu(h_1,\eta)} \\[.5ex]
&& h_0
\end{array}\)
\end{infig}

Note that if 
$\means{\self:B\proves S_0:\CON}\mu(h,\eta) $ or 
$\means{\self:C\proves S}\mu(h_1,\eta)$  
is $\bot$ then so is $\means{\self:C\proves S:\CON}\mu(h,\eta)$.
The result $\bot$ is possible due to $\NIL$ dereferences and cast failures
but not divergence (because there are no method calls or cyclic constructor
dependencies).  

\begin{infign}{Semantics of method declaration}
Suppose $M$ is a method declaration in $CT(C)$, with
$ M = \T\; m(\ol{\T}\;\ol{x})\{\S\} $.
Its meaning $\means{M}$ is the total function $\means{\menv}\to
\means{C,\ol{x},\ol{\T}\TO\T}$ defined by  
\[ 
\begin{array}{l}
\means{M}\mu(h,\eta) 
= \begin{array}[t]{l}
\mylet{\eta_1 = \ext{\eta}{\result}{\mathit{default}}} \\[.5ex]
\mylet{(h_0,\eta_0) = \means{\ol{x}:\ol{\T},\self:C,\result:\T\proves S
  }\mu(h,\eta_1)}\\[.5ex] 
 (h_0,\eta_0\:\result)
\end{array}
\end{array}
\]
\end{infign}

For precision in the semantics of a method inherited in $C$ from $B$ we
make an explicit definition for the domain-restriction of a method meaning
in  $\means{B,\ol{x},\ol{T}\TO\T}$ to the global states  $(h,\eta)$ 
in $\means{\heap\otimes \ol{x}:\ol{\T},\self:C}$.

\begin{definition}[{\bf {\boldmath $\restr$}}]\label{def:rest}
For $d\in \means{B,\ol{x},\ol{T}\TO\T}$ and $C\leq B$, define 
$\restr(d,C)$, an element of $\means{C,\ol{x},\ol{T}\TO \T}$, 
by  $\restr(d,C)(h,\eta) = d(h,\eta)$.\qed
\end{definition}

\begin{infign}{Semantics of class table and its approximation
    chain $\mu_j$}  
The semantics of a well formed  class table $CT$, written $\means{CT}$, is
the least upper bound of the ascending chain $\mu\in \nats\to\means{\menv}$
defined as follows.  
\[
\begin{array}{lcll}
\mu_0\,C\,m &=& \lam{(h,\eta)}{\bot} 
&\mbox{if $m$ is declared or inherited in $C$}\\[.5ex]
\mu_{j+1}\,C\,m &=& \means{M}\mu_j 
&\mbox{if $m$ is declared as $M$ in $C$}\\[.5ex]
\mu_{j+1}\,C\,m &=& \restr((\mu_{j+1}\,B\,m),C) 
&\mbox{if $m$ is inherited in $C$ from $B$}
\end{array}
\]
\end{infign}

\begin{remark}[On proofs]
We give some proofs in considerable detail.
To avoid repetition, we use the same identifiers as in the relevant
semantic definition for each case ---often different from those in the statement of the
result being proved--- taking care to avoid ambiguity.  This saves explicit
introduction of the identifiers or mention of the ranges and scopes of
quantification.  But it requires the reader to keep an eye on the semantic
clauses.  Often, without remark, we consider only
the case where the outcome and various intermediate values are non-$\bot$, as the
$\bot$ cases are straightforward.  
\end{remark}

\begin{lemma}[semantics is well defined and typed]\label{lem:semty}
Let $CT$ be well formed.  
  \begin{enumerate}
  \item\label{stya} If $C\leq B$ then for any $\Gamma$ with
    $\self\not\in\dom\,\Gamma$ we have  
$\means{\heap\otimes \Gamma,\self:C}\subseteq\means{\heap\otimes\Gamma,\self: B}$.

\item\label{styb}
If  $\Gamma  \proves e: \T$ then
$ \means{\Gamma\proves e:\T} \in
\means{\heap\otimes\Gamma}\to\means{\T_\bot}$.

\item\label{styd}
If $(h,\eta)\in\means{\heap\otimes\Gamma}$ and 
$d= \means{\Gamma; C\proves e:\T}(h,\eta)$ with $d\neq\bot$ then 
$(h,d)\in \means{\heap\otimes T}$.

\item\label{styc}
If  $\Gamma \proves S $ 
then 
$\means{\Gamma \proves S } \in \means{\menv}\to\means{\heap\otimes\Gamma}\to
\means{(\heap\otimes\Gamma)_\bot}$.

\item\label{stye}
$\means{CT}$ is well defined.
  \end{enumerate}
\end{lemma}

\begin{proof} 
(\ref{stya}) follows easily from the fact that 
$C\leq B$ implies $\means{C}\subseteq\means{B}$.

For (\ref{styb}),  inspection of the definitions shows that 
$ \means{\Gamma; C\proves e:\T}(h,\eta)$ is in $\means{\T_\bot}$.
It is property $(h,d)\in \means{\heap\otimes T}$, i.e., (\ref{styd}),  that
we need explicitly in 
some proof steps.  This holds because $(h,\eta)$ is closed
and no expression creates fresh locations.

Property (\ref{styc}) requires a straightforward but not entirely trivial check
that, for any $\mu$, $\means{\Gamma; C\proves S }\mu(h,\eta)$ is in
$\means{(\heap\otimes\Gamma)_\bot}$.  For example, in the case of method
call $\assign{x}{\mcall{e}{m}{\ol{e}}}$ 
we need the fact that $\mu C m$ is in 
$\means{C,\pars(m,C),\mtype(m,C)}$ regardless of whether $m$ is
declared or inherited in $C$.  
The store $\eta_1$ is passed to the method meaning 
$\mu(\loctype\,\ell)m$ determined by the type, $\loctype\,\ell$, of
the target.  Note that $\eta_1\,\self=\ell$ and 
$\mu(\loctype\,\ell)m$ is from a declaration in $\loctype\,\ell$ or a
superclass, so $\eta_1$ is in its domain by (\ref{stya}).
Of course the call aborts if $\ell=\NIL$.  

For (\ref{stye}), acyclicity of $\leq$ ensures that the semantics of
the class table is well founded on inheritance depth.  
And (\ref{stya})  ensures that the definition $\mu_{j+1}\,C\,m$
for an inherited method yields a value in the semantic domain
$\means{C,\pars(m,C),\mtype(m,C)}$. 
We only take fixpoints for method environments, which form a complete partial
order with bottom.  
The fixpoint is well defined because the meaning $\means{M}$  of a
method declaration $M$ is a continuous functions of the method
environment.  This is because  
each $\means{\Gamma\proves S }$ is a continuous function on method
environments ---which in turn depends on the fact that the semantic
definitions for commands are continuous in their constituent commands and
expressions.  

The semantics of object construction commands ($\NEW$) is mutually dependent on 
the semantics of constructors.
This is resolved as follows.

First, the semantics of constructors is defined by well founded recursion
on the order $\well$ on classes. 
For semantics of $\self:B\proves \constr\:B:\CON$ we use both (a) the
constructor semantics of $\self:(\super\,B)\proves \constr(\super\,B):\CON$
and (b) the command semantics for $\constr\:B$.  For (a), note that
$\super\,C\well C$ by Lemma~\ref{lem:cdep}.
For (b), note that if $\constr\:C$ uses $\new{B}{}$ for
other classes $B$, we have $B\cdep C$ by a condition on well formed class tables; then
$B\well C$ by Lemma~\ref{lem:cdep}.  Note that there is no dependence on
the method environment. 

Finally, for semantics of methods we need all constructors as there is no
restriction on which objects can be constructed.
The semantics of methods is by structural recursion on method bodies, using 
the semantics of constructors.
\qed\end{proof}


\section{Confinement ramified}\label{sect:conf}

Our aim is to support reasoning where simulations are specified on a
per-island basis, where an island consists of a single owner and its
reps.\footnote{\label{fn:frame}In particular, this entails describing how a 
  simulation 
is established by an owner constructor acting on a single owner object.
As constructors have no parameters, one could define the semantics in terms
of constructors applied to a single object and yielding a small heap.
But such a constructor will in fact be executed in a larger heap.  
Suppose $(h,\eta)\in\means{\heap\otimes\Gamma}$, so that everything reachable
from $\eta$ is already in $h$.  If $h'$ is a heap, not necessarily
closed, such that $h'*h$ is in $\means{\heap}$, then it is immediate from the
definitions that $(h'*h,\eta)$ is in $\means{\heap\otimes\Gamma}$. 
For any $\S$ and $\mu$ we have $\means{\Gamma\proves S}\mu(h,\eta) = \bot$ iff
$\means{\Gamma\proves S}\mu(h'*h,\eta) = \bot$, 
as can be shown using the fact that  
$\means{\Gamma\proves e:T}(h'*h,\eta) = \means{\Gamma\proves e:T}(h,\eta)$.
(Strictly speaking, this depends on $\mu$ having the property; and then one
shows that $\means{CT}$ has the property.) 
What is not true is the following: 
if $(h_0,\eta_0)=\means{\Gamma\proves S}\mu(h,\eta)$ then 
$\means{\Gamma\proves S}\mu(h'*h,\eta) = (h'*h_0,\eta_0)$.
The reason is that the allocator $\fresh$ depends on the domain of the
entire heap, and we have made no assumptions to relate its behavior on
$h$ and $h'*h$.

We have not checked the details but it seems clear that 
if $(h_0,\eta_0)= \means{\Gamma\proves S}\mu(h'*h,\eta)$ then
there is $h'_0$ such that $h_0 = h'*h'_0$ and  $(h'_0,\eta_0)
\in\means{\heap\otimes \Gamma}$.
Also, for $S$ without method calls and satisfying
the dependency condition for constructors (Def.~\ref{def:wfct}),
if $h_0= \means{\Gamma\proves S:\CON}\mu(h'*h,\eta)$ then
there is $h'_0$ such that $h_0 = h'*h'_0$.  
But to be useful for our purposes this property would have to be
strengthened to take partitions into account. 
}
This section formalizes a semantic notion of confinement suited to
this purpose.  In particular, it takes into account the limited access
to reps allowed for owner subclasses, which is discussed further in
Sect.~\ref{sect:osub}. 

\subsection{Confinement of states}

As discussed in Sect.~\ref{sect:sconf} we assume that class names $Own$
and $Rep$ are given, such that $Own\ncomp Rep$ and thus
$\means{Own}\intersect \means{Rep} = \{\NIL\}$.  
As an abbreviation, we write $\locsOR$ for $locs(Own\subclasses)\union
locs(Rep\subclasses)$.    

We say heaps $h_1$ and $h_2$ are \dt{disjoint} if 
$\dom\,h_1\intersect \dom\,h_2 =\Empty$.    Let $h_1*h_2$ be the union of $h_1$ and $h_2$ if they are disjoint,
and undefined otherwise.  

We shall partition the heap as  $h=\Ch*\ldots$ where $\Ch$ contains client
objects and the rest is partitioned into islands of the form $\Oh*\Rh$ consisting of a
singleton heap $\Oh$ with an owner object and a heap $\Rh$ of its
representation objects.  In such a partition, the heaps $\Ch$, $\Oh$, and $\Rh$ need
not be closed. 
An example is  Fig.~\ref{fig:confineObs} in
Sect.~\ref{ssect:conf}; the general scheme is depicted in Fig.~\ref{fig:part}.
Our use of the word ``partition'' is slightly non-standard: we allow the
blocks $\Rh_i$ and $\Ch$ to be empty.

\begin{narrowfig}{20em}
  \begin{center}
\psfrag{xxCh}{$Ch$}
\psfrag{xxOhj}{$Oh_j$}
\psfrag{xxOhi}{$Oh_i$}
\psfrag{xxRhj}{$Rh_j$}
\psfrag{xxRhi}{$Rh_i$}
\includegraphics{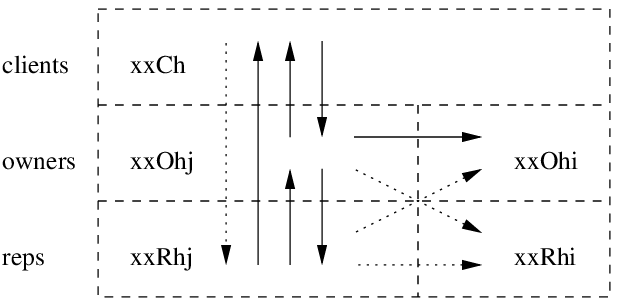}     
  \end{center}
    \caption{Confinement scheme for island $j$.
Dashed boxes are partition blocks.
Solid lines indicate allowed references and 
dotted lines indicate prohibited ones.
There is no restriction within blocks.
}
\label{fig:part}
\end{narrowfig}

\begin{definition}[{\bf admissible partition}]
An \dt{admissible partition} of heap $h$ is a set of pairwise disjoint heaps 
$\Ch,\Oh_1,\Rh_1,\ldots,\Oh_k,\Rh_k$, for $k\geq 0$,   with 
\[ h= \Ch * \Oh_1 * \Rh_1 * \ldots * \Oh_k * \Rh_k \]
and for all $i$ ($1\leq i\leq k$)
\begin{itemize}
\item $\dom\,\Oh_i\subseteq locs(Own\subclasses)\mbox{ and }size(\dom\,\Oh_i) = 1$
\hfill (owner blocks)
\item $\dom\,\Rh_i\subseteq locs(Rep\subclasses)$ \hfill (rep blocks)
\item $\dom\,\Ch\intersect \locsOR = \Empty $
\hfill (client blocks)
\end{itemize}
\end{definition}

\begin{definition}[{\bf confined heap, confining partition, $\npoints$}]\label{def:hconf}
To say that no object in $h_1$ contains a reference to an object in $h_2$, we 
define $\npoints$ by
\[ h_1\npoints h_2 \iff
\all{\ell\in \dom\,h_1}{rng(h_1\,\ell) \intersect \dom\,h_2 = \Empty}
\epunc .\]
To say that no object in $h_1$ contains a reference to an object in $h_2$
\dt{except via a field in $\ol{f}$},
we define $\npointsx{\ol{f}}$ by 
\[ h_1\npointsx{\ol{f}} h_2 \iff
\all{\ell\in \dom\,h_1}{
rng((h_1\,\ell)\mathord{\downharpoonright}\ol{f})\intersect\dom\, h_2 =
\Empty }
\epunc .
\] 
A heap $h$ \dt{is confined}, written $\conf{h}$, iff it has a confining partition.  
A \dt{confining partition} is an admissible partition such that for all $j,i$ with
$j\neq i$ we have 
\begin{enumerate}
\item\label{hconfa} $\Ch \npoints \Rh_j$ \hfill (clients do not point to reps)
\item\label{hconfb} $\Oh_j\npoints \Rh_i$ \hfill (owners do not share reps)
\item\label{hconfc} $\Oh_j\npointsx{\ol{g}}\Rh_j$
where $\ol{g} = \dom(\dfields(Own))$ \hfill (reps are private to $Own$)
\item\label{hconfd} $\Rh_j\npoints \Oh_i*\Rh_i$ \hfill (reps are confined
  to their islands)
\end{enumerate}
\end{definition}

A heap may have several admissible partitions, because there is no
inherent order on islands and because unreachable reps can be put in
any island.  The definitions and results do not depend on choice of partition. 
We have not found a workable formulation that determines unique partitions.  
To describe the effect of confined commands on partitions we use the following.

\begin{definition}[{\bf extension of confining partition, $\hext$}]\label{def:hext}
Define $h\hext h_0$ iff $h$ is confined and 
for any confining partition of $h$, 
\[ h= \Ch * \Oh_1 * \Rh_1 * \ldots * \Oh_k * \Rh_k \quad(k\geq 0),\]
there is a confining partition of $h_0$, 
\[ h_0= \Ch^0 * \Oh^0_1 * \Rh^0_1 * \ldots * \Oh^0_n * \Rh^0_n 
\epunc ,\]
that is an extension in the sense that it satisfies the following:
\begin{itemize}
\item $n\geq k$
\item $\dom(\Ch)\subseteq \dom(\Ch^0)$
\item $\dom(\Oh_j)\subseteq \dom(\Oh^0_j)$ for all $j\leq k$
\item $\dom(\Rh_j)\subseteq \dom(\Rh^0_j)$ for all $j\leq k$  \qed
\end{itemize}
\end{definition}

Confinement of a store depends on the class in which it may occur.
For owners and reps it depends on the domain of the heap as well.  
\begin{definition}[{\bf confined store, global state}]\label{def:envconf}
Let $h$ be a confined heap and $\eta$ be a store in $\means{\Gamma,\self:C}$ for some
$\Gamma$. We say $\eta$ is \dt{confined in $h$ for $C$} iff  
\begin{enumerate}
\item \label{envconfa}
$C\nleq Rep \land C\nleq Own  \implies
rng\;\eta\intersect locs(Rep\subclasses) = \Empty $
\item\label{envconfb}
$  C \leq Own  \implies 
\begin{array}[t]{l}
rng\;\eta\intersect locs(Rep\subclasses) \subseteq\dom(\Rh_j)\\ 
\mbox{for some confining partition and $j$ with $\eta\,\self\in \dom(\Oh_j)$ }
\end{array}$
\item\label{envconfc}
$  C\leq Rep \implies 
\begin{array}[t]{l}
rng\;\eta\intersect \locsOR \subseteq\dom(\Oh_j * \Rh_j)\\ 
\mbox{for some confining partition and $j$ with $\eta\,\self\in \dom(\Rh_j)$ }
  \end{array}$
\end{enumerate}
A global state $(h,\eta)$ is \dt{confined}, written 
$\etaconf{C}{\eta}{h}$, iff $h$ is confined and $\eta$ is confined in
$h$ for $C$.\qed
\end{definition}
Apropos the examples in Sect.~\ref{sect:bool}, take $Rep$ to be $\code{Bool}$ 
and suppose the sequence
\textsf{z\assym \NEW\ OBool;  w\assym z.bad()}
occurs in a method of some client class.  
Executed in a confined initial state, the state after assignment of a new
\code{OBool} to \code{z} is still confined. The assignment to \code{w}
then yields a state where the heap is confined but the client's store
is not.

\subsection{Confinement of commands and methods}

A confined command is one that preserves confinement of global states.  
Because command meanings depend on the method environment and expression
meanings, confinement for those is formalized first.
We need to ensure that a method call yields a heap confined for the caller.
This is achieved using the condition $h\hext h_0$ in the following.

\begin{definition}[{\bf confined method environment}]\label{def:muconf}
Method environment $\mu$ is confined, written $\conf{\mu}$, if and
only if the following holds for all $C$ and  $m$ with $\mtype(m,C)$
defined.
Let $\mtype(m,C)=\ol{\T}\TO\T$ and $\pars(m,C)=\ol{x}$.  
For all  $(h,\eta)\in \means{\heap\otimes \ol{x}:\ol{T},\self:C}$,  
if  $\etaconf{C}{\eta}{h}$ and $\mu C m (h,\eta) \neq \bot$ then   
\begin{enumerate}
\item\label{muconfa}
$C\nleq Rep  \implies 
\etaconf{C}{\eta}{h_0} \land h\hext h_0 \land  d\not\in locs(Rep\subclasses) $
\item\label{muconfb}
$C\leq Rep \implies 
\begin{array}[t]{l}
\etaconf{C}{\eta}{h_0} \land h\hext h_0 \\
\land\: (d\in \locsOR \implies
    d\in\dom(\Oh_j * \Rh_j)) \\
\quad \mbox{for some confining partition $h_0 = \Ch*\Oh_1*\Rh_1\ldots$}\\
\quad \mbox{and $j$ with $\eta\,\self\in\dom(\Rh_j)$} 
\end{array}$
\end{enumerate}
where $(h_0,d)=\mu C m (h,\eta)$.\qed
\end{definition}
Condition (\ref{muconfa}) fails for method \code{bad} of the
example  in Sect.~\ref{sect:bool}, regardless of whether the return
type of \code{bad} is taken to be \OBJECT\ or \code{Bool}.

The conditions for confinement of expressions are like those for
confined stores ---after all, a store provides the
meaning for the expression $x$.  The conditions are somewhat different
for confined method environments, because methods are public and can
be called both by clients and from within an owner island.
(In Sect.~\ref{sect:osub}, Def.~\ref{def:muconf} is refined to allow
module-scoped owner methods to return reps.)
Also, confinement of commands does not explicitly require heap extension
$h\hext h_0$ like Def.~\ref{def:muconf} does, because it is a
consequence of the other conditions (see Lemma~\ref{lem:c}).  

\begin{definition}[{\bf confined expression}]\label{def:econf}
Let $C=\Gamma\,\self$.
Expression $\Gamma\proves e:T$ is confined iff 
for any $(h,\eta)$,
if $\etaconf{C}{\eta}{h}$ and $\means{\Gamma\proves
  e:T}(h,\eta)  \neq \bot$ then the following hold, where
$d= \means{\Gamma\proves e:T}(h,\eta)$.
\begin{enumerate}
\item \label{econfa}
$C\nleq Rep \land C\nleq Own  \implies d\not\in locs(Rep\subclasses)$

\item
$ C \leq Own  \implies 
\begin{array}[t]{l}
(d\in locs(Rep\subclasses) \implies d\in \dom(\Rh_j))\\ 
\mbox{for some confining partition and $j$ with $\eta\,\self\in \dom(\Oh_j)$ }
\end{array}$
\item\label{econfc}
$  
C\leq Rep \implies 
\begin{array}[t]{l}
(d\in \locsOR \implies d\in \dom(\Oh_j * \Rh_j))\\ 
\mbox{for some confining partition and $j$ with $\eta\,\self\in \dom(\Rh_j)$ \qed}
  \end{array}$ 
\end{enumerate}
\end{definition}

\begin{definition}[{\bf confined command}]\label{def:cconf}
Let $C=\Gamma\,\self$.
Command $\Gamma\proves \S $ is confined iff 
\begin{itemize}
\item $\:\conf{\mu}\land\etaconf{C}{\eta}{h} \land \means{\Gamma\proves
  \S }\mu(h,\eta) \neq \bot \implies \etaconf{C}{\eta_0}{h_0}$, 
for any $\mu$ and any $(h,\eta)$,  
where $(h_0,\eta_0) = \means{\Gamma\proves \S }\mu(h,\eta)$
\item if $S$ is a method call then it has confined arguments (see below). \qed
\end{itemize}
\end{definition} 

Confinement of arguments means that the store $\eta_1$
passed in the semantics of method call is confined for the callee.

\begin{definition}[{\bf confined arguments}]\label{def:argconf}
Let $C=\Gamma\,\self$.
A call $\Gamma\proves \assign{x}{\mcall{e}{m}{\ol{e}}} $
has \dt{confined arguments} provided the following holds.
Suppose $\ol{U}$ is the static type of $\ol{e}$ and $D$ the static
type of $e$.   
For any  $(h,\eta)$ with $\etaconf{C}{\eta}{h}$, 
let 
\[ 
\ol{d} = \means{\Gamma\proves \ol{e}:\ol{U}}(h,\eta) \qquad
\ell = \means{\Gamma\proves e:D}(h,\eta) \qquad
\eta_1 = [\ol{x}\mapsto\ol{d},\self\mapsto \ell] 
\epunc .
\]  
If $\ell\neq\bot$, $\ell\neq\NIL$, and $\ol{d}\neq \bot$ then 
$\etaconf{(\loctype\,\ell)}{\eta_1}{h}$. 

A super-call $\Gamma\proves \assign{x}{\mcall{\SUPER}{m}{\ol{e}}} $
has \dt{confined arguments} provided the following holds.
Suppose $\ol{U}$ is the static type of $\ol{e}$.
For any  $(h,\eta)$ with  $\etaconf{C}{\eta}{h}$, 
let 
\[ 
\ol{d} = \means{\Gamma\proves \ol{e}:\ol{U}}(h,\eta) \qquad
\ell = \eta\,\self \qquad
\eta_1 = [\ol{x}\mapsto\ol{d},\self\mapsto \ell]\epunc .\]  
If $\ol{d}\neq \bot$ then $\etaconf{(\super\,C)}{\eta_1}{h}$. 
\qed
\end{definition}

A purely semantic formulation would call class table $CT$ confined
just if $\means{CT}$ is a confined method environment.  But under simple
restrictions, confinement of $\means{CT}$ follows from confinement of
method bodies and constructors.  Thus we choose the following.

\begin{definition}[{\bf confined class table}]\label{def:tconf}
Class table $CT$ is confined iff for every $C$ and every $m$ 
with  $\mtype(m,C)=\ol{T}\TO T$ the following hold.  
\begin{enumerate}
\item\label{tconfa}
 If $m$ is declared in $C$ by
  $\T\;m(\ol{T}\;\ol{x})\{\S\}$ then 
$S$ and all its constituents are confined.
\item\label{tconfe} If the constructor declaration in $C$ is $\construct{S}$ 
then $S$ and all its constituents are confined.  
\item\label{tconfc} If  $C\leq Own$ then   $T\ncomp Rep$.
\item\label{tconfb} If $m$ is inherited in  $Own$ from some $B > Own$
  then $\ol{T}\ncomp Rep$.
\item\label{tconfd} No $m$ is inherited in $Rep$ from any $B> Rep$.
\qed
\end{enumerate}
\end{definition}

In Sect.~\ref{sect:sabs} we add module-scoped methods on which 
condition (\ref{tconfc}) need not be imposed.
This condition ensures that owner methods do not return reps, which is
not ensured by confinement of the method body.
Condition (\ref{tconfd}) is needed because confinement of a method inherited
from $B>Rep$ depends on the arguments, including $\self$, being confined at
$B$ where reps are disallowed.   Invocation of such a method on an object
of type $Rep$ (or a subclass) would yield a store with $\self$ a
rep.  A more refined restriction is to disallow inheritance into $Rep$ only
for methods which leak $\self$; see Sect.~\ref{sect:disc}.

\begin{example}
Condition (\ref{tconfc}) precludes the
\code{bad} method of Sect.~\ref{sect:bool}, for both return types
$\OBJECT$ and $\code{Bool}$.  Except for this, all examples in 
Sect.~\ref{sect:rif} yield confined class tables (e.g., a well formed
class table is obtained by combining Figs.~\ref{fig:obs} and
\ref{fig:obsCli}). 
One way to prove confinement for these examples is to check that they are
safe according to the static analysis of Sect.~\ref{sect:static}.
For this one uses the desugarings of Remark~\ref{rem:sugar}.
\qed
\end{example}

\subsection{Properties of confinement}

We need a number of results about confinement.  The most important is that
the semantics of a confined class table is a confined method environment 
(Theorem~\ref{thm:cfix}).  This depends on Lemma~\ref{lem:c} which says
that confined commands extend heap partitions, provided that method
meanings have this property.  

\begin{lemma}\label{lem:nlec}
If $T$ is $\BOOL$ or $\UNIT$, then every $\Gamma\proves e:T$ is confined.
\end{lemma}
\begin{proof}
Direct from the definitions: confinement only pertains to locations.
\qed\end{proof}

\begin{lemma}\label{lem:a}
Suppose $rng\,\eta\intersect locs(Rep\subclasses) = \Empty$ and $C\leq B$.
Then for any $h$ and any $\eta\in\means{\Gamma,\self:C}$ we have 
$\etaconf{C}{\eta}{h}$ iff $\etaconf{B}{\eta}{h}$.
\end{lemma}
\begin{proof} Straightforward.  See Appendix.
\end{proof}

\begin{lemma}\label{lem:b}
  If $\etaconf{C}{\eta}{h}$ and $h\hext h_0$ then $\etaconf{C}{\eta}{h_0}$.
\end{lemma}
\begin{proof} Straightforward. See Appendix.
\qed\end{proof}

Although confining partitions are not unique, a given confining partition
of an initial state can be extended to one on the final state for any
command.  
This is Lemma~\ref{lem:c} below, which depends on the analogous property
for constructors, Lemma~\ref{lem:cy}.  From the proof of the latter, we
factor out the induction step as a somewhat complicated separate result,
Lemma~\ref{lem:cx}, because it is also used in Sect.~\ref{sect:static}
to show soundness of the static analysis.   Skip on first reading!

\begin{lemma}\label{lem:cx}
Let $\mu$ be a method environment.
Suppose we have the following:
\begin{enumerate}
\item \label{cxa} $\self:C\proves S$ is a confined command. 
\item \label{cxb} for any $B$ with an occurrence of $\new{B}{}$ in $S$ we have
$B\cdep C$ and moreover no method calls occur in $S$.
\item \label{cxc}  for any $B$ with an occurrence of $\new{B}{}$ in $S$,
and also for $B=\super\,C$ unless $\super\,C=\OBJECT$,
the following holds for any $(h,\eta)$ with $\etaconf{B}{\eta}{h}$: 
\[ \means{\self:B\proves S_0:\CON}\mu(h,\eta)\neq \bot \implies
h\hext h_1 \epunc,\]  
where $S_0 = \constr\:B$ and $h_1 = \means{\self:B\proves S_0:\CON}\mu(h,\eta)$.
\end{enumerate}
Then for any $(h,\eta)$ with $\etaconf{C}{\eta}{h}$, 
if $\means{\self:C\proves S:\CON}\mu(h,\eta)\neq \bot$ then 
$h\hext h_0$ where $h_0= \means{\self:C\proves S:\CON}\mu(h,\eta)$.
\end{lemma}
\begin{proof}
Assume (\ref{cxa}--\ref{cxc}) hold.
To show the conclusion for the non-$\bot$ case, consider any $(h,\eta)$
with $\etaconf{C}{\eta}{h}$ and let  
$h_1$ be as in the semantics of $S$ as a constructor.  
If $\super\,C = \OBJECT$ then $h_1=h$ and thus $h\hext h_1$.  Otherwise,
$h_1 = \means{\self:\super\,C\proves \constr(\super\,C):\CON}\mu(h,\eta)$
and $h\hext h_1$ holds by hypothesis (\ref{cxc}).     
Now by semantics, $h_0 = \means{\self:C\proves S}\mu(h_1,\eta)$.  

To show that $h\hext h_0$, we can argue by induction on the structure of
$S$.   Note that $S$ has no method calls, 
by hypothesis (\ref{cxb}), so $\mu$ is not
relevant.  Moreover, for any object construction the result holds by
hypothesis (\ref{cxc}).  
We omit the rest of the argument, which uses hypothesis (\ref{cxa}): it is exactly the same
as in the proof of Lemma~\ref{lem:c} below, 
except for appealing to hypothesis (\ref{cxb}) for the case of \NEW, where   
that proof appeals to Lemma~\ref{lem:cy}.
\end{proof}

\begin{lemma}[extension by constructors] \label{lem:cy}
Suppose $\self:C\proves \constr\:C$ is confined, for all $C$.
Then for any $(h,\eta)$ with $\etaconf{C}{\eta}{h}$ we have 
\[ \means{\self:C\proves S:\CON}\mu(h,\eta)\neq \bot \implies
h\hext h_0 \quad\mbox{where } h_0= \means{\self:C\proves S:\CON}\mu(h,\eta)
\epunc .\]
\end{lemma}
\begin{proof}
This is exactly the conclusion of Lemma~\ref{lem:cx}.  We prove it by well
founded induction on $C$, using 
$\well$ which is well founded by Lemma~\ref{lem:cdep}(\ref{cdepc}).  
For any $C$ and $S$, it suffices to show that the hypotheses
(\ref{cxa}--\ref{cxc}) of 
Lemma~\ref{lem:cx} hold for classes smaller than $C$ with respect to $\well$.
First, (\ref{cxa}) holds by hypothesis of the present Lemma.
By well formedness of the class table, there are no method calls in
$\constr\:C$, and moreover if $\new{B}{}$ occurs in $S$ then $B\cdep C$;
this is hypothesis (\ref{cxb}).  
Now from Lemma~\ref{lem:cdep} and well formedness of the class table 
we have that $B\well C$ for every 
$\new{B}{}$ that occurs in $S$ and also $\super\,C\well C$.  Thus by
the induction hypothesis we have (\ref{cxc}).
\end{proof}

\begin{lemma}[extension by commands] \label{lem:c}
Suppose $\Gamma\proves S $ is confined and all its constituents are 
confined.  
Suppose moreover that $\self:B\proves \constr\:B$ is confined, for all $B$.
Let $C=\Gamma\,\self$.
For any $\mu,h,\eta$ with $\conf{\mu}$ and  $\etaconf{C}{\eta}{h}$ 
\[ \means{\Gamma\proves S }\mu(h,\eta)\neq \bot \implies h\hext h_0
\quad\mbox{where } (h_0,-)=\means{\Gamma\proves S }\mu(h,\eta)
\epunc .\]
\end{lemma}
\begin{proof}
By structural induction on $S$.
Let $C=\Gamma\,\self$.
We assume a confining partition
$ h= \Ch * \Oh_1 * \Rh_1 * \ldots * \Oh_k * \Rh_k$ is
given ($k$ may be 0, i.e., there need not be any islands).
We show how to construct confining partition 
$ h_0= \Ch^0 * \Oh^0_1 * \Rh^0_1 * \ldots $ that extends the given one.

\medskip\Case\ $\Gamma\proves\assign{\faccess{e_1}{f}}{e_2}$.
From $\means{\Gamma\proves\assign{\faccess{e_1}{f}}{e_2}}\mu \eta
h\neq \bot$
and Lemma~\ref{lem:semty}(\ref{styd})  we have that $\ell\in\dom\,h$ where
$\ell=\means{\Gamma\proves e_1:C} (h,\eta)$.  
By semantics, $h_0 =\ext{h}{\ell}{\ext{h\ell}{f}{d}}$.
We partition $h_0$ using the given partition for $h$.
That is, the domain for each block of the updated heap $h_0$ is the same as the
corresponding block for $h$.
Clearly this extends the partition for $h$.  
To show that this partition is confining for $h_0$, it suffices to show that the update
of $h \ell f$ to $d$ satisfies the confinement property for $\ell$.  
We argue by cases on $\loctype\,\ell$
\begin{itemize}
\item 
$\loctype\,\ell\nleq Own\land\loctype\,\ell\nleq Rep$.
Then Def.~\ref{def:hconf}(\ref{hconfa}) applies; it requires $d\not\in
locs(Rep\subclasses)$.  By typing, $\loctype\,\ell\leq C$, 
so $C\nleq Own\land C\nleq Rep$.  Thus by confinement of $e_1$ (a
constituent of $\assign{\faccess{e_1}{f}}{e_2}$ and therefore confined by
hypothesis), 
we have by Def.~\ref{def:econf}(\ref{envconfa}) that $d\not\in
locs(Rep\subclasses)$.  
\item $\loctype\,\ell\leq Own$.
Def.~\ref{def:hconf}(\ref{hconfb}) and (\ref{hconfc}) apply here.
Letting $j$ be the index 
of the island with $\{\ell\} = \dom(\Oh^0_j) = \dom(\Oh_j)$, we must show 
both $\Oh^0_j\npoints \Rh^0_i$ (for $i\neq j$) and 
$\Oh^0_j\npointsx{\ol{g}}\Rh^0_j$.  
By typing, $\loctype\,\ell\leq C$, so $C\leq Own$ or $Own\leq C$ by the
tree property of $\leq$. 
We argue by cases on $C$.
\begin{itemize}
\item  $Own<C$.
By $Own\ncomp Rep$, we have $C\nleq Rep$ so confinement of $e_2$ at
$C$ yields $d\not\in locs(Rep\subclasses)$.  Thus 
$\Oh^0_j\npoints \Rh^0_i$  and $\Oh^0_j\npointsx{\ol{g}}\Rh^0_j$.

\item $C\leq Own$.  
By confinement of $e_2$, if $d\in locs(Rep\subclasses)$ then $d\in
\dom(\Rh^0_j)$ so $\Oh^0_j\npoints \Rh^0_i$ for $i\neq j$.  
If $C=Own$ then, by the typing rule for field update, $f$ is in the private
fields $\ol{g}$ of $Own$, so the update cannot violate
$\Oh^0_j\npointsx{\ol{g}}\Rh^0_j$.   If $C<Own$ then 
$d\not\in locs(Rep\subclasses)$ because if $d$ is a rep then there would be
no confining partition, contradicting confinement of $h_0$ which holds by
confinement of $S$.
\end{itemize}

\item $\loctype\,\ell\leq Rep$.
Def.~\ref{def:hconf}(\ref{hconfd}) applies in this case:
we need to show $\Rh^0_j\npoints \Oh^0_i*\Rh^0_i$ where $i\neq j$ and $j$
is the island for $\ell$ in the partition of $h$.  
By typing, $\loctype\,\ell\leq C$, hence $C\leq Rep$ or $Rep <C$.  
But if $Rep<C$ then $C\ncomp Own$ and the confinement condition for $e_1$
(Def.~\ref{def:econf}(\ref{econfa})) at $C$ contradicts 
$\loctype\,\ell\leq Rep$, so we have $C\leq Rep$.  Now  confinement of
$e_2$ yields  $d\in \locsOR \implies d\in \dom(\Oh_j*\Rh_j)$.  
This proves $\Rh^0_j\npoints \Oh^0_i*\Rh^0_i$, because 
$\dom(\Oh_j*\Rh_j) = \dom(\Oh^0_j*\Rh^0_j)$.
\end{itemize}

\Case\ $\Gamma\proves\assign{x}{\new{B}{\ }}$.
In the semantic definition, 
$h_1 =\ext{h}{\ell}{[\fields\,B\mapsto\mathit{defaults}]}$ where $\ell = \fresh(B,h)$.  
Define $Bh = [\ell\mapsto [\fields\,B\mapsto\mathit{defaults}]]$ so $h_1 = h* Bh$.  
Let $\eta_1 = [\self\mapsto \ell]$.  
Next, we argue that $h\hext h_1$ and $\etaconf{B}{\eta_1}{h_1}$.
Because $h$ is closed, $\ell$ is not in the range of any object state in
$h$. To construct an extending partition it suffices to deal with the
new object, as its addition cannot violate confinement of existing
objects.  (This would not be the case if dangling pointers were allowed,
unless further restrictions are imposed on $\fresh$.)
We define the extension and argue by cases on $B$.
\begin{itemize}
\item $B\nleq Own\land B\nleq Rep$.
For a confining partition of $h_1$ we extend that for $h$ by 
defining $\Ch^0 = \Ch * Bh$ and using the given partition of owner islands.
Because $\mathit{defaults}$ contains no locations, this is a confining
partition and we have $\etaconf{B}{\eta_1}{h_1}$. 

\item $B\leq Own$.  We extend the partition by adding an island 
$\Oh^0_{k+1}*\Rh^0_{k+1}$ with 
$\Oh^0_{k+1} =  Bh$ and $\Rh^0_{k+1}=\Empty$.  This is a confining partition because
$\mathit{defaults}$ has no locations and we have $\etaconf{B}{\eta_1}{h_1}$
because $rng\,\eta_1$ has no reps.   

\item $B\leq Rep$.  We can obtain a confining extension by adding 
$Bh$ to any of the $\Rh_i$, as $\mathit{defaults}$ has no locations.
As $rng\,\eta_1= \{\ell\}$, we have $\etaconf{B}{\eta_1}{h_1}$ 
by definition.\footnote{In this case we have $C\leq Own$ or $C\leq Rep$, 
as otherwise the command would not be confined.  
To show $\etaconf{C}{\eta}{h_1}$ in this case, we would have to 
we put $\ell$ in $\Rh_j$, choosing $j$ such that $\eta\,\self$ is in the
$j$th island.  But we are only showing the extension of the partition for
this lemma.  For soundness of the static analysis, we do have to show
$\etaconf{C}{\eta}{h_1}$.}
\end{itemize}
This concludes the argument for $h\hext h_1$ and
$\etaconf{B}{\eta_1}{h_1}$.
These let us  apply 
 Lemma~\ref{lem:cy} for $\constr\:B$ to get $h_1\hext
h_0$ where $h_0=\means{\self:B\proves \constr\:B:\CON}\mu(h_1,\eta_1)$.  
Then $h\hext h_0$ by transitivity of $\hext$.

\medskip\Case\ $\Gamma \proves \assign{x}{\mcall{e}{m}{\ol{e}}}$. 
As $ \mcall{e}{m}{\ol{e}}$ is confined, its argument values are confined.  
Thus we can obtain the result directly from confinement of $\mu$, which
explicitly stipulates $h\hext h_0$, and semantics of $
\mcall{e}{m}{\ol{e}}$. 

\medskip
The remaining cases are straightforward.  See Appendix.\qed
\end{proof}

\begin{theorem}\label{thm:cfix}
Suppose that $CT$ is confined.
Then the semantics $\means{CT}$ is confined, as is each $\mu_j$ in
the approximation chain used to define it.
\end{theorem}

The proof uses fixpoint induction, which is only sound for inclusive predicates, i.e., those
closed under limits of ascending chains.
For confinement of method environments the definition is given
pointwise, ultimately unfolding to the property that the semantics of each
method body preserves confinement.  This definition, as well as the one for the
simulation $\REL$ later, is in the usual form of logical relations.  
By the structure of the definition, and continuity of the semantics, 
the property is an inclusive predicate.\footnote{See Ploto's notes.}

\begin{proof}
Confinement of $\means{CT}$ follows by fixpoint induction
from confinement of $\mu_i$ for all $i$, which we show by induction on
$i$.  The base case holds because $\mu_0 C m = \lam{(h,\eta)}{\bot}$, 
for any $C,m$, and this is confined by definition. 

For the induction step, suppose $\conf{\mu_i}$, to show
$\conf{\mu_{i+1}}$. Consider an arbitrary $m$.  
We argue for all $C$ with $\mtype(m,C)$ defined, by induction on
method depth (Def.~\ref{def:dep}) of $C$ for $m$.
The base case is $C$ such that $depth(m,C)=0$.  In this case, 
$CT(C)$ has a declaration 
\[  \T\; m(\ol{\T}\;\ol{x})\{\S\} \epunc .\]
Suppose $\etaconf{C}{\eta}{h}$ and $\mu_{i+1} C m (h,\eta) \neq \bot$.
Let $(h_0,d)=\mu_{i+1} C m (h,\eta)$, which by definition of $\mu_{i+1}$ is obtained as
\begin{eqnarray*}
\eta_1 &=& \ext{\eta}{\result}{\mathit{default}} \\
(h_0,\eta_0) &=& \means{\ol{x}:\ol{T},\self:C,\result:\T\proves S }\mu_i (h,\eta_1)\\  
d &=& \eta_0\,\result
\end{eqnarray*}
Default values do not violate confinement so $\etaconf{C}{\eta_1}{h}$.  
As $CT$ is confined, $S$ and its constituents are confined.
By Lemma~\ref{lem:c} we have $h\hext h_0$, so by Lemma~\ref{lem:b} we have
$\etaconf{C}{\eta}{h_0}$.  
To show the confinement condition for $\mu_{i+1} C m $ it remains to deal with the
result value $d$.  
We have $\etaconf{C}{\eta_0}{h_0}$ by confinement of $S$. We argue by cases on $C$.  
\begin{itemize}
\item $C\nleq Own \land C\nleq Rep$.  We need $d\not\in
  locs(Rep\subclasses)$, for Def.~\ref{def:muconf}(\ref{muconfa}), 
 and this follows from 
 $\etaconf{C}{\eta_0}{h_0}$ by Def.~\ref{def:envconf}(\ref{envconfa}).
\item $C \leq Own$.  We need $d\not\in locs(Rep\subclasses)$, and since by
  typing we have $d\in\means{T_\bot}$, 
Def.~\ref{def:tconf}(\ref{tconfc}) ensures $T\ncomp Rep$ and hence $d\not\in
locs(Reps\subclasses)$.
(Note that semantic confinement of $\eta_0$ at $C\leq Own$ allows reps, so
it is not enough for this case).
\item $C\leq Rep$.  Then we need $d\in locs(Own\subclasses,
  Rep\subclasses)$ to imply that $d$ is in the domain of $\Oh_j*\Rh_j$
  for some partition and island $j$ such that
  $\eta\,\self\in\dom(\Oh_j*\Rh_j)$.  This follows from 
 $\etaconf{C}{\eta_0}{h_0}$ by Def.~\ref{def:envconf}(\ref{envconfc}).
\end{itemize}
This concludes the base case of the induction on depth.

For the induction step, i.e., $depth(m,C) > 0$, 
$m$ may be inherited or declared in $C$.  If it is
declared in $C$ the argument is the same as for the case
$depth(m,C)=0$ above. Suppose $m$ is inherited in $C$ from $B$.  Now 
$\mu_{i+1} C m = \restr((\mu_{i+1} B m),C)$ by definition of $\mu_{i+1}$. 
By induction on depth $\mu_{i+1} B m$ satisfies the confinement
condition for $m,B$.
To show the condition for $\mu_{i+1} C m$, suppose $\etaconf{C}{\eta}{h}$.
We claim that $\etaconf{B}{\eta}{h}$.   
Using the claim, we argue as follows. 
If $\mu_{i+1} B m (h,\eta)\neq \bot$, let 
$(h_0,d) = \mu_{i+1} B m (h,\eta)$.  
By induction on depth we have $\etaconf{B}{\eta}{h_0}$ and $h\hext h_0$.
By Lemma~\ref{lem:b} we obtain $\etaconf{C}{\eta}{h_0}$.  
It remains to show the confinement condition for $d$ and to prove the
claim.  We argue by cases on $C$.

In the following non-rep cases, the claim holds by Lemma~\ref{lem:a}.
To apply the Lemma, we just need to show that 
$rng\,\eta\intersect locs(Rep\subclasses) = \Empty$. 
\begin{itemize}
\item $C\nleq Own\land C\nleq Rep$. 
In this case, 
we have $rng\,\eta\intersect locs(Rep\subclasses) = \Empty$ by
confinement of $\eta$ at $C$, Def.~\ref{def:envconf}(\ref{envconfa}).
\item $C\leq Own<B$.  Then $Own$ inherits $m$ from $B>Own$, so by
  confinement of the class table,
  Def.~\ref{def:tconf}(\ref{tconfb}), we have $\ol{T}\ncomp 
  Rep$.  Also, $Own\ncomp Rep$, so by Lemma~\ref{lem:ncomp} we have no
  reps in $rng\,\eta$. 
\end{itemize}
In the preceding cases, the condition imposed on $d$ by
Def.~\ref{def:muconf}(\ref{muconfa}) for class $C$  
is $d\not\in locs(Rep\subclasses)$.  But this same condition is
imposed for class $B$, and it holds by induction on depth.
For the remaining cases we prove the claim $\etaconf{B}{\eta}{h}$ as follows.
\begin{itemize}
\item $C< B\leq Own$.  Both $B$ and $C$ impose the same condition
(Def.~\ref{def:envconf}(\ref{envconfb})).    
\item $C<B\leq Rep$.  
Both $C$ and $B$ impose the same conditions on $\eta$
(Def.~\ref{def:envconf}(\ref{envconfc})).    
\end{itemize}
In these two cases the requirement for $d$ at $C$,
Def.~\ref{def:muconf}(\ref{muconfb}) or (\ref{muconfa}), is
the same as for $B$, so it holds by induction on depth.

The case $C\leq Rep<B$ cannot occur in a confined class table.
If $m$ is inherited in $C\leq Rep$ from $B$ then it is inherited in $Rep$
from $B$, and this is explicitly disallowed in
Def.~\ref{def:tconf}(\ref{tconfd}).
\qed\end{proof}


\section{First abstraction theorem}\label{sect:abs}

This section formulates and proves the central result of the paper.
First, we make precise the idea of comparing two class tables that differ
only in their implementation of class $Own$.  Then we define basic
coupling: a relation between single instances of class
$Own$ for the two implementations.  This induces the coupling relations for
other data types, for heaps containing multiple instances of
$Own$, and for method meanings.  Related method meanings have the
simulation property: if initial states are coupled, then so are outcomes.  
The main theorem says that if methods of $Own$ have the simulation
property, then so do all methods of all classes.  

\subsection{Comparing class tables}

We compare two implementations of a designated class $Own$.  They  can have
completely different declarations, so long as methods of the same
signatures are present ---declared or inherited--- in both.  They can use
different reps, distinguished 
by class name $Rep$ for one implementation and $Rep'$ for the other. 
We allow $Rep = Rep'$. 
For simplicity, we assume that both $Rep$ and $Rep'$ are in each of the two 
compared class tables.\footnote{An alternative formulation would consider
  different declarations of $Own$ together with associated class tables in
  which $Rep$ or $Rep'$ but not both are declared.  But these could be
  combined into class tables fitting our formulation.}  

\begin{definition}[{\bf comparable class tables, non-rep classes}]\label{def:comp}
Suppose class names $Own,Rep,Rep'$ are given, such that 
$Own\ncomp Rep$ and $Own\ncomp Rep'$.
We say $C$ is a \dt{non-rep} class iff $C\nleq Rep$ and $C\nleq Rep'$. 
Well formed class tables $CT$ and $CT'$ are \dt{comparable} provided
the following hold.
\begin{enumerate}
\item\label{simb} $CT$ and $CT'$ are identical except for their values
  on $Own$.  (In particular, $CT(Rep)=CT'(Rep)$ and 
$CT(Rep')=CT'(Rep')$.)

We write $\proves,\proves'$ for the typing relations determined by $CT,CT'$
respectively, and similarly for the auxiliary functions, such as
$\mtype,\mtype'$.
We also write $\means{-},\means{-}'$ for the respective semantics and
assume that the same allocator, $\fresh$, is used for both 
$\means{-}$ and $\means{-}'$.
\item
$\super\,Own = \super'\,Own$.
\item\label{simd}
For any $m$, either $\mtype(m,Own)$ and $\mtype'(m,Own)$ are both undefined
or both are defined and equal. \qed
\end{enumerate}
\end{definition}

\begin{example}
Let $CT$ be given by Figs.~\ref{fig:obs} and \ref{fig:obsCli}.
Let $CT'$ be
given by Figs.~\ref{fig:obsaa} and \ref{fig:obsCli} together with
\code{Observer} from Fig.~\ref{fig:obs}.  These are comparable. \qed 
\end{example}


Instead of condition (\ref{simd}), one could require that $CT(Own)$ and
$CT'(Own)$ declare the same methods.  But that would disallow some
situations that occur in practice.
Suppose class $C$ extends $B$ by adding a method $m$ implemented using
calls to methods inherited from $B$.
This might be the easiest way to achieve desired functionality for $m$, but
there could be an alternative data structure that is more efficient for $m$
and for the methods of $B$.  An alternative implementation of $C$ could 
add that data structure and override the methods of $B$ to use it.
One can argue that the program is poorly designed, e.g., because space for
attributes of $B$ is wasted in $C$ objects.
Better designs are possible in languages with interface types separate from
classes. Nonetheless, 
such examples do arise in practice.  Allowing them complicates the
proof of Theorem~\ref{thm:fat} but none of the other results.
The main  consequence we need from condition (\ref{simd}) is the following.

\begin{lemma}\label{lem:eqdep}
If $\mtype(m,C)$ is defined then $depth(m,C) = depth'(m,C)$.  
\end{lemma}
\begin{proof} Straightforward. See Appendix.
\qed\end{proof}

One can imagine a theory in which an owner subclass $C< Own$ has
different declarations in $CT$ and $CT'$.  But we are concerned with 
an abstraction provided by a single class rather than by a
collection of classes, so $CT(C)=CT'(C)$ here.  In
Sect.~\ref{ssect:osub} we impose a restriction on owner
subclasses that is needed for the first abstraction theorem.  
The issue is explored in Sect.~\ref{sect:app} and the
restriction lifted in Sect.~\ref{sect:sabs}.  

\subsection{Coupling relations and simulation}\label{ssect:coup}

The definitions are organized as follows.  A \dt{basic coupling} $R$ is a
suitable relation on islands.  This induces a family of \dt{coupling relations},
$\REL\:\theta$ for each $\theta$.   Then comes the definition of
\dt{simulation}, a coupling that is preserved by all methods of $Own$ and
established by the constructor.

\begin{definition}[{\bf basic coupling}]\label{def:psim}
Given comparable class tables, a \dt{basic coupling} is a binary 
relation $R$ on heaps ---not necessarily closed--- such that 
the following holds:
For any $h,h'$, if $R\:h\:h'$ then there is a location $\ell$
 with $\loctype\,\ell \leq Own$ and partitions 
$h=\Oh *\Rh$ and $h'=\Oh'*\Rh'$ such that  
\begin{enumerate}
\item\label{simea}
 $\dom\,\Oh  = \{\ell\} = \dom\,\Oh'$
\item
 $\dom(\Rh) \subseteq locs(Rep\subclasses)$ 
and   $\dom(\Rh') \subseteq locs(Rep'\subclasses)$ 
\item\label{simec} $h\ell f = h'\ell f $
for all $(f:T)\in \dom(\fields(\loctype\,\ell))$ with $f\not\in
\ol{g}$ and $f\not\in \ol{g}'$,
where $\ol{g}= \dom(\dfields(Own))$ and 
$\ol{g}'= \dom(\dfields'(Own))$
 \qed
\end{enumerate}
\end{definition}
Example~\ref{ex:obsSim} below shows why we allow $R$ to act on heaps that are not closed.

Although $R$ is unconstrained for the private fields and reps, condition
(\ref{simec}) determines it for fields of subclasses of $Own$.
Once we have defined the induced relation $\REL$, item
(\ref{simec}) will be equivalent to the condition
$ \rel{(\type(f,\loctype\,\ell))}{(h\ell f)}{(h'\ell f)}$.

Because $CT$ and $CT'$ are well formed, the declared field names $\ol{g}$ and $\ol{g}'$
do not occur as fields of subclasses or superclasses of $Own$.  
In (\ref{simec}), $f$ ranges over fields of both subclasses and superclasses; except
for $\ol{g}$ and $\ol{g}'$, we have $\fields\,C=\fields'\,C$ for all $C$.
The typing relations $\proves$ and $\proves'$ are also the same except for class
$Own$.

\begin{example}\label{ex:OBsim}
Sect.~\ref{sect:repe} discusses this coupling relation:
\[ (\code{o.g} = \NIL = \code{o}'.\code{g}) \lor (\code{o.g} \neq \NIL \neq
\code{o}'.\code{g} \land \code{o.g.f} = \neg(\code{o}'.\code{g.f})) \epunc .\]
For this example we take both $Rep$ and $Rep'$ to be \code{Bool}, and $Own$ to be
\code{OBool}.  
The displayed formula can be interpreted as relation $R$ which relates 
$h$ to $h'$ just if either 
$ h=[\ell_1\mapsto [g\mapsto \NIL]]$ and $h'=[\ell_1\mapsto [g\mapsto
\NIL]]$ or else 
\[
\begin{array}{l}
h=[\ell_1\mapsto [g\mapsto \ell_2], \ell_2\mapsto [f\mapsto d]] \mbox{ and }
h'=[\ell_1\mapsto [g\mapsto \ell_3], \ell_3\mapsto [f\mapsto \neg d]] 
\end{array}
\] 
for some boolean $d$ and locations $\ell_1$ in $locs\,\code{OBool}$
and $\ell_2,\ell_3$ in $locs\,\code{Bool}$. 
We assume that the class table contains only
\code{Bool}, \code{OBool}, and some client classes. 
If \code{OBool} had subclasses, the relation on their fields would be
determined by condition (\ref{simec}) above.
\qed\end{example}

\begin{example}\label{ex:msim}
Sect.~\ref{sect:mse} uses the formula $\code{o.g} =
\code{o}'.\code{g} \land \code{o.g}\mathbin{\keyword{mod}} 2 = 0 $.
This can  be interpreted as the 
basic coupling $R$ that relates $h$ to $h'$ just if there is some $\ell$
with $\loctype\;\ell \leq \code{A}$, $h$ and $h'$ have domain $\{\ell\}$,
and $h\,\ell\,g = h'\,\ell\,g = 2\times m$ for some integer $m\geq 0$.  
\qed\end{example}

\begin{example}\label{ex:obsSim}
The \code{Observer} examples show why we allow $R$ to relate non-closed heaps.
Consider the version in Fig.~\ref{fig:obs}.
Here $Rep$ is \code{Node}, $Own$ is \code{Observable}, and there is a
client class \code{Observer}.    
Fig.~\ref{fig:confineObs} illustrates two instances of this simple data structure.  
Fig.~\ref{fig:obsaa} gives code for an alternative version which uses an
extra node as  sentinel for the list.  The sentinel does not point to an
\code{Observer}. 
Fig.~\ref{fig:basicSimObs} depicts a corresponding pair of heaps for the
two alternatives, using arrows without destination objects to indicate
dangling pointers.  
\begin{narrowfig}{20em}
  \begin{center}
\psfrag{xxObservable}{\code{Observable}}
\psfrag{xxNode}{\code{Node}}
\psfrag{xxNode2}{\code{Node2}}
\psfrag{xxl}{$\ell$}
\psfrag{xxl1}{$\ell_1$}
\psfrag{xxl2}{$\ell_2$}
\psfrag{xxl3}{$\ell_3$}
\psfrag{xxl4}{$\ell_4$}
\psfrag{xxl0p}{$\ell'_0$}
\psfrag{xxl1p}{$\ell'_1$}
\psfrag{xxl3p}{$\ell'_3$}
\includegraphics{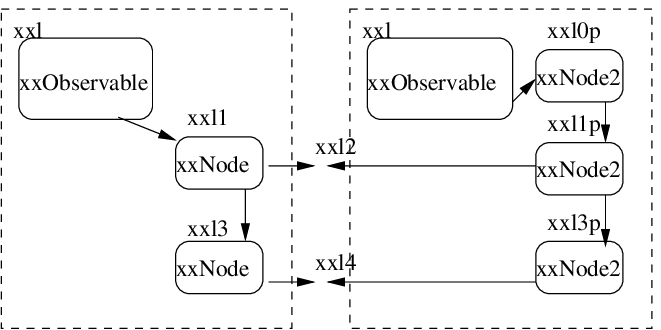}     
  \end{center}
    \caption{Basic coupling example.  Labels indicate locations as
      described in Example~\ref{ex:obsSim}. Note dangling pointers $\ell_2$ and $\ell_4$ and
      sentinel node $\ell'_0$.}
    \label{fig:basicSimObs}
\end{narrowfig}
Upon initialization of an \code{Observable}, there are no installed
\code{Observer}s, so for the version of Fig.~\ref{fig:obs} we should have
\code{fst} = $\NIL$.  But in the alternative version, this should
correspond to \code{snt} holding the location of a \code{Node} with
\code{ob} = $\NIL$.  This is established by the constructor in 
Fig.~\ref{fig:obsaa}.
An attempt at formalizing the correspondence is as follows:
\[ (\code{o.fst}=\NIL=\code{o}'.\code{snt.nxt}) \lor
(\code{o.fst}\neq\NIL\neq\code{o}'.\code{snt.nxt} 
 \land \alpha(\code{o.fst}) = \alpha'(\code{o}'.\code{snt.nxt}) \]
where $\alpha,\alpha'$ are functions that yield the list of locations in the
 \code{ob} fields of successive nodes.  But how should this formula be
 interpreted if, say, $\code{o'.snt}=\NIL$ or there is sharing such as a
 chain with cyclic tail?  Separation logic \cite{ReynoldsPtrs} 
offers a precise way to formulate such  definitions 
but its development is at an early stage.
 We simply sketch the coupling in terms of semantics:
$R\,h\,h'$ iff 
either $h$ and $h'$ have the form
\[ 
 \begin{array}{l}
h = [\: \ell \mapsto [\code{fst}\mapsto \NIL]] 
\\
h' = [\: \ell \mapsto [\code{snt}\mapsto \ell'_0],
\; \ell'_0 \mapsto [\code{ob}\mapsto \NIL,\code{nxt}\mapsto\NIL]]
 \end{array}
\]
or they have the form 
\[ 
 \begin{array}{l}
h = [\: \ell \mapsto [\code{fst}\mapsto \ell_1],
\; \ell_1 \mapsto [\code{ob}\mapsto \ell_2, \code{nxt}\mapsto\ell_3], 
\; \ell_3 \mapsto [\code{ob}\mapsto \ell_4, \code{nxt}\mapsto \ldots],
        \ldots ] 
\\
h' = [\: \ell \mapsto [\code{snt}\mapsto \ell'_0],
\; \ell'_0 \mapsto [\code{ob}\mapsto \NIL,\code{nxt}\mapsto\ell'_1], \\
\hspace*{9em} \ell'_1 \mapsto [\code{ob}\mapsto \ell_2,\code{nxt}\mapsto\ell'_3], 
\; \ell'_3 \mapsto [\code{ob}\mapsto \ell_4, \code{nxt}\mapsto \ldots],
        \ldots ] 
 \end{array}
\]
for some locations 
$\ell$ in $locs(\code{Observable})$, 
$\ell_1,\ell_3,\ldots$ in $locs(\code{Node})$, 
$\ell'_0,\ell'_1,\ell'_3,\ldots$ in $locs(\code{Node2})$,
and $\ell_2,\ell_4,\ldots$ in $locs(\code{Observer}\subclasses)$.

Note that the owners are at the same location, $\ell$, as are the
referenced client objects at $\ell_2,\ell_4,\ldots$. No 
correspondence is required between locations 
$\ell_1,\ell_3,\ldots$ and $\ell'_0,\ell'_1,\ell'_3,\ldots$ of reps.
\qed\end{example}

A basic coupling induces a relation $\REL\:\heap$ on arbitrary heaps by requiring
that they  have confining partitions such that islands can be put in
correspondence so that pairs are related by $R$.  
The formal definition uses the induced relation 
$\REL\:(\state{C})$ for object states of non-rep classes $C\nleq Own$,
and this in turn is defined in terms of 
$\REL\:C$ for non-rep classes $C\nleq Own$.  
For uniformity, we give the definition of $\REL$ for all $\theta$, 
but forcing the case for $\theta=\state{Own}$ to be false, as the compared states
have different fields.  
Aside from the ramifications of heap confinement, the definition 
is induced in the standard way for logical relations. 

\begin{definition}[{\bf coupling relation, $\REL\;\theta$}]\label{def:R}
In the context of a basic coupling with given relation $R$, we define
for each $\theta$ a relation
$\REL\:\theta \subseteq \means{\theta}\times\means{\theta}'$ as follows.  

For heaps $h, h'$, we define  $\rel{\heap}{h}{h'}$
iff there are confining partitions of $h,h'$, 
with the same number $n$ of owner islands,
such that 
\begin{itemize}
\item 
$ R \; (\Oh_i * \Rh_i) \; (\Oh'_i * \Rh'_i)  $ for all $i$ in $1..n$
\item $\dom(\Ch)=\dom(\Ch')$
\item   $\rel{(\state{(\loctype\;\ell)})}{(h\ell)}{(h'\ell)}$ 
for all $\ell\in\dom(\Ch)$ 
\end{itemize}
For other categories $\theta$ we define 
$\REL\:\theta$ as follows.
\[\begin{array}{lcl}
\rel{\BOOL}{d}{d'}&\Leftrightarrow& d=d'\\[.5ex]
\rel{\UNIT}{d}{d'}&\Leftrightarrow& d = d' \\[.5ex]
\rel{C}{d}{d'}&\Leftrightarrow& d=d'\\[.5ex]
\rel{\Gamma}{\eta}{\eta'}& \Leftrightarrow&
\all{x\in\dom\,\Gamma}{\rel{(\Gamma x)}{(\eta x)}{(\eta' x)}}\\[.5ex]
\rel{(\state{C})}{s}{s'}&\Leftrightarrow& \\[.5ex]
\multicolumn{3}{l}{\qquad C\nleq Own \land 
\all{ f\in\dom(\fields\,C) }{ \rel{(\type(f,C))}{(s\;f)}{(s'\;f)} }}\\[.5ex]
\rel{(\theta_\bot)}{\alpha}{\alpha'} &\Leftrightarrow& 
(\alpha=\bot=\alpha')\vee(\alpha\neq\bot\neq\alpha'\land\rel{\theta}{\alpha}{\alpha'})
\\[.5ex] 
\rel{(\heap\otimes\Gamma)}{(h,\eta)}{(h',\eta')} &\Leftrightarrow& 
\rel{\heap}{h}{h'}\land \rel{\Gamma}{\eta}{\eta'}
\\[.5ex] 
\rel{(\heap\otimes\T)}{(h,d)}{(h',d')} &\Leftrightarrow& 
\rel{\heap}{h}{h'}\land \rel{\T}{d}{d'}
\\[.5ex] 
\rel{(C,\ol{x},\ol{T}\TO T)}{d}{d'}&\Leftrightarrow&
\all{ (h,\eta)\in\means{\heap\otimes\Gamma},(h',\eta')\in\means{\heap\otimes\Gamma}'\:}{}
 \\[.5ex]
\multicolumn{3}{l}{\qquad \rel{(\heap\otimes\Gamma)}{(h,\eta)}{(h',\eta')}
\land  \etaconf{C}{\eta}{h}\land \etaconf{C}{\eta'}{h'} 
} 
\\[.5ex]
\multicolumn{3}{l}{ \qquad 
\implies \rel{(\heap\otimes\T)_\bot}{(d (h,\eta))}{(d'(h',\eta'))} }
\\[.5ex] 
\multicolumn{3}{l}{
\qquad  \mbox{where }\Gamma= [\ol{x}\mapsto \ol{T},\self\mapsto C] } \\[.5ex]
\rel{\menv}{\mu}{\mu'}&\Leftrightarrow& \all{ C,m \:}{
(C\mbox{ is non-rep})\land(\mtype(m,C)\mbox{ is defined})}
\\[.5ex]
\multicolumn{3}{l}{
\qquad   \implies
\rel{(C,\pars(m,C),\mtype(m,C))}{(\mu\, C\, m)}{(\mu'\, C\, m)} }  
\end{array}
\]
\end{definition}

The gist of the abstraction theorem is that if methods of $Own$ are related
by $\REL$ then all methods are.  We can now express this conclusion as 
$\rel{\menv}{\means{CT}}{\means{CT'}'}$.  
To express the antecedent, note that the relation applicable to a method $m$ of $Own$ is
$\REL\:(Own,\ol{x},\ol{\T}\TO\T)$ where $\mtype(m,Own) = \ol{\T}\TO\T$
and $\pars(m,Own)=\ol{x}$.
The definition of $\REL\:(C,\ol{x},\ol{T}\TO T)$ 
quantifies over confined initial states but does not require confinement of
outcomes.\footnote{One might think that $\REL\:\heap$  could be
defined in terms of admissible partitions without the assumption of
confinement.  But because partitions are not unique this leads to
difficulties: a heap could be confined with respect to one partition but
related with respect to another.}
The antecedent will also take into account that methods may be declared or
inherited.

Although the definition is technically intricate, the core idea is the
extension of a basic coupling, for a single owner instance, to a heap
containing potentially many owners.  This idea is given straightforward
expression using heap partitions.  By contrast, sharing of representations
between owners would require a more complicated form of extension (see
Sect.~\ref{sect:disc}).

We aim to define per-instance simulations, and in particular the
establishment of such a relation by a constructor of class $Own$ on a
single island.  But to formulate this semantically we  describe the
constructor's action on a heap in which other islands may be present.
The reason is that there is not an easy way to connect a constructor's
action on a small heap with its action on a larger one (see Footnote~\ref{fn:frame}).

\begin{definition}[{\bf simulation}]\label{def:sim}
A simulation is a coupling $\REL$ such that the following hold.
\begin{enumerate}
\item \label{simce}
(constructors of $Own$ establish $\REL$)
For any $\ell\in locs(Own\subclasses)$, any $h,h'$ with $\rel{\heap}{h}{h'}$,  
and any $\mu,\mu'$, let 
\[ 
\begin{array}{l}
h_1 = \ext{h}{\ell}{[\fields(\loctype\,\ell)\mapsto
  \mathit{defaults}]} \\
h'_1 = \ext{h'}{\ell'}{[\fields(\loctype\,\ell')\mapsto
  \mathit{defaults}']} \\
h_0 =\means{\self:(\loctype\,\ell)\proves
  \constr(\loctype\,\ell):\CON}\mu(h_1,[\self\mapsto\ell]) \\
h'_0 =\means{\self:(\loctype\,\ell)\proves'
  \constr(\loctype\,\ell):\CON}\mu'(h'_1,[\self\mapsto\ell]) 
\end{array}
\]
Then $R\;h_0\;h'_0$. 
\item \label{simp}
(methods of $Own$ preserve $\REL$)
Let $\mu\in \nats\to\means{\menv}$ (resp.\ $\mu'\in \nats\to\means{\menv}'$) 
be the approximation chain in the definition of $\means{CT}$ (resp.\ $\means{CT'}'$). 
For every $m$ with $\mtype(m,Own)$ defined, 
the following implications hold for every $i$,
where $\ol{x}= \pars(m,Own)$ and $\ol{T}\TO\T = \mtype(m,Own)$.  
 \begin{enumerate}
 \item\label{simA} $ \rel{\menv}{\mu_i}{\mu'_i} \implies 
\rel{(Own,\ol{x},\ol{\T}\TO\T )}{(\means{M}\mu_i)}{(\means{M'}'\mu'_i)}  $ \\[.5ex]
if $m$ has declaration $M$ in $CT(Own)$ and  $M'$ in $CT'(Own)$ 
\item\label{simB} $ \rel{\menv}{\mu_i}{\mu'_i} \implies 
\rel{(Own,\ol{x},\ol{\T}\TO\T )}{(\means{M}\mu_i)}{(\restr(\means{M_B}'\mu'_i,Own))}  $\\[.5ex]
if $m$ has declaration $M$ in $CT(Own)$ and is inherited from $B$ in
$CT'(Own)$, with $M_B$  the declaration of $m$ in $B$
\item\label{simC} $ \rel{\menv}{\mu_i}{\mu'_i} \implies 
\rel{(Own,\ol{x},\ol{\T}\TO\T )}{(\restr(\means{M_B}\mu_i,Own))}{(\means{M'}'\mu'_i)}  $\\[.5ex]
if $m$ has declaration $M'$ in $CT'(Own)$ and is inherited from $B$ in
$CT(Own)$, with $M_B$  the declaration of $m$ in $B$ 
\end{enumerate}
\end{enumerate}
\end{definition}
In the case where constructors in $Own$ and its subclasses are \SKIP,
condition (\ref{simce})  simply says that the default values are related.
Note that it also precludes aborting constructors, as $R$
applies to heaps but not to $\bot$; this is convenient but not necessary.

The following properties are straightforward consequences of the definition.

\begin{lemma}\label{fact:foo}
For all $h, h'$ and all locations $\ell\not\in
locs(Rep\subclasses,Rep'\subclasses)$,  if $\rel{\heap}{h}{h'}$ then
$\ell\in\dom\,h\iff\ell\in\dom\,h'$. \qed
\end{lemma}

\begin{lemma}\label{lem:sty}
$\means{T}=\means{T}'$ for all $T$, and 
$\means{\Gamma}=\means{\Gamma}'$ for all $\Gamma$.\qed
\end{lemma}

\begin{lemma}\label{fact:id}
For any data type $T$, 
$\REL\;T$ is the identity relation on $\means{T}$
and $\REL\;T_\bot$ is the identity relation on $\means{T_\bot}$.\qed
\end{lemma}

\begin{lemma}\label{fact:id:c}
If $\ol{U}\leq\ol{T}$ and $\rel{\ol{U}}{\ol{d}}{\ol{d'}}$ then
$\rel{\ol{T}}{\ol{d}}{\ol{d'}}$.\qed
\end{lemma}

\subsection{Restricting reps in owner subclasses}\label{ssect:osub}

The preceding properties express a strong connection between locations for related
heaps.  To ensure that this connection is preserved by object construction,
we shall assume the allocator is parametric.  But it is not reasonable to
require that related heaps have the same rep locations, so parametricity
cannot be exploited for reps.  As a result, the present form of simulation
is not adequate for construction of reps in subclasses of $Own$, although
such construction is allowed by confinement.  The first abstraction theorem
depends on an assumption expressed in the following terms.  

\begin{definition}[{\bf new rep in sub-owner}]
We say $CT$ \dt{has a new rep in a sub-owner} if, for some $B\leq Rep$ or
$B\leq Rep'$, an object construction $\new{B}{}$  occurs in some
method declaration in a class $C<Own$.
\end{definition}
If $CT$ has no new reps in sub-owners then neither does a comparable $CT'$. 
The examples in Sects.~\ref{sect:rif} and \ref{sect:rifa} have no new
reps in sub-owners; examples which do are given in Sect.~\ref{sect:app}.

In the rest of Sect.~\ref{sect:abs} we make the following assumption.  It is used
in the proof of Lemma~\ref{lem:cpres} on which the first abstraction
theorem depends.
For the second abstraction theorem the second sentence will be dropped.
\begin{assumption}\label{ass:first}
First, $CT$ and $CT'$ are confined class tables for which a simulation
$\REL$ is given.  
Second, $CT$ has no new reps in sub-owners and the allocator is parametric
in the sense of Def.~\ref{def:param}.      
\end{assumption}

\subsection{Identity extension}

A typical formulation of identity extension is that $\REL\:\T$ is the
identity on any type $\T$  
for which  it is the identity on all primitive types $b$ that occur in $\T$.
The reason is that  no value of type $b$ can occur in a value of
type $\T$ if $b$ does not occur in $\T$ ---but this fails with extensible records
and structural subtyping, and with procedures that may have global variables
\cite{sdr}.  It can be made to work using name-based (declaration) subclassing
\cite{CNfm02}: in the context of a complete class table, one can consider
the classes that have no attributes with subclasses in which $b$
occurs. For our purposes here it is enough to deal with the heap.   

In our language, $\REL\;T$ is the identity for every data type $\T$ (Lemma~\ref{fact:id}),
but that is only because  the interesting data is in the heap
---which is not typed at all.\footnote{Nor would we want to impose a typing system on
  the heap, as it would likely preclude unbounded data structures \cite{Grossman00}.}
In general, $\means{\state{Own}} \neq\means{\state{Own}}'$ and $\REL(\state{Own})$ is
not the identity. 
Related heaps can contain owner objects with different states that may point
to completely different rep objects. 
But consider executing a method on an object $o$ from whose fields no $Own$
objects are reachable, i.e., $Own$ objects are not part of the representation of $o$.
The resulting heap may contain $Own$ objects that were assigned to local variables, but
if the method is confined then those objects are unreachable in the final state.

\begin{definition}[{\bf garbage collection, Own-free}]\label{def:gc}
For a set or list $\ol{d}$ of values, define the heap $gc(\ol{d},h)$ to be the
restriction of $h$ to cells reachable from $\ol{d}$.  
For $(h,\eta)\in\means{\heap\otimes\Gamma}$, define 
$collect(h,\eta) = (gc(rng\,\eta,h),\eta)$.
Extend $collect$ to $\means{(\heap\otimes\Gamma)_\bot}$ by $collect\,\bot = \bot$.

Say $h$ is $Own$-\dt{free} just if $\dom\,h \intersect locs(Own\subclasses)
= \Empty$ and $\eta$ is $Own$-\dt{free} just if $rng\,\eta\intersect
locs(Own\subclasses)=\Empty$.  \qed
\end{definition}

\begin{lemma}[{\bf identity extension}]\label{lem:idex}
Suppose $\rel{(\heap\otimes\Gamma)}{(h,\eta)}{(h',\eta')}$ and 
$\Gamma\,\self$ is non-rep.
Let $(h,\eta)$ and $(h',\eta')$ be confined at $\Gamma\,\self$.
If $collect(h,\eta)$ and $collect(h',\eta')$ are $Own$-free
then $collect(h,\eta) = collect(h',\eta')$.
\end{lemma}
\begin{proof}
In confined heaps, reps are only reachable from owners.
Now the argument is a straightforward induction using the definition of $\REL$.
\qed\end{proof}

\begin{lemma}\label{lem:hh}
For any $\REL$ given by Def.~\ref{def:R} from a basic coupling, if 
$h\in\means{\heap}$ is $Own$-free then $\rel{\heap}{h}{h}$.       
If, in addition, $(h,\eta)\in\means{\heap\otimes\Gamma}$ and 
$rng\,\eta$ is $Own$-free then 
$\rel{(\heap\otimes\Gamma)}{(h,\eta)}{(h,\eta)}$.       
\end{lemma}
\begin{proof}
If $h$ is $Own$-free and confined then it has no reps; its admissible
partition is a single block, the clients.  
For such a heap it is immediate from the definition of $\REL$ that 
$\rel{\heap}{h}{h}$.
If $rng\,\Gamma$ is $Own$-free then $\rel{\Gamma}{\eta}{\eta}$ is also
direct from the definition.
\end{proof}

\subsection{Abstraction theorem}

The main theorem says that if methods of $Own$ 
preserve the coupling relation then so do all
methods.\footnote{Readers familiar with \citeN{Reynolds84} may expect
  that, as our language has fixpoints, the result only holds for
  couplings that are $\bot$-strict and join-complete.  But our basic
  couplings have this property, trivially, because heaps are ordered
  by equality.  The induced coupling is strict and join-complete by construction.}
The proof depends on lemmas for constructors and commands.
These are given following the theorem.
The other main ingredient for the proof is the following connection between 
$\REL$ and the semantics of inherited methods.

\begin{lemma}\label{lem:inh}
Suppose $C$ and all class names in $\ol{T}$ are non-rep, and 
$B<C$. If  $\rel{(C,\ol{x},\ol{T}\TO T)}{d}{d'}$ then
$  \rel{(B,\ol{x},\ol{T}\TO T)}{(\restr(d,B))}{(\restr(d',B))} $ 
where $\restr$ is the restriction to global states of $B$ (see Def.~\ref{def:rest}).
\end{lemma}
\begin{proof} Straightforward, using Lemma~\ref{lem:semty}(\ref{stya}). See Appendix.
\qed\end{proof}

\begin{theorem}[{\bf abstraction}]\label{thm:fat}
$\rel{\menv}{\means{CT}}{\means{CT'}'}$.
\end{theorem}
\begin{proof}
We show that the relation holds for each step in 
the approximation chain in the semantics of class tables.  That is, we show
by induction on $i$ that 
\[  \rel{\menv}{\mu_i}{\mu'_i} \quad\mbox{for every }i\in\nats \epunc .\] 
The result $\rel{\menv}{\means{CT}}{\means{CT'}'}$ then 
follows by fixpoint induction, as $\means{CT}$ and $\means{CT'}'$ are
defined to be the fixpoints of these ascending chains.
Admissibility of fixpoint induction is discussed preceding the proof of 
Theorem~\ref{thm:cfix}.  

For the base case, we have 
$\rel{(C,\pars(m,C),\mtype(m,C))}{(\mu_0\,C\,m)}{(\mu'_0\,C\,m)}$
for every $m,C$ because $\lam{(h,\eta)}{\bot}$ relates to itself.

For the induction step, suppose 
\begin{trivlist}
\item  \hfill $\rel{\menv}{\mu_i}{\mu'_i}$ \epunc.
\hfill $(*)$
\end{trivlist}
We must show $\rel{\menv}{\mu_{i+1}}{\mu'_{i+1}}$, that is,
for every non-rep $C$ and every $m$ with $\mtype(m,C)$ defined:  
\begin{trivlist}
\item 
\hfill
$\rel{(C,\ol{x},\ol{\T}\TO\T)}{(\mu_{i+1}\,C\,m)}{(\mu'_{i+1}\,C\,m)}$
\hfill $(\dagger)$ 
\end{trivlist}
where $\ol{x}=\pars(m,C)$ and $\ol{\T}\TO\T=\mtype(m,C)$.
For arbitrary $m$ we show $(\dagger)$ for all $C$ with $\mtype(m,C)$
defined, using a secondary induction on $depth(m,C)$.  
We have $depth'(m,C) = depth(m,C)$ (Lemma~\ref{lem:eqdep}). 

The base case is the unique $C$ with $depth(m,C) = 0$; here  
$m$ is declared in both $CT(C)$ and $CT'(C)$.  We go by cases on $C$.
If $C=Own$, we get ($\dagger$) from the assumption that $\REL$ is a
simulation.  In detail: 
Using $(*)$ and Def.~\ref{def:sim}(\ref{simA}) we get 
\[
\rel{(Own,\ol{x},\ol{\T}\TO\T)}{(\means{M}\mu_i)}{(\means{M'}'\mu'_i)}
\epunc,\]
whence $(\dagger)$ by definition of $\mu_{i+1}$ and $\mu'_{i+1}$.
The other case is $C$  a non-rep class different from $Own$.  
Then by Def.~\ref{def:comp}(\ref{simb}) of comparable class
tables we have $CT(C)=CT'(C)$ and in particular both class tables have the
same declaration 
\[ \T\;m(\ol{\T}\;\ol{x})\;\{\S\} \epunc .\] 
To show ($\dagger$), suppose $\etaconf{C}{\eta}{h}$, $\etaconf{C}{\eta'}{h'}$, 
$\rel{\heap}{h}{h'}$, and $\rel{\Gamma}{\eta}{\eta'}$,
where $\Gamma=\ol{x}:\ol{T},\self:C$.
Then by Lemma~\ref{lem:cpres} below, using $\rel{\menv}{\mu_i}{\mu'_i}$, 
the results from $S$ are related.  That is, either 
$\means{\Gamma \proves S }\mu_i(h,\eta) =\bot
=\means{\Gamma\proves' S }'\mu'_i(h',\eta')$ or neither is $\bot$. In
the latter case, $(h_0,\eta_0)$ is related to $(h'_0,\eta'_0)$ 
where $(h_0,\eta_0)=\means{\Gamma\proves S }\mu_i(h,\eta) $ and 
$(h'_0,\eta'_0)=\means{\Gamma\proves' S }'\mu'_i(h',\eta')$.
Then, by definition of $\REL\,\Gamma$,
$\rel{\Gamma}{\eta_0}{\eta'_0}$ implies
$\rel{\T}{(\eta_0\,\result)}{(\eta'_0\,\result)}$.
Thus $(\dagger)$ holds by definition of $\mu_{i+1}$ and $\mu'_{i+1}$.
This concludes the base case of the secondary induction.  

For the induction step, suppose $depth(m,C) > 0$.
By induction on depth we have, by definition of $depth$, 
\begin{trivlist}
\item 
\hfill $
\rel{(C,\ol{x},\ol{\T}\TO\T)}{(\mu_{i+1}\,(\super\, C)\,m)}{(\mu'_{i+1}\,(\super\, C)\,m)}$
\epunc. \hfill $(\ddagger)$ 
\end{trivlist}
If $m$ is declared in both $CT(C)$ and $CT'(C)$ then the argument is the
same as in the base case of the secondary induction.
If $m$ is inherited in both $CT(C)$ and $CT'(C)$ then $(\ddagger)$ follows
from $(\dagger)$  because the semantics defines $\mu_{i+1}\,C\,m$ by
restriction from $(\mu'_{i+1}\,(\super\, C)\,m)$ and restriction preserves
simulation.  (This is Lemma~\ref{lem:inh}, which 
is applicable because if $B>Own$ and $m$ is inherited
in $Own$ from $B$ then $\ol{T}\ncomp Rep$ and 
$\ol{T}\ncomp Rep'$ by confinement of $CT,CT'$,
Def.~\ref{def:tconf}(\ref{tconfb}).)
The remaining possibility is that $m$ is declared in $CT(C)$ and inherited
in $CT'(C)$ from some $B$ (or the other way around).  Then $C=Own$, by
comparability of $CT$ and $CT'$.
Using Def.~\ref{def:sim}(\ref{simB}) and $(*)$ we get 
\[ \rel{(Own,\ol{x},\ol{\T}\TO\T)}{(\means{M}\mu_i)}{(\restr(\means{M_B}\mu'_i,Own))} \]
and thus $(\dagger)$ by definition of $\mu_{i+1}$ and $\mu'_{i+1}$.
\qed\end{proof}

\begin{lemma}[{\bf establishment by constructors}]\label{lem:cxpres}
Let $\mu$ and $\mu'$ be any method environments.
Then the following holds for any non-rep class $C\neq Own$.

For all $(h,\ell)\in\means{\heap\otimes C}$ and 
        $(h',\ell')\in\means{\heap\otimes C}$,
if  $\etaconf{C}{\eta_1}{h}$,
$\etaconf{C}{\eta'_1}{h'}$ and 
$\rel{(\heap\otimes C)}{(h,\ell)}{(h',\ell')}$ then 
$\rel{\heap_\bot}{h_0}{h'_0}$, 
where
\[  
\begin{array}{l}
\eta_1= [\self\mapsto\ell] \qquad
h_0 = \means{\self:C\proves \constr\:C:\CON}\mu(h,\eta_1) \\
\eta'_1= [\self\mapsto\ell'] \qquad
h'_0 = \means{\self:C\proves' \constr\:C:\CON}\mu'(h',\eta'_1) 
\end{array}
\]
\end{lemma}
\begin{proof}
By well founded induction on $C$ with respect to $\well$.
Suppose $\etaconf{C}{\eta}{h}$, $\etaconf{C}{\eta'}{h'}$, and
 $\rel{(\heap\otimes C)}{(h,\ell)}{(h',\ell')}$.
Let $h_1= \means{\self:C\proves S:\CON}\mu(h,\eta)$ be as in the semantics
of $S$ as a constructor, and similarly for $h'_1$.   
If $\super\,C = \OBJECT$ then $h_1=h$ and thus $\rel{\heap}{h_1}{h'_1}$ by
hypothesis.
Otherwise, $h_1 = \means{\self:\super\,C\proves
  \constr(\super\,C):\CON}\mu(h,\eta)$ and  we get 
$\rel{\heap}{h_1}{h'_1}$ by the induction hypothesis noting that $\super\,C\well C$ 
by Lemma~\ref{lem:cdep}.
It follows that  $\rel{(\heap\otimes C)}{(h_1,\ell)}{(h'_1,\ell')}$.

It remains to show $\rel{\heap_\bot}{h_0}{h'_0}$, where $h_0,h'_0$ are
obtained by applying the command semantics of $\constr\:C$ to $(h_1,\ell)$
and $(h'_1,\ell)$.
This holds because, taking $S$ to be $\constr\:C$ in the claim below, we get
$\rel{(\heap\otimes C)_\bot}{(h_0,\eta_0)}{(h'_0,\eta'_0)}$ and thus 
either both outcomes are $\bot$ or $\rel{\heap}{h_0}{h'_0}$.  

\textbf{Claim:}
For all $\self:C\proves S$ such that $S$ has no method calls 
and every $\new{B}{}$ in $S$ has $B\cdep C$, for all 
$(h,\eta)$ and $(h',\eta')$, if 
$\etaconf{C}{\eta}{h}$, $\etaconf{C}{\eta'}{h'}$, 
and $\rel{(\heap\otimes\Gamma)}{(h,\eta)}{(h',\eta')}$ then
\[ 
\rel{(\heap\otimes\Gamma)_\bot}{(\means{\Gamma\proves S }\mu(h,\eta))}{
(\means{\Gamma\proves' S }'\mu'(h',\eta'))} 
\epunc .\]
Proof of the claim is by induction on the structure of 
$S$.   Note that by hypothesis $S$ has no method calls, so $\mu$ is not
relevant.  The argument is exactly the same as in the proof of
Lemma~\ref{lem:cpres} below,  except for the case of $\NEW$.
In the proof of Lemma~\ref{lem:cpres}, the case of $\NEW$ appeals to
the present Lemma for constructors.  To prove the claim for this case, the
argument is the same except for appealing to the main induction hypothesis;
this is sound because the claim includes the hypothesis that if $\new{B}{}$
occurs in $S$ then $B\cdep C$ and thus $B\well C$ by Lemma~\ref{lem:cdep}. 
\end{proof}

\begin{lemma}[{\bf preservation by expressions}]\label{lem:epres}
For any non-rep class $C\neq Own$ and any constituent expression
$\Gamma\proves e:T$ of a method declared in  $C$, the following
holds:
For all $(h,\eta)\in\means{\heap\otimes\Gamma}$ and 
$(h',\eta')\in\means{\heap\otimes\Gamma}'$, if 
$\rel{(\heap\otimes\Gamma)}{(h,\eta)}{(h',\eta')}$ then
\[ \rel{(T_\bot)}{(\means{\Gamma\proves e:T}(h,\eta))}{(\means{\Gamma\proves' e:T}'(h',\eta'))}  
\epunc .\] 
\end{lemma}
\begin{proof}By induction on the derivation of $\Gamma\proves e:T$.  
For each case of $e$, we give an argument assuming that 
$\Gamma,C,T,\eta,\eta',h,h'$ satisfy the hypotheses of the Lemma.

\medskip
\Case\ $\Gamma\proves \cast{B}{e}: B$.
Induction on $e$ yields that $\rel{D_\bot}{\ell}{\ell'}$ (or else both
denotations of $e$ are $\bot$). 
By confinement of $e$, as $C\neq Own$ and $C$ is non-rep, we have 
$\ell\not\in locs(Rep\subclasses)$ and   
$\ell'\not\in locs(Rep'\subclasses)$.  
Thus, $\ell'=\ell$ by Lemma~\ref{fact:id}.
Hence either both semantics yield $\ell$, whence $\rel{B_\bot}{\ell}{\ell}$, or
both yield $\bot$ and again $\rel{B_\bot}{\bot}{\bot}$.

\medskip
\Case\ $\Gamma\proves\is{e}{B}:\BOOL$.
The argument is similar to that for type cast.

\medskip
\Case\  $\Gamma \proves \faccess{e}{f}:T$. 
By induction on $e$ we have $\rel{C_\bot}{\ell}{\ell'}$, hence
$\ell=\ell'$ by Lemma~\ref{fact:id}.  
In the non-$\bot$ case, $\ell\neq\NIL$.  
By closure of the heaps, $\ell\in \dom\,h$ and $\ell\in \dom\,h'$.
We consider cases on whether $C<Own$.   
Consider confining partitions $(\Ch * \Oh_1 * \Rh_1 \ldots) = h$ and 
$(\Ch'* \Oh'_1 * \Rh'_1\ldots) = h'$ that have corresponding islands
as in the definition of $\REL\;\heap$.
In the case $C<Own$, we have  $\ell\in locs(Own\subclasses)$ and hence
$\ell$ in some $\dom(\Oh_i)$.  
From $\rel{\heap}{h}{h'}$ we have 
\[ R\;(\Oh_i * hRep_i)\;(\Oh'_i * hRep'_i) \]
and thus $\ell\in\dom(\Oh'_i)$ by basic coupling
Def.~\ref{def:psim}(\ref{simea}).    
Since $C\neq Own$, we know by visibility that $f$ is not in the private
fields $\ol{g}$ of $Own$.
Thus, as $\type(f, \loctype\,\ell)) = T$, we have
$\rel{T}{(h \ell f)}{(h' \ell f)} $ by Def.~\ref{def:psim}(\ref{simec}) and
Lemma~\ref{fact:id}.  

Finally, in the case  $C\nleq Own$ (recall that $C$ is non-rep and $C\neq
Own$ by hypothesis, 
we have $\ell\in\dom(\Ch)$ and hence $\ell\in\dom(\Ch')$ by definition
$\REL\;\heap$. 
Hence $\rel{(\state{(loctype\:\ell)})}{(h\ell)}{(h'\ell)}$ and thus
$\rel{T}{(h\ell f)}{(h'\ell f)}$ by definition of
$\REL\;(\state{(loctype\:\ell)})$.

\medskip
The remaining cases are straightforward.  See Appendix.\qed
\end{proof}

\begin{lemma}[{\bf preservation by commands}]\label{lem:cpres}
Suppose $\REL$ is a simulation, and moreover
$\mu$ and $\mu'$ are confined method environments such that $\rel{\menv}{\mu}{\mu'}$. 
Then the following holds for any non-rep class $C\neq Own$.
For any constituent command $\Gamma\proves S $ 
in a method declaration in  $CT(C)$ and any 
$(h,\eta)$ and $(h',\eta')$, if $\etaconf{C}{\eta}{h}$, $\etaconf{C}{\eta'}{h'}$, 
and $\rel{(\heap\otimes\Gamma)}{(h,\eta)}{(h',\eta')}$ then
\[ 
\rel{(\heap\otimes\Gamma)_\bot}{(\means{\Gamma\proves S }\mu(h,\eta))}{
(\means{\Gamma\proves' S }'\mu'(h',\eta'))}
\epunc .\]
\end{lemma}
\begin{proof}
For any $C$, the proof is by structural induction on $\Gamma\proves S$.

\medskip
\Case\ $\Gamma \proves\assign{x}{e} $.
As $CT$ is confined, constituent $e$ of the assignment is confined.
So by Lemma~\ref{lem:epres} we have $\rel{T_\bot}{d}{d'}$.  Hence, by 
$\rel{\Gamma}{\eta}{\eta'}$ and definition of $\REL\;\Gamma$, we have
$\rel{\Gamma}{\ext{\eta}{x}{d}}{\ext{\eta'}{x}{d'}}$ whence the result.

\medskip
\Case\ $\Gamma \proves\assign{\faccess{e_1}{f}}{e_2} $.
By Lemma~\ref{lem:epres} for $e_1$ we have 
$\rel{C}{\ell}{\ell'}$, hence $\ell=\ell'$ by Lemma~\ref{fact:id}.
By Lemma~\ref{lem:epres} for $e_2$ we have $\rel{U}{d}{d'}$ and hence
$\rel{T}{d}{d'}$  by Lemma~\ref{fact:id:c}, where $(f:T)\in\dfields\,C$
as in the typing rule.    
To conclude the argument it suffices to show
\begin{trivlist}
\item 
\hfill $\rel{\heap}{
  \ext{h}{\ell}{\ext{h\ell}{f}{d}}}{\ext{h'}{\ell}{\ext{h'\ell}{f}{d'}} }$
\epunc. \hfill $(*)$ 
\end{trivlist}
Consider confining partitions $(\Ch * \Oh_1 * \Rh_1 \ldots) = h$ and 
$(\Ch'* \Oh'_1 * \Rh'_1\ldots) = h'$ that correspond as in the definition of
$\rel{\heap}{h}{h'}$. 
We argue by cases on $C$.

\begin{itemize}
\item $C<Own$: Then $\loctype\,\ell\leq C < Own$.
From typing we have $e_1:C$ and hence there is some $i$ with
$\{\ell\} =\dom(\Oh_i)$ and  by
$\rel{\heap}{h}{h'}$ we get
\[ R\;(\Oh_i * \Rh_i)\;(\Oh'_i * \Rh'_i) \]
and so $\{\ell\} =\dom(\Oh'_i)$.
By typing and $C\neq Own$, field $f$ is not in the private fields $\ol{g}$ of $Own$.  So
$(*)$ follows from $\rel{\heap}{h}{h'}$ and $\rel{T}{d}{d'}$.
\item $C\nleq Own$:
As $C$ is non-rep,   
we have $\ell\in\dom\,\Ch$ and thus 
$\ell\in\dom\,\Ch'$ by hypothesis $\rel{\heap}{h}{h'}$.
Moreover,  
$\rel{(\state{(loctype\,\ell)})}{(h\ell)}{(h'\ell)}$ and so by 
$\rel{T}{d}{d'}$ we get
$\rel{(\state{(loctype\,\ell)})}{\ext{h \ell}{f}{d}}{\ext{h'
    \ell}{f}{d'}}$. Hence $(*)$.
\end{itemize}

\Case\ $\Gamma\proves\assign{x}{\new{B}{\ }} $.
By confinement of $CT$, this command is confined and hence the final
states are confined: $\etaconf{C}{\eta_0}{h_0}$ and
$\etaconf{C}{\eta'_0}{h'_0}$.
We have $C\nleq Rep$ and $C\neq Own$.
In the case $C\not < Own$
confinement of $\eta_0$ and $\eta'_0$ implies 
$rng\,\eta_0\intersect locs(Rep\subclasses)=\Empty
= rng\,\eta'_0\intersect locs(Rep'\subclasses)$.  
So $\ell\not\in locs(Rep\subclasses)$
and $\ell'\not\in locs(Rep'\subclasses)$,
hence by typing $B$ is non-rep.
In the case $C<Own$, we have $B$ non-rep by Assumption~\ref{ass:first} (no 
reps in sub-owners).  Either way, $B$ is non-rep so Lemma~\ref{fact:foo}
applies, to yield $\dom\,h\intersect locs\,B = \dom\,h'\intersect locs\,B$.
Thus by parametricity of $fresh$ we have
$\ell = \fresh(B,h) = \fresh(B,h') = \ell'$.  
So, by Lemma~\ref{fact:id} and $\rel{\Gamma}{\eta}{\eta'}$ we have
$\rel{\Gamma}{\eta_0}{\eta'_0}$. 

It remains to show $\rel{\heap_\bot}{h_0}{h'_0}$ in order to get the final
result $\rel{(\heap\otimes\Gamma)_\bot}{(h_0,\eta_0)}{(h'_0,\eta'_0)}$.
We argue by cases on $B$.
\begin{itemize}
\item $B\nleq Own$:   
Writing $\fields'$ for the fields given by $CT'$, we have
$\fields\,B=\fields'\,B$ and thus 
$\rel{(\state{B})}{[\fields\,B\mapsto \mathit{defaults}]}{
                         [\fields'\,B\mapsto \mathit{defaults}]}$.
So, as $B$ is non-rep and $B\neq Own$, we can add $\ell$ to $\Ch$ and $\Ch'$ to get
partitions that witness $\rel{\heap}{h_1}{h'_1}$.
We also have $\etaconf{B}{\eta_1}{h_1}$  and $\etaconf{B}{\eta'_1}{h'_1}$ 
because $\conf{h}$, $\conf{h'}$, and defaults do not contain any locations. 
Now by Lemma~\ref{lem:cxpres} we get 
$\rel{\heap_\bot}{h_0}{h'_0}$.  Combining this with what was shown above we have
$\rel{(\heap\otimes\Gamma)_\bot}{(h_0,\eta_0)}{(h'_0,\eta'_0)}$.

\item $B\leq Own$:   
By basic coupling, Def.~\ref{def:sim}, we have
$\rel{\heap_\bot}{h_0}{h'_0}$.
\end{itemize}

\Case\ $\Gamma  \proves \assign{x}{\mcall{e}{m}{\ol{e}}} $. 
By Lemma~\ref{lem:epres} for $e$ we have $\rel{D_\bot}{\ell}{\ell'}$, hence
$\ell=\ell'$ by Lemma~\ref{fact:id}.  
Let $\eta_1=[\self\mapsto\ell,\ol{x}\mapsto\ol{d}]$
and $\eta'_1=[\self\mapsto\ell,\ol{x}\mapsto\ol{d}']$.  
By confinement of $\assign{x}{\mcall{e}{m}{\ol{e}}}$ (Def.~\ref{def:cconf}) we
have confined arguments, i.e., $\etaconf{(\loctype\,\ell)}{\eta_1}{h}$ and
$\etaconf{(\loctype\,\ell)}{\eta'_1}{h'}$.
By Lemma~\ref{lem:epres} for $\ol{e}$ we have
$\rel{\ol{U}_\bot}{\ol{d}}{\ol{d}'}$ and hence
$\rel{\ol{U}}{\ol{d}}{\ol{d}'}$ as we are considering the non-$\bot$
case. Thus $\rel{[\ol{x}:\ol{T},\self:\loctype\,\ell]}{\eta_1}{\eta'_1}$  by
Lemma~\ref{fact:id:c}.  
From $\rel{\menv}{\mu}{\mu'}$ we get
\[ \rel{(\loctype\,\ell,\ol{x},\ol{\T}\TO\T)}{(\mu(\loctype\,\ell)m)}{(\mu'(\loctype\,\ell)m)}
\]
hence, as $h,h',\eta_1,\eta'_1$ are confined and related, 
$\rel{(\heap\otimes\T)_\bot}{(h_1,d_1)}{(h'_1,d'_1)}$, where 
$(h_1,d_1) =\mu(\loctype\,\ell)m(h,\eta)$ and  
$(h'_1,d'_1) =\mu'(\loctype\,\ell)m(h',\eta')$.
Thus $\rel{T}{d_1}{d_1}$ and $\rel{\heap}{h_1}{h_1}$.
It remains to show that the updated stores 
$\ext{\eta}{x}{d_1}$ and 
$\ext{\eta'}{x}{d'_1}$ are related for $\Gamma$. This follows from 
$\rel{T}{d_1}{d'_1}$ and $T\leq \Gamma\,x$, using Lemma~\ref{fact:id:c}.

\medskip
The remaining cases are similar.  See Appendix. \qed
\end{proof}


\section{Applications and further examples}
\label{sect:app} 

In this section we use the abstraction theorem to show some program
equivalences for the examples discussed in Sections~\ref{sect:rif} and
\ref{sect:rifa}.  
Then we discuss further variations on the observer pattern.

To establish the hypothesis of the abstraction theorem for the examples we use the
couplings given as examples in Sect.~\ref{ssect:coup}.  Both the theorem and
these couplings are defined in terms of the semantics. To show that the couplings are 
simulations we argue directly in terms of the semantics.  For practical purposes in
program verification, the abstraction theorem would be expressed syntactically as a
proof rule and rules for program constructs would be used to
establish the simulation property
\cite{ReynoldsCraft,Jones:Sys,Morgan:dr,deRdataref}.
Adequate proof rules for a language like ours remains an open
challenge (see Sect.~\ref{sect:disc}).  

\subsection{Program equivalence}\label{ssect:peq}

We take \dt{program} to mean a well formed class table $CT$ together
with a command $\Gamma \proves S $.
We consider the object states reachable from variables of $\Gamma$ to be the inputs
and outputs of the program.
For example, if $S$ is the body of method \code{main} in Sect.~\ref{sect:bool} then 
$\Gamma$ is $\code{\self:Main}$ and what can be reached is $\self$ and the string
$\code{\self.inout}$.   
We restrict attention to confined programs, meaning that $CT$ and
$\Gamma\proves S $ are confined. Thus, by Theorem~\ref{thm:cfix} the method
environment $\means{CT}$ is confined. 
To prove program equivalence using the abstraction theorem, we need to both
introduce and eliminate a simulation.  Elimination is by identity
extension Lemma~\ref{lem:idex} and introduction is by the related
Lemma~\ref{lem:hh}.  
There is a small technicality: To establish the hypothesis of Lemma~\ref{lem:cpres},   
we require w.o.l.o.g.\ that $S$ occurs in some method of $CT$. 

We compare programs only for class tables $CT,CT'$ that are comparable in
the sense of Def.~\ref{def:comp}, and with commands in the same
context $\Gamma$.  
As commands denote functions on global states, the obvious notion of
equivalence is that $\means{\Gamma \proves S }$ and
$\means{\Gamma \proves' S' }'$ are equal as functions.  
By Lemma~\ref{lem:sty}, $\means{\Gamma}=\means{\Gamma}'$ for any $\Gamma$,
but in general the semantic domains differ for owner object states which
may have different private fields. A global state
$(h,\eta)\in\means{\heap\otimes\Gamma}$ for $CT$ need not be an element of  
$\means{\heap\otimes\Gamma}'$ for $CT'$.
However, an $Own$-free heap in $\means{\heap}$ is also an element of
$\means{\heap}'$. So we compare command meanings on the $Own$-free states.

\begin{definition}[{\bf client program equivalence}]\label{def:peq}
Suppose programs $CT,(\Gamma \proves S )$ and $CT',(\Gamma \proves' S' )$
are such that $CT,CT'$ are comparable and confined, and moreover $S$
(resp.\ $S'$) occurs in $CT$ (resp.\ $CT'$).  
The programs are \dt{equivalent} iff
\[ collect(\means{\Gamma \proves S }\hat{\mu}(h,\eta))
= collect(\means{\Gamma \proves' S' }'\hat{\mu}'(h,\eta)) \]
for all confined and $Own$-free $(h,\eta)\in\means{\heap\otimes\Gamma}$, 
where $\hat{\mu}= \means{CT}$ and $\hat{\mu}'= \means{CT'}'$.
\qed
\end{definition}
If $\Gamma\,\self\leq Own$ then $\eta$ cannot be $Own$-free. The 
resulting vacuous quantification makes the definition equate all commands
for such $\Gamma$.  But we are only interested in using the definition for
clients.  Simulation is the relation of interest between owners.

The static analysis for confinement Sect.~\ref{sect:static}
can be used to show that each of the following examples is confined for the
appropriate $Own$ and $Rep$.

\begin{example}\label{ex:OBsimx}
Consider the command $S$ comprising the body of method
\code{main} of class \code{Main} in Sect.~\ref{sect:bool}
and take
$\Gamma = (\self:\code{Main})$.  
As $CT$ we take the declarations of \code{Main}, \code{Bool}, and the first version of
\code{OBool}.  For $CT'$ we use the second version of \code{OBool}.  
Let $Rep$ and $Rep'$ be \code{Bool} and $Own$ be \code{OBool}.

To show that $CT,(\Gamma\proves\S)$ is equivalent to $CT',(\Gamma\proves\S)$, recall the
basic coupling of Example~\ref{ex:OBsim} and let $\REL$ be the induced
coupling. Let $(h,\eta)$ be any confined state for $\Gamma$, noting that
$\code{Main}\ncomp Own$ so $\eta$ is $Own$-free. 
Let $\hat{\mu}= \means{CT}$ and $\hat{\mu}'= \means{CT'}'$.
To show
\begin{trivlist}
\item 
\hfill $collect(\means{\Gamma \proves S }\hat{\mu}(h,\eta))
= collect(\means{\Gamma \proves' S' }'\hat{\mu}'(h,\eta)) $
\epunc, \hfill $(*)$ 
\end{trivlist}
note first that 
$\rel{(\heap\otimes\Gamma)}{(h,\eta)}{(h,\eta)}$ by Lemma~\ref{lem:hh}.
It is straightforward to show that $\REL$ is established by the constructors
and preserved by the methods of 
\code{OBool}; thus $\REL$ is a simulation.
The abstraction theorem yields $\rel{\menv}{\hat{\mu}}{\hat{\mu}'}$.   
This in turn justifies application of the preservation Lemma~\ref{lem:cpres} to
command $S$, as its context $Main$ is a non-rep class $\neq Own$.
Thus the outcomes $\means{\Gamma \proves S }\hat{\mu}(h,\eta)$
and $\means{\Gamma \proves' S' }'\hat{\mu}'(h,\eta)$ are related by
$\REL$.  By definition of $\REL$, either both outcomes are $\bot$, in which
case $(*)$ holds by definition of $collect$, or the outcomes are non-$\bot$
states $(h_0,\eta_0)$ and $(h'_0,\eta'_0)$ with 
$\rel{(\heap\otimes\Gamma)}{(h_0,\eta_0)}{(h'_0,\eta'_0)}$.
Note that $h_0$ and $h'_0$ each contains at least one owner, the one constructed in $S$.  But 
$Main\ncomp Own$, so $rng\,\eta_0$ and $rng\,\eta'_0$ are $Own$-free.
Moreover, the owners were reached only by variable \code{z} which is local
in $S$;  they are not reachable via fields of the objects $h_0(\eta\,\self)$
or $h'_0(\eta'\,\self)$.  That is, both 
$collect(h_0,\eta_0)$ and $collect(h'_0,\eta'_0)$ are $Own$-free.
Thus by identity extension Lemma~\ref{lem:idex} we have
$collect(h_0,\eta_0) = collect(h'_0,\eta'_0)$ which concludes the proof of 
$(*)$.
\qed\end{example}

This proof depends on parametricity of the allocator, because
that is needed for the abstraction theorem.  The same argument will go
through, however, for the second abstraction theorem in the sequel which
drops parametricity of the allocator.

\begin{example}\label{ex:ms}
Recall the Meyer-Sieber-O'Hearn example from Sect.~\ref{sect:mse}, and
in particular the command
\medskip

\begin{sf}
\noindent
\medskip
\hspace*{2em}
C y\assym \NEW\ C \IN\ A x\assym \NEW\ A \IN\ x.callP(y)  \hfill ($\ddagger$)\\
\end{sf}
Take ($\ddagger$) to be the body of method \code{main} in 
\begin{quote}\sf
\begin{tabbing}
  \CLASS\ Main\ \EXT\ \OBJECT\ \{ \UNIT\ main()\{ \ldots \} \}
\end{tabbing}
\end{quote}
To be very precise we need to include a class 
\begin{quote}\sf
\begin{tabbing}
  \CLASS\ Rep\ \EXT\ \OBJECT\ \{ \}
\end{tabbing}
\end{quote}
so we can take $Rep$ and $Rep'$ to be \code{Rep} which is not comparable to
the classes \code{C} and \code{A} of interest.  
Let $Own$ be \code{A}.  
Let $CT$ consist of the declarations of \code{A}, \code{Rep}, \code{Main},
and an arbitrary class 
\begin{quote}\sf
\begin{tabbing}
  \CLASS\ C\ \EXT\ \OBJECT\ \{ \UNIT\ P(A\,z)\{ \ldots \} \ldots \}
\end{tabbing}
\end{quote}
such that methods of $C$ satisfy the confinement conditions.  
Then $CT$ and $CT'$ are confined, because 
no reps are constructed or manipulated.
We use the basic coupling of Example~\ref{ex:msim}.
To appeal to the abstraction theorem, we must argue that $\REL$ is a simulation.
The constructors are $\SKIP$ and the default value 0 for field \code{g} establishes
the relation.  Preservation by \code{inc} is straightforward because both versions
have the same code and it makes no method calls.  
We give the details for preservation by \code{callP}.  The relevant condition is
Def.~\ref{def:sim}(\ref{simA}).  To show it for \code{callP}, suppose $i\geq 0$ 
and $\rel{\menv}{\mu_i}{\mu'_i}$.  Note that $\mu_i$ and $\mu'_i$ are confined, by
Theorem~\ref{thm:cfix}.  
Suppose that $\rel{(\heap\otimes y:C,\self:A)}{(h,\eta)}{(h',\eta')}$ with
$\etaconf{A}{\eta}{h}$ and $\etaconf{A}{\eta'}{h'}$.
In both versions of \code{callP}, the body is a sequence and the first command is
\code{y.P(\self)}.  
Let $\eta_1=[z\mapsto \eta\,\self,\self\mapsto \eta\,y]$
and $\eta'_1=[z\mapsto \eta'\,\self,\self\mapsto \eta'\,y]$ be the environments 
for semantics of this call.
By definition of $\REL$ we get
$\rel{(\heap\otimes z:A,\self:C)}{(h,\eta_1)}{(h',\eta'_1)}$.
From the hypothesis $\etaconf{A}{\eta}{h}$ we get $\etaconf{C}{\eta_1}{h}$ and
likewise $\etaconf{C}{\eta'_1}{h'}$.  Applying the hypothesis 
$\rel{\menv}{\mu}{\mu'}$ to these environments we get that either
$\mu C P (h,\eta_1) = \bot = \mu C P (h,\eta_1)$ or neither are $\bot$ and 
$\rel{(\heap\otimes\UNIT)}{(h_0,\IT)}{(h'_0,\IT)}$ where 
$(h_0,\IT) = \mu C P (h,\eta_1)$ and $(h'_0,\IT) = \mu' C P (h',\eta'_1)$.
The call is desugared to an assignment of the result value to a local but the value
is discarded for both versions, so the states following the calls are
$(h_0,\eta)$ and $(h'_0,\eta')$ and we have 
$\rel{(\heap\otimes y:C,\self:A)}{(h_0,\eta)}{(h'_0,\eta')}$.  In these states we
have $h_0\ell g = h_0'\ell g \land h_0\ell g \mathbin{\textbf{mod}} 2
= 0$.  So the command  
\begin{quote}\sf
      \IF\ \self.g\ \keyword{mod}\ 2 = 0 \THEN\ \ABORT\  \ELSE\ \SKIP\  \FI 
 \end{quote}
aborts, as does its counterpart which is simply \ABORT.
This concludes the argument that the bodies of \code{callP} are related.

Having established the antecedents of the abstraction theorem, we conclude
that the command ($\ddagger$) preserves $\REL$.
By semantics of the second version of \code{A} we know \code{callP}
aborts, so both interpretations of ($\ddagger$) abort.  The programs are equivalent.
\qed\end{example}

This example is handled  without using the identity extension
Lemma~\ref{lem:idex}, but that is only because the example uses abortion.
In subsequent examples the proof needs all the steps of the one 
for Example~\ref{ex:OBsimx}.  The steps are not spelled out in detail; only
the interesting bits are highlighted.

\begin{example}\label{ex:obs}
We consider the observer pattern, taking $Own$ to be \code{Observable}.
Let $CT$ be given by the first version, Fig.~\ref{fig:obs}, together with
the client given in Fig.~\ref{fig:obsCli}.  Let $CT'$ be given by the sentinel
version of \ref{fig:obsaa} together with Fig.~\ref{fig:obsCli}.  
We consider equivalence for the command
$\code{\self:Main,ob:AnObserver}\proves S$ where $S$ is the body of
\code{Main.main}.  
Because \code{obl} is local to $S$, no owners are reachable in the final state.

Taking $Rep,Rep'$ to be \code{Node,Node2}, we use the coupling relation
of Example~\ref{ex:obsSim}.
Clearly the constructors establish the relation.  
To show that method \code{add} preserves it, note that the bodies of these methods
are both sequential compositions; both construct a new node and then set its \code{ob} field
to the value passed as a parameter.  The next step is to add it to the beginning of
the list; the difference between the two versions is that \code{\self.snt.nxt} is
assigned in Fig.~\ref{fig:obsaa} whereas \code{\self.fst} is assigned in
Fig.~\ref{fig:obs}.  
Both versions of \code{add} then invoke methods on the new node \code{n}.
In practice one would argue in terms of the behavior of those methods.
Note that they need not preserve the relation; it is just that their behavior is used 
to maintain the relation.  
To give a precise argument in terms of the semantics, we consider cases on $i$.  
For $i=0$, both $\mu_i$ and $\mu'_i$ make every method abort, in which case the body
of \code{add} aborts due to method calls.  As the methods in class \code{Node} and
class \code{Node2} are not recursive, their semantics is already completely defined
for $i=1$, so for $i>0$ the behavior of \code{add} is to insert nodes at the head of
the list, maintaining the relation.  

The remaining owner method is \code{notifyAll}.  Again, the two versions are similar
except for skipping over the sentinel node.
To argue that the calls to \code{getNext} act correctly one considers cases as in the 
proof for \code{add}.  For the calls to \code{notify} on the \code{Observer} objects, 
recall that by the relation, the related lists contain the same \code{Observer}
pointers in the same order.  The two versions thus make the same series of
invocations of \code{notify}.  Each of those calls preserves the relation by
hypothesis $\rel{\menv}{\mu_i}{\mu'_i}$. 
\qed\end{example}

The last step of the argument, concerning invocations of \code{notify}, is like
reasoning about invocations of \code{P} in Example~\ref{ex:ms}.
This example has the additional complication of calls to objects within the owner
island.  
The case distinction between $i=0$ and $i>0$ is needed because our argument is purely 
in semantic terms.  In a practical proof system, one would reason only in terms of
the actual semantics of the methods involved rather than its approximants.  

Strictly speaking, use of Lemma~\ref{lem:cpres} depends on desugaring the
examples, and the desugarings  Remark~\ref{rem:sugar} do not include
loops.  We return to this issue in Sect.~\ref{sect:osub}. 

\begin{example} Suppose we change the client of Fig.~\ref{fig:obsCli} to
  use the following. 
  \begin{quote}\sf 
\CLASS\ AnObserver \EXT\ Observer \{  \UNIT\  notify()\{ \SKIP\ \} \} 
  \end{quote}
Then in Fig.~\ref{fig:obsaa} we can replace the body of 
\code{Observable.notifyAll} by \SKIP\  and still have equivalence with
the implementation of Fig.~\ref{fig:obs}.
What changes with respect to Example~\ref{ex:obs} 
is that the two implementations do not make the corresponding calls to
\code{notify}.  But because \code{AnObserver.notify} is \SKIP, calling it
has the same effect as not calling it; in particular, the relation is preserved.  

The argument here is not modular: by contrast with the preceding example,
here we reason directly in terms of the client code.
\qed\end{example}

\subsection{Further variations on observer}

Fig.~\ref{fig:obsc} gives another implementation of \code{Observable},
using a singly linked list but with most of the work delegated to methods
of \code{Node1}.
Method \code{add} of class \code{Node1} in the Figure is an example of
class-based visibility: The private fields of object \code{n} are both
assigned and read.  

Unlike the example of Sect.~\ref{sect:mse}, where method \code{P} is
called once by \code{callP}, method \code{Observable.notifyAll} invokes 
\code{notify} on multiple objects ---and multiple times if some of those are
aliases.  
By sharing state, it is possible for multiple observers to detect the
order in which they are notified.  
In our versions of \code{Observable}, method \code{add} maintains the
set in last-in order.  In Fig.~\ref{fig:obsc}, method \code{add} in
\code{Node1} shuffles pointers to maintain the last-in order.

\begin{figure}[t] 
\sf\begin{tabbing}
\CLASS\ Node1 \EXT\ \OBJECT \{ // rep for Observable \\
\hspace*{1em}\= Observer ob; \+\\
  Node1 nxt;\\
  \UNIT\ setOb(Observer o)\{ \self.ob\assym o \}\\
  \UNIT\ add(Node1 n)\{ \\
\hspace*{1em} \= Observer o\assym n.ob; n.ob\assym \self.ob; \self.ob\assym o; 
      n.nxt\assym \self.nxt; \self.nxt\assym n \} \\
  \UNIT\ notifyAll()\{ \self.ob.notify(); \IF\ \self.nxt $\neq$ \NULL\ 
  \THEN\ \self.nxt.notifyAll() \ELSE\ \SKIP\  \FI\ \} \} \- \\
\CLASS\ Observable \EXT\ \OBJECT \{ // owner \+ \\
  Node1 fst;\\
  \UNIT\  add(Observer ob)\{ \+ \\
Node1 n\assym \NEW\ Node1; n.setOb(ob); 
\IF\ \self.fst = \NULL\  \THEN\ \self.fst\assym n \ELSE\ \self.fst.add(n) \FI
\}  \- \\
  \UNIT\ notifyAll()\{ \IF\ \self.fst $\neq$ \NULL\ \THEN\ 
  \self.fst.notifyAll() \ELSE\ \SKIP\  \FI
  \} \}
\end{tabbing}
\caption{Version of the observer pattern in object-oriented style: nodes are active.}
\label{fig:obsc}
\end{figure}

\begin{figure}[t] 
\sf\begin{tabbing}
\CLASS\ Node3 \EXT\ \OBJECT\ \{ // rep for Observable \\
\hspace*{1em}\=    Node3 nxt; \+ \\
   \UNIT\ notif()\{ \SKIP\ \} \\
   \UNIT\ notifyAll()\{ \self.notif(); \IF\ \self.nxt $\neq$ \NULL\
   \THEN\ \self.nxt.notifyAll() \ELSE\ \SKIP\ \FI\ \} \\
   \UNIT\ add(Observer ob)\{ 
NodeO n\assym \NEW\ NodeO; n.setOb(ob); n.nxt\assym \self.nxt;
\self.nxt\assym n \} \} \- \\
\CLASS\ NodeO \EXT\ Node3 \{ // rep subclass \\
\hspace*{1em}\= Observer ob; \+ \\
   \UNIT\ setOb(Observer o)\{ \self.ob\assym o \} \\
   \UNIT\ notif()\{ \self.ob.notify() \} \} \- \\
\CLASS\ Observable \EXT\ \OBJECT  \{ // owner \+ \\
   Node3 snt; \\
   \CON\{ \self.snt\assym \NEW\ Node3 \} \\
   \UNIT\ add(Observer ob)\{ \self.snt.add(ob) \}\\
   \UNIT\ notifyAll()\{ \self.snt.notifyAll() \} \}
\end{tabbing}
\caption{Sentinel in object-oriented style.
In class \code{Node3}, method \code{add} constructs an object of the subclass
\code{NodeO} and method \code{notifyAll} uses dynamic dispatch of \code{notif}.
}
\label{fig:obsca}
\end{figure}
A less awkward version, using a sentinel node, is given in Fig.~\ref{fig:obsca}.

The following example indicates the limits of what can be proved using the abstraction 
theorem.  For this discussion, instead of treating loops as syntactic
sugar we assume they are in the language.  The semantic clause would
use a fixpoint but this is separate from the fixpoint of the
approximation chain used for method meanings.  Thus for each $i>0$ the
full semantics of a loop is defined in $\mu_i$. 

\begin{example}\label{ex:ver}
Consider the versions given by Figs.~\ref{fig:obs} and \ref{fig:obsc}.  
The data structures are very similar; essentially the identity coupling can be used.
(It is not literally the identity, because 
because \code{Node} and \code{Node1} are distinct classes and thus the sets
$\means{\code{Node}}$ and $\means{\code{Node1}}$ have no non-$\NIL$ location in
common.  But that is just the reflection of a coding trick in our formalization of
semantics.)  
The bodies of \code{add} and \code{notifyAll} in the 
two versions have significant differences in the calling graph, and in particular 
\code{notifyAll} in one version uses a loop whereas in the other it calls a recursive 
method in \code{Node1}.
To reason about these would require proving a loop invariant and
verifying specifications for methods \code{add} and \code{notifyAll} in \code{Node1}.
But for this one wants the final semantics of the program, not the approximate one
given by $\mu_i$ and $\mu'_i$.  For given $i$, the semantics of \code{notifyAll} is
only defined up to recursion depth $i$; for a list longer than that, the loop
in Fig.~\ref{fig:obs} works correctly but the recursion in Fig.~\ref{fig:obsc}
aborts.

By contrast, equivalence between Figs.~\ref{fig:obsaa} and \ref{fig:obsc} can be
shown by an argument similar to that in Example~\ref{ex:obs}.  They do not have the
same method call graph, but the called methods are not recursive so one can argue by
cases for $i=0$ and $i>0$.

If the loop in Fig.~\ref{fig:obs} is treated as syntactic
sugar for a method call then the equivalence has a complicated proof
in terms of corresponding unfoldings of the semantic approximations.
But this is an accidental feature of the example. 
\qed\end{example}

Example~\ref{ex:ver} might lead one to wonder whether there is a flaw
in the definition of simulation. Instead of requiring that owner
methods preserve the relation given any approximating and related
environments $\mu_i,\mu'_i$, perhaps it should be enough to consider
the final semantics $\means{CT},\means{CT'}'$.  But this is not a
sufficiently strong induction hypothesis to prove the abstraction theorem.  
In fact the example reflects a limitation in most theories of simulation and logical
relations: what can be shown equivalent are programs with the same structure in some
sense; see Sect.~\ref{sect:disc}.

\begin{example}
Equivalence between the versions given by Figs.~\ref{fig:obsc} and
\ref{fig:obsca} can be shown by an argument similar to that in
Example~\ref{ex:obs}.  The basic coupling is like that of
Example~\ref{ex:obsSim} with minor changes: $Rep,Rep'$ are named
\code{Node1},\code{Node3} and the sentinel is at location $\ell'_0\in
locs(\code{Node3})$ whereas the locations $\ell'_1,\ell'_3,\ldots$ following it are in
$locs(\code{NodeO})$.  
The method call graphs are not identical for the two versions and
dynamic dispatch is used in the second version for \code{Node3.notif}.
But the differences involve non-recursive methods and it suffices, as in 
Example~\ref{ex:obs}, to consider two cases for $\mu_i,\mu'_i$,
namely $i=0$ and $i>0$. 
\qed\end{example}


\section{Owner subclassing: the protected interface}\label{sect:osub}

This section considers examples involving subclasses of the owner
class.  Rather than formalizing the ``protected'' construct of Java,
we address the issues using a module construct.
We augment the syntax to designate certain methods as having
module scope, meaning that they cannot be called by clients. 
The confinement conditions for these methods are relaxed.

\subsection{Owner subclassing and module scope}\label{ssect:osm}

The code for \code{notifyAll} in \code{Observable} of Fig.~\ref{fig:obs}
uses a loop.  Here is an equivalent version using a tail recursive helper
method \code{doNotif}.  
\begin{quote}\sf
\begin{tabbing} 
  \UNIT\  notifyAll()\{ doNotif(\self.fst) \}  \\
  \UNIT\  doNotif(Node n)\{ \\
\qquad \IF\ n $\neq$ \NULL\ \THEN\ n.getOb().notify(); doNotif(n.getNext()) \ELSE\
  \SKIP\ \FI\ \}
\end{tabbing} 
\end{quote}
In a language with nested method declarations, \code{doNotif} could be
declared within \code{notifyAll}.  Absent that, it could be given private
scope, allowing its calls only in \code{Observable}. 
But the language of Sects.~\ref{sect:syn}--\ref{sect:app} has only public methods.  To apply our abstraction theorem
to the desugared version we would have to include a suitable implementation
of \code{doNotif} in every version.  This can be done for the 
examples in this paper, but it is awkward.  

In Sect.~\ref{ssect:mscope} we add module-scoped methods to the
language.  These suffice for desugaring loops and for interactions
between reps and owners.  In the sequel we focus on their use in
subclasses of $Own$. 

Fig.~\ref{fig:obsd} is a variation on the observer pattern in which class
\code{Observable} has subclass \code{ObservableAcc}.  For accounting
purposes it keeps track of the number of times each observer has been
notified. To this end, the rep class \code{NodeAcc} overrides method
\code{notifyAll} of the client class \code{Node4}.    
Such examples have led to our treatment of owner subclasses:
They are distinguished from clients in that their methods may manipulate
reps, but unlike $Own$ they cannot store reps in fields.

\begin{figure}[t] 
\sf\begin{tabbing}
\CLASS\  Node4 \EXT\  \OBJECT\ \{ // rep for Observable\\ 
\hspace*{1em}\=
   Observer ob;\+\\
   Node4 nxt;\\
   \UNIT\ setOb(Observer o)\{ \self.ob\assym o \} \\
   \UNIT\ setNext(Node4 n)\{ \self.nxt\assym n \}\\
   Observer getOb()\{ \result\assym \self.ob \}\\
   Node4 getNext()\{ \result\assym  \self.nxt \}\\
   Node4 getNextPri()\{ \result\assym  \self.nxt \}\\
   \UNIT\ notifyAll()\{ \self.ob.notify(); \IF\ \self.nxt $\neq$ \NULL\
     \THEN\ \self.nxt.notifyAll() \ELSE\ \SKIP\  \FI\ \} \} 
\- \\
\CLASS\  NodeAcc \EXT\  Node4 \{   \+ \\
   \keyword{int}  notifs;\\
   \UNIT\  notifyAll()\{ \self.notifs\assym \self.notifs+1; \SUPER.notifyAll() \} \\
   \keyword{int}  notifications(Observer o)\{  \result\assym 0; \\ 
\hspace*{1em}\=
\IF\ \self\ = \self.getOb() \THEN\ \result\assym notifs  \+ \\
\ELSE\ \IF\ \self.getNext() $\neq$ \NULL\  \THEN\  \result\assym
(NodeAcc)(\self.getNext()).notifications(o) \\
\ELSE\ \SKIP\  \FI\ \}\} 
\- \- \\
\CLASS\  ObservableSup \EXT\ \OBJECT\ \{ // superclass of owner;
"abstract" class \+ \\
   \UNIT\  add(Observer ob)\{ \ABORT\ \} \\
   \UNIT\  notifyAll()\{ \ABORT\ \} 
\- \\
\CLASS\  Observable \EXT\ ObservableSup \{ // owner \+ \\
   Node4 fst;               // first node of list \\
   Node4 getFirst()\{ \result\assym  \self.fst \}  // module scope \\
   \UNIT\  add(Observer ob)\{ Node4 n\assym \NEW\  Node4; \self.addn(ob,n) \} \\
   \UNIT\  addn(Observer ob, Node4 n)\{ n.setNext(\self.fst); n.setOb(ob);
   \self.fst\assym n \} // module scope\\
   \UNIT\  notifyAll()\{ \self.fst.notifyAll() \} \- \\
\CLASS\ ObservableAcc \EXT\  Observable \{ \+ \\
   \UNIT\ add(Observer ob)\{ Node4 n\assym \NEW\  NodeAcc(); \self.addn(ob,n) \}\\
   \keyword{int}  notifications(Observer ob)\{ 
\result\assym ((NodeAcc)(\self.getFirst())).notifications(ob) \} \}
\end{tabbing}
\caption{Version with owner and rep subclasses and super-call.
The owner also has a superclass.  The two versions of \code{getNext}
in \code{Node4} are needed for later examples.
}
\label{fig:obsd}
\end{figure}


Method \code{addn} has been added to \code{Observable}, so
that \code{ObservableAcc} can construct reps of the subtype
\code{NodeAcc} and install although \code{fst} is a private field not
visible in \code{ObservableAcc}.  
Method \code{Observable.getFirst} is also added for this purpose.  But
\code{getFirst} leaks a rep; it cannot be allowed in the public interface.  
One possibility is to treat \code{getFirst} and \code{addn} as 
visible only in subclasses of \code{Observable}.   Instead, we give
them module scope, meaning that calls to \code{getFirst} and
\code{addn} are allowed in subclasses of both \code{Observable} and \code{Node4}. 

Method \code{add} in class \code{ObservableAcc} constructs a rep, violating the
condition ``no reps in sub-owners'' in Assumption~\ref{ass:first}.
That assumption is needed for the first abstraction theorem because 
methods of an owner subclass are like clients in that they must 
preserve the induced relation. That means in particular that they
manipulate related ---i.e., equal--- rep locations.  
(By contrast, methods of $Own$ preserve the basic coupling which need
not impose a correspondence on rep locations.)
But if we compare two versions, one with sentinel node and one without,
the parametricity condition for $\fresh$ will not apply and the new objects in
\code{ObservableAcc.add} will be at different locations.  
The solution, given in Sect.~\ref{sect:sabs}, is to relax equality to bijection.

This relaxation is needed anyway, to avoid unobservable distinctions.
As an example, suppose we add to class \code{Observable} in
Fig.~\ref{fig:obs} the following method:
\begin{quote}\sf
\begin{tabbing} 
  String\  version()\{ \= result\assym \NEW\  String(``vsn 0'') \}
\end{tabbing} 
\end{quote}
Consider an alternative that is identical in every way except for the following:
\begin{quote}\sf
\begin{tabbing} 
  String\  version()\{ \= result\assym \NEW\  String(``trash'');  
result\assym \NEW\  String(``vsn 0'') \}
\end{tabbing} 
\end{quote}
According to Def.~\ref{def:sim}, the induced relation for
locations of type \code{String} is equality.
But, even if the allocator is parametric, the locations returned by
these two methods are not equal.  
(So condition (\ref{simA}) fails in Def.~\ref{def:sim} of simulation.)
But they cannot be distinguished; this claim is justified by the
generalized theory of 
Sect.~\ref{sect:sabs}, where  the induced relation allows an arbitrary
bijection between locations of client types like \code{String}.
For this example, the bijection would be extended to relate the
returned results from the two versions.

Returning to the example in Fig.~\ref{fig:obsd}, the interface
betweeen \code{Observable} and its subclass \code{ObservableAcc} is awkwardly
designed.  An improvement is to use the factory 
pattern \cite{DesPat} so that \code{add} itself can be inherited. In Fig.~\ref{fig:obsf},
we add method \code{makeNode}, which should have module scope, and
remove \code{addn}.
\begin{figure}[t] 
\sf\begin{tabbing}
\CLASS\  Observable \EXT\  ObservableSup \{  \\
\hspace*{1em}\=
   Node4 fst; \+ \\
   Node4 getFirst()\{ \result\assym  \self.fst \} // module scope\\
   Node4 makeNode()\{ \result\assym \NEW\  Node4 \} // module scope\\
   \UNIT\  add(Observer ob)\{ 
      Node4 n\assym makeNode(); n.setNext(\self.fst); n.setOb(ob); \self.fst\assym n \}\\
   \UNIT\  notifyAll()\{ \self.fst.notifyAll() \} \}  \- \\
\CLASS\  ObservableAcc \EXT\  Observable \{ \+ \\
   Node4 makeNode()\{ \result\assym \NEW\  NodeAcc \} // module scope\\
   \keyword{int}  notifications(Observer ob)\{ 
\result\assym ((NodeAcc)(\self.getFirst())).notifications(ob) \} \}
\end{tabbing}
\caption{Variation on Fig.~\ref{fig:obsd} using factory pattern.   
\code{Node4} and \code{NodeAcc} are as in Fig.~\ref{fig:obsd}.
}
\label{fig:obsf}
\end{figure}

To illustrate that owners may reference each other, let us add a
method \code{allNotifications} 
which reports the number of times a given observer has been notified by any
observable in a group thereof.  In the code of Fig.~\ref{fig:obsg},
groups are represented by cyclic lists.  An \code{ObservableAccG} is
initially in a singleton group; groups grow using method
\code{joinGroup}. 

\begin{figure}[t] 
\sf\begin{tabbing}
\CLASS\  ObservableAccG \EXT\  ObservableAcc \{  \\
\hspace*{1em}\=
     ObservableAccG peer; \+ \\
     \CON\{ \self.peer\assym \self\ \}  \\
     \UNIT\  joinGroup(ObservableAccG o)\{ 
      // pre: \self.peer=\self\ and o.peer is cyclic list of length $\geq$ 1  \\
\hspace*{1em}\=  \self.peer\assym o.peer; o.peer\assym \self\ \}  \\
      \keyword{int} allNotifications(Observer ob)\{ \+ \\
      \result\assym \self.notifications(ob); 
      ObservableAccG p\assym \self.peer;  \\
      \keyword{while} p $\neq$ \self\ \DO\ \result\assym \result\ + p.notifications(ob);
      p\assym p.peer \OD\ \} \}
\end{tabbing}
\caption{Extension of Fig.~\ref{fig:obsd} or Fig.~\ref{fig:obsf} with
  grouped owners.}
\label{fig:obsg}
\end{figure}

These examples show subclasses of reps and owners.  There is
inheritance into the owner but not into the rep.  
Inheritance into reps is
disallowed by our definition of confined class table, because to
handle it requires a more sophisticated analysis to prevent leaks via
\self; a suitable analysis of ``anonymous methods'' is discussed in Sect.~\ref{sect:disc}.
Inheritance into owners also needs restriction; we have chosen a
simple restriction that nonetheless allows the preceding examples.

Finally, let us consider an alternative version of
Fig.~\ref{fig:obsf} to illustrate the consequences of allowing the
owner class, but not its subclasses, to differ in comparable class
tables.  In Fig.~\ref{fig:obsf} the subclass \code{ObservableAcc} manipulates reps,   both
constructing a new \code{NodeAcc} and invoking method
\code{notifications} declared in \code{NodeAcc}.  Although an
alternative version of \code{Observable} could use an entirely
different type of nodes internally, it has to provide method
\code{getFirst} with return type \code{Node4}.  Because clients can
manipulate objects of class \code{ObservableAcc}, methods of that
class must preserve the relation and this only holds if methods they
invoke preserve the relation.  So coupling must be preserved not only
by public methods of \code{Observable} but also by those module scope
methods that are invoked in \code{ObservableAcc}. 
As a simple example, Fig.~\ref{fig:obsh} gives an alternative that uses
\code{Node4} and differs from Fig.~\ref{fig:obsf} only in using a
sentinel node.  

\begin{figure}[t] 
\sf\begin{tabbing}
\CLASS\  Observable \EXT\  ObservableSup \{  \\
\hspace*{1em}\=
   Node4 snt; \+ \\
\CON\{ snt\assym \NEW\ Node4 \} \\
   Node4 getFirst()\{ \result\assym  \self.snt.getNextPri() \}  // module scope \\
   Node4 makeNode()\{ \result\assym \NEW\  Node4 \} // module scope\\
   \UNIT\  add(Observer ob)\{  \\
\hspace*{1em} \= Node4 n\assym makeNode();
      n.setNext(\self.snt.getNextPri()); n.setOb(ob); \self.snt.setNext(n) \}  \\ 
   \UNIT\  notifyAll()\{ \self.snt.getNextPri().notifyAll() \} \} \- \\
\end{tabbing}
\caption{Variation on Fig.~\ref{fig:obsf} using sentinel.}
\label{fig:obsh}
\end{figure}

\subsection{On behavioral subclassing}\label{ssect:beh}

Behavioral subclassing \cite{LiskovWing} is very useful for reasoning about specific examples. 
However, as mentioned earlier, it is not required in general for representation
independence.  Client, rep, or owner subclasses may fail to exhibit
behavioral subclassing.  
To illustrate the point let us consider two
revisions of Fig.~\ref{fig:obsd}, both of which violate behavioral
subclassing.
For the first example, we add an overriding declaration to \code{NodeAcc}:
\begin{quote}\sf
   Node4 getNext()\{ \ABORT\ \}  
\end{quote}
This causes \code{NodeAcc} to fail to be a behavioral subclass of
\code{Node4} by most definitions.   
(It also prevents the intended 
functioning of the added method \code{NodeAcc.notifications} and its callers).
Nonetheless, there is still a simulation between
Figs.~\ref{fig:obsf} and 
\ref{fig:obsh}.\footnote{Here we consider a class table comprised of
  \code{Node4}, \code{NodeAcc}, and \code{ObservableSup} from
  Fig.~\ref{fig:obsd}, along  with the overriding declaration
  \code{NodeAcc.getNext} and also \code{Observable} and
  \code{ObservableAcc} from Fig.~\ref{fig:obsf}.  The alternative 
    class table is the same except for using \code{Observable} from
    Fig.~\ref{fig:obsh}. 
}
Making this true is the reason Fig.~\ref{fig:obsh} uses
\code{getNextPri} instead of \code{getNext}.  

The second revision makes malicious use of a type test.
We add nothing to
\code{NodeAcc}, but rather revise \code{Node4} as follows:
\begin{quote}\sf
  \begin{tabbing}
   \UNIT\ notifyAll()\{ \= \IF\ \self\ \IS\ NodeAcc \THEN\ \ABORT\ \ELSE\ 
   \self.ob.notify() \FI; \+ \\ 
\IF\ \self.nxt $\neq$ \NULL\ \THEN\ \self.nxt.notifyAll() \ELSE\ \SKIP\  \FI\ \}  
  \end{tabbing}
\end{quote}
Method \code{notifyAll} in \code{NodeAcc} now fails to behave properly.  
In some sense, the revised \code{Node4} is non-monotonic with respect
to subclassing.
Again, there is still a simulation between Figs.~\ref{fig:obsf} and
\ref{fig:obsh}.  Method \code{notifyAll} aborts for
\code{ObservableAcc} objects in both versions.  

\subsection{Formalization of module-scoped methods}\label{ssect:mscope}

In Sect.~\ref{sect:app} we saw the need for methods that are
effectively private to $Own$, for desugaring loops, and also for
methods in $Own$ that cannot be called by clients but can be called in
subclasses of $Own$.  There is also a need for methods of owners and
reps that can be called by each other but not by clients.  For
simplicity, we address these needs with a simple notion:
$Own$, $Rep$, and their subclasses are considered to be inside a
module, and methods may be designated as being visible only inside the
module.

To avoid belaboring the formalization, we make no change to the
concrete syntax.  We assume that a class table designates the class
names $Own$ and $Rep$ and is equipped with a predicate  $\mscope$
with the interpretation that $\mscope(m,C)$ means this method has package scope.
The following changes are made to the definitions of preceding sections.
\begin{enumerate}
\item For a well formed class table, $\mscope$ must satify conditions
  that reflect what in practice would be achieved by declaring $Rep$,
  $Own$, and their subclasses inside the module.  
If $\mscope(m,C)$ then 
\begin{itemize}
\item $C\leq Own$ or $C\leq Rep$,
\item $\mtype(m,B)$ is undefined for $B>Own$ and $B>Rep$, and 
\item $B\leq C$ implies $\mscope(m,B)$.  
\end{itemize}

\item The typing rule for method call has an added restriction that
  module-scoped methods are only visible within the module:
\[ 
\Rule{
\begin{array}{c}
\Gamma \proves e:D \quad
\mtype(m,D) = \ol{\T}\TO \T \\ 
\Gamma  \proves\ol{e}:\ol{U} \quad \ol{U}\leq\ol{T} \quad x\neq \self
\quad T\leq \Gamma\,x \\
\mscope(m,D) \implies \Gamma\,\self\leq Own \lor \Gamma\,\self\leq Rep 
\end{array}}
{\Gamma \aproves \assign{x}{\mcall{e}{m}{\ol{e}}} }
\] 
\item For method environments, the confinement condition of
  Def.~\ref{def:muconf}(\ref{muconfa}) is replaced by the 
  following:
  \begin{itemize}
\item $C\leq Own \land \mscope(m,C) \implies 
\etaconf{C}{\eta}{h_0} \land 
h\hext h_0 \land (d\in locs(Rep\subclasses)\implies d\in \dom(Rh_j))$
for some confining partition and $j$ with $\eta\,\self\in\dom(\Oh_j)$
  \item $C\nleq Rep\land (C\nleq Own\lor \neg\mscope(m,C))  \implies 
\etaconf{C}{\eta}{h_0} \land h\hext h_0 \land  d\not\in
locs(Rep\subclasses) $
  \end{itemize}

\item For confinement of class tables, the restriction of 
  Def.~\ref{def:tconf}(\ref{tconfc}) is only 
  applied to methods with $\neg(\mscope(m,C))$.

\item \label{protdef}
For simulation, Def.~\ref{def:gsimu} in the sequel revises
  Def.~\ref{def:sim}(\ref{simp}) to require 
  preservation of the relation only for public methods, that is, if
  $\neg(\mscope(m,Own))$.  But those module-scoped methods that are
  called in sub-owners must also preserve the relation.  

To formalize this, we define $\protected(m,C)$ just if $C\leq Own$,
  $\mscope(m,Own)$, and there is a call to $m$ in some subclass of
  $Own$.   

\item Comparable class tables must agree on the public and protected
  methods of $Own$.  Def.~\ref{def:comp}(\ref{simb}) is extended to
  require that $\mscope(m,C)=\mscope'(m,C)$ for all $C\neq Own$.
Moreover, if $\mtype(m,Own)$ is defined then the following hold (and
\emph{mutatis mutandis} for $\mtype'$):
\begin{itemize}
\item 
$\neg\mscope(m,Own)$ implies 
$\mtype'(m,Own)=\mtype(m,Own)$ and $\neg\mscope'(m,Own)$, and 
\item $\protected(m,Own)$ implies
$\mtype'(m,Own)=\mtype(m,Own)$ and $\mscope'(m,Own)$
(which in turn implies $\protected'(m,Own)$).
\end{itemize}
\end{enumerate}

\begin{example}
Method \code{doNotif} in Sect.~\ref{ssect:osm} can be given module
scope.  It would not be called in owner subclasses, so it is not
required to be present in a comparable class table.  Method
\code{getFirst} of \code{Observable} in Fig.~\ref{fig:obsd} is called
in subclass \code{ObservableAcc}, so 
$\protected(\code{getFirst},\code{Observable})$ holds and
\code{getFirst} must be present in a comparable class table (and be simulated). \qed
\end{example}

Results of Sections~\ref{sect:sem} and \ref{sect:conf} hold for the extended
language; the only proof affected by the changes is that of 
Theorem~\ref{thm:cfix} which says that $\means{CT}$ is
confined if $CT$ is confined.  The result holds for the revised
definitions; the necessary revisions for the proof are as follows:
\begin{itemize}
\item In the base case of the induction on depth, the argument proving
  confinement of $\mu_{i+1} C m$ for the 
  result value $d$ goes by cases on $C$. 
  The argument for the case $C\leq Own$ still holds for $m$ with 
$\neg\mscope(m,C)$.  For the case $C\leq Own$ and $\mscope(m,C)$, the revised
definition requires the result value $d$ to satisfy
$d\in locs(Rep\subclasses)\implies d\in \dom(Rh_j)$
for some confining partition and $j$ with
$\eta\,\self\in\dom(\Oh_j)$.  This follows by definition from
$\etaconf{C}{\eta_0}{h_0}$.
\item In the step of the induction on depth, there is case analysis on
  $C$ and $B$, proving claim $\etaconf{B}{\eta}{h}$ and confinement of the result
  value $d$.  For the case $C\leq Own< B$, the argument still holds,
  noting that $\neg\mscope(m,C)$ because in a well formed class table
  module-scoped methods do not occur outside owner and rep classes.  
For the cases $C<B\leq Own$ and $C<B\leq Rep$, the arguments still
hold, noting that the restrictions on $\mscope$ ensure 
$\mscope(m,B)=\mscope(m,C)$ so the relevant conditions are the same.
\end{itemize}


\section{Second abstraction theorem}\label{sect:sabs}

This section improves the first abstraction theorem in two ways.
First, the result applies to the language extended with modules (see
Sect.~\ref{ssect:mscope}). The module-scoped methods of the two
versions of $Own$ can be different unless they are used in subclasses
of $Own$.  
The second improvement is that parametricity of the allocator is no
longer required (cf. Sect.~\ref{ssect:osub}).
To compare behaviors of two versions of a program we use a bijection
between locations rather than equality.  This can be seen as expressing
that the language is parametric in locations, which would fail if the language had
pointer arithmetic.  
As discussed in Sect.~\ref{ssect:osm}, allowing bijection handles
the problem with new reps in sub-owners that necessitates
Assumption~\ref{ass:first}.  
Moreover, it allows coarsening of the notion of
equivalence for commands and method meanings so that, for example, the
bodies of the two versions of method \code{version} in
Sect.~\ref{ssect:osm} are equivalent.

These extensions are enough to treat all the examples in
Sect.~\ref{ssect:osm} in addition to those of Sect.~\ref{sect:app} 
(except Example~\ref{ex:ver}, for reasons discussed there). 

\begin{definition}[{\bf typed bijection}]
A \dt{typed bijection} is finite bijective function  $\sigma$ from $Locs$
to $Locs$ such that $\sigma\,\ell=\ell'$ implies 
$loctype\,\ell = loctype\,\ell'$.  \qed
\end{definition}
Throughout the section we let $\sigma$ range over typed bijections and
sometimes omit the word ``typed''.  
To express how  bijections cut down to bijections on blocks of partitions,
we use the notation $\sigma(X)$ for the direct image of $X$ through
$\sigma$.

\begin{definition}[{\bf basic coupling}]\label{def:ssim}
Given comparable class tables, a basic coupling 
is a function $G$ that assigns to each typed bijection a binary relation 
$G\,\sigma$ on heaps (not necessarily closed heaps) that satisfies the
following.
For any $\sigma,h,h'$, if $G\:\sigma\:h\:h'$ then 
there are partitions $h=\Oh *\Rh$ and $h'=\Oh'*\Rh'$ and locations $\ell$
and $\ell'$  in $locs(Own\subclasses)$ such that
\begin{enumerate}
\item\label{ssimea}
$\sigma\,\ell=\ell'$ and  
 $\{\ell\} = \dom\,\Oh$ and 
 $\{\ell'\} = \dom\,\Oh'$
\item
 $\dom(\Rh) \subseteq locs(Rep\subclasses)$ 
and   $\dom(\Rh') \subseteq locs(Rep'\subclasses)$ 
\item\label{ssimec} 
$\srel{\sigma}{(\type(f,\loctype\,\ell))}{(h\ell f)}{(h'\ell' f)} $
for all $(f:T)\in \dom(\fields(\loctype\,\ell))$ with 
$f\not\in\ol{g}= \dom(\dfields(Own))$  
and $f\not\in \ol{g}' = \dom(\dfields'(Own))$.\qed
\end{enumerate}
\end{definition}
Item (\ref{ssimec}) uses the induced coupling $\SREL$ defined below; it is a
harmless forward reference because the definition of $\SREL$ for data types
does not depend on $\SREL$ (or $G$) for heaps.
Note that we do not require $\dom\,\sigma$ to include the reps, nor do we
disallow that it includes some of them.

\begin{definition}[{\bf coupling relation, $\SREL$}]\label{def:sR}
In the context of a basic coupling with given relation $G$,
and for each typed bijection $\sigma$,
relations 
$\SREL\:\sigma\:\theta \subseteq \means{\theta}\times\means{\theta}'$ as
follows.  
(Note that in the case of method meanings and method environments there is
no dependence on $\sigma$.)

For heaps $h, h'$, we define  $\srel{\sigma}{\heap}{h}{h'}$
iff there exist confining partitions of $h,h'$, 
with the same number $n$ of owner islands,
such that 
\begin{itemize}
\item $\dom\,\sigma\subseteq \dom\,h$ and $rng\,\sigma\subseteq\dom\,h'$
\item 
$ G\;\sigma \; (\Oh_i * \Rh_i) \; (\Oh'_i * \Rh'_i)  $ for all $i$ in $1..n$
\item $\sigma(\dom(\Ch))=\dom(\Ch')$,
i.e., $\sigma$ restricts to a bijection between $\dom(\Ch)$ and
  $\dom(\Ch')$
\item   $\srel{\sigma}{(\state{(\loctype\;\ell)})}{(h\ell)}{(h'\ell')}$ 
for all $\ell,\ell'$ with $\ell\in\dom(\Ch)$ and $\sigma\,\ell\,\ell'$
\end{itemize}
For other categories $\theta$ we define 
$\srel{\sigma}{\theta}{}{}$ as follows.
\[\begin{array}{lcl}
\srel{\sigma}{\BOOL}{d}{d'}&\Leftrightarrow& d=d'\\[.5ex]
\srel{\sigma}{\UNIT}{d}{d'}&\Leftrightarrow& d = d' \\[.5ex]
\srel{\sigma}{C}{d}{d'}&\Leftrightarrow& \sigma\,d=d' \lor d=\NIL=d'\\[.5ex]
\srel{\sigma}{\Gamma}{\eta}{\eta'}& \Leftrightarrow&
\all{x\in\dom\,\Gamma}{\srel{\sigma}{(\Gamma x)}{(\eta x)}{(\eta' x)}}\\[.5ex]
\srel{\sigma}{(\state{C})}{s}{s'}&\Leftrightarrow& \\[.5ex]
\multicolumn{3}{l}{\qquad C\nleq Own \land 
\all{ f\in\dom(\fields\,C) }{ \srel{\sigma}{(\type(f,C))}{(s\;f)}{(s'\;f)} }}\\[.5ex]
\srel{\sigma}{(\theta_\bot)}{\alpha}{\alpha'} &\Leftrightarrow& 
(\alpha=\bot=\alpha')\vee(\alpha\neq\bot\neq\alpha'\land\srel{\sigma}{\theta}{\alpha}{\alpha'})
\\[.5ex] 
\srel{\sigma}{(\heap\otimes\Gamma)}{(h,\eta)}{(h',\eta')} &\Leftrightarrow& 
\srel{\sigma}{\heap}{h}{h'}\land \srel{\sigma}{\Gamma}{\eta}{\eta'}
\\[.5ex] 
\srel{\sigma}{(\heap\otimes\T)}{(h,d)}{(h',d')} &\Leftrightarrow& 
\srel{\sigma}{\heap}{h}{h'}\land \srel{\sigma}{\T}{d}{d'}
\\[.5ex] 
\srell{(C,\ol{x},\ol{T}\TO T)}{d}{d'}&\Leftrightarrow&
\all{\sigma,\,
  (h,\eta)\in\means{\heap\otimes\Gamma},(h',\eta')\in\means{\heap\otimes\Gamma}'\:}{} 
 \\[.5ex]
\multicolumn{3}{l}{\qquad \srel{\sigma}{(\heap\otimes\Gamma)}{(h,\eta)}{(h',\eta')}
\land  \etaconf{C}{\eta}{h}\land \etaconf{C}{\eta'}{h'} 
} 
\\[.5ex]
\multicolumn{3}{l}{ \qquad \implies
\some{\sigma_0\supseteq \sigma}{
 \srel{\sigma_0}{(\heap\otimes\T)_\bot}{(d (h,\eta))}{(d'(h',\eta'))}
} }
\\[.5ex] 
\multicolumn{3}{l}{
\qquad  \mbox{where }\Gamma= [\ol{x}\mapsto \ol{T},\self\mapsto C] } \\[.5ex]
\srell{\menv}{\mu}{\mu'}&\Leftrightarrow& \all{ C,m \:}{} \\[.5ex]
\multicolumn{3}{l}{
\qquad (\neg\mscope(m,C)\lor \protected(m,C)) \land 
(C\mbox{ is non-rep})\land(\mtype(m,C)\mbox{ is defined})}
\\[.5ex]
\multicolumn{3}{l}{
\qquad   \implies
\srel{}{(C,\pars(m,C),\mtype(m,C))}{(\mu\, C\, m)}{(\mu'\, C\, m)}
\qed }  
\end{array}
\]
\end{definition}
(Recall that $\protected$ is defined in (\ref{protdef}) of Sect.~\ref{ssect:mscope}.)

As an example, the body of \code{makeNode} in \code{ObservableAcc}
(Fig.~\ref{fig:obsf}) returns a new rep.  
Consider a coupling with a version using a sentinel.
Given a bijection $\sigma$
and related heaps $h,h'$, the location $\ell=\fresh(\code{Node4},h)$ may be
different from $\ell'=\fresh(\code{Node4},h')$ even if $\fresh$ is
parametric, because $h'$  has extra reps, the sentinels.
But $\sigma$ can be extended with the pair $(\ell,\ell')$.

The following facts are straightforward consequences of the definition.
The first says that if $h$ and $h'$ are related by $\SREL$ at
$\sigma$, then $\sigma$ is a bijection between the domains of $h$ and $h'$
except for reps. 
\begin{lemma}\label{fact:sfoo}
For all $\sigma,h, h'$ and all  $\ell,\ell'$ not in
$locs(Rep\subclasses,Rep'\subclasses)$,   
if $\srel{\sigma}{\heap}{h}{h'}$ then 
$\sigma(
(\dom\,h)\downharpoonright(locs(Rep\subclasses,Rep'\subclasses)))
= (\dom\,h')\downharpoonright(locs(Rep\subclasses,Rep'\subclasses))$.
\qed\end{lemma}

\begin{lemma}\label{fact:id:sc}
If $\ol{U}\leq\ol{T}$ and $\srel{\sigma}{\ol{U}}{\ol{d}}{\ol{d'}}$ then
$\srel{\sigma}{\ol{T}}{\ol{d}}{\ol{d'}}$.\qed
\end{lemma}

For equivalence of values and states, we define a family of relations indexed on
categories $\theta$.  To streamline the notation, we say
``$\veq{\sigma}{x}{x'}$ in $\means{\theta}$'' here, and simply use the
symbol $\veq{\sigma}{}{}$ later.

\begin{definition}[{\bf value equivalence}]\label{def:veq}
For any $\sigma$, we define a relation $\veq{\sigma}{}{}$ for data values,
object states, heaps, and stores, as follows.
\[
\begin{array}{llcl}
  \veq{\sigma}{\ell}{\ell'} & \mbox{ in } \means{C}
& \iff & \sigma\,\ell=\ell' \lor \ell=\NIL=\ell' \\
  \veq{\sigma}{d}{d'} &\mbox{ in } \means{T}
& \iff & d=d' \quad\mbox{for primitive types $T$} \\
  \veq{\sigma}{s}{s'} &\mbox{ in } \means{\state{C}} 
& \iff & \all{f\in\fields\,C}{ \veq{\sigma}{s f}{s' f} } \\
  \veq{\sigma}{\eta}{\eta'} &\mbox{ in } \means{\Gamma} 
& \iff & \all{x\in\dom\,\Gamma}{ \veq{\sigma}{\eta\,x}{\eta'\,x} } \\
  \veq{\sigma}{h}{h'} &\mbox{ in } \means{\heap} 
& \iff & \sigma(\dom\,h) = \dom\,h'
\land \all{\ell\in\dom\,h}{\veq{\sigma}{h\,\ell}{h'(\sigma\,\ell)} } \\
 \veq{\sigma}{(h,\eta)\!}{\!(h',\eta')} \!\!& \mbox{ in } \means{\heap\otimes\Gamma}\!\!
&\iff& \veq{\sigma}{h}{h'} \land \veq{\sigma}{\eta}{\eta'} \\
  \veq{\sigma}{d}{d'} & \mbox{ in } \means{\theta_\bot}
& \iff & d=\bot=d' \lor (d\neq\bot\neq d' \land \veq{\sigma}{d}{d'}\mbox{ in }\means{\theta})
\end{array}
\]
\end{definition}

\begin{lemma}[{\bf identity extension}]\label{lem:sidex}
Suppose $\srel{\sigma}{(\heap\otimes\Gamma)}{(h,\eta)}{(h',\eta')}$ 
and $\Gamma\,\self$ is non-rep.
Let $(h,\eta)$ and $(h',\eta')$ be confined at $\Gamma\,\self$.
If both $collect(\eta,h)$ and $collect(\eta',h')$ are $Own$-free
then $\veq{\sigma}{collect(\eta,h)}{collect(\eta',h')}$.\qed
\end{lemma}

The reader may care to check that in the case that $\sigma$ is equality, 
the relations $\SREL\:\sigma\:\theta$ coincide with $\REL\:\theta$ and
$\veq{\sigma}{}{}$ is just equality.

\begin{definition}[{\bf client program equivalence}]\label{def:speq}
Suppose programs $CT,(\Gamma \proves S )$ and $CT',(\Gamma \proves' S' )$
are such that $CT,CT'$ are comparable and confined, and moreover $S$
(resp.\ $S'$) occurs in $CT$ (resp.\ $CT'$).  
The programs are equivalent iff for all confined, $Own$-free $(h,\eta)$
and $(h',\eta')$ in 
$\means{\heap\otimes\Gamma}$  and all $\sigma$ with  
$\veq{\sigma}{(h,\eta)}{(h',\eta')}$, there is some $\sigma_0\supseteq
\sigma$ with 
\[ \veq{\sigma_0}{collect(\means{\Gamma \proves S }\hat{\mu}(h,\eta))}{
collect(\means{\Gamma \proves' S' }'\hat{\mu}'(h',\eta'))} 
\epunc ,\]
where $\hat{\mu}= \means{CT}$ and $\hat{\mu'}= \means{CT'}'$.\qed
\end{definition}

\begin{lemma}\label{lem:sinh}
Suppose $C$ and all class names in $\ol{T}$ are non-rep, and 
$B<C$. If  $\srel{}{(C,\ol{x},\ol{T}\TO T)}{d}{d'}$ then
$\srel{}{(B,\ol{x},\ol{T}\TO T)}{(\restr(d,B))}{(\restr(d',B))} $
where $\restr$ is the restriction to global states of $B$ (see
Def.~\ref{def:rest}).
\qed\end{lemma}

As discussed in Sect.~\ref{sect:osub}, the relation must be preserved
not only by public methods but also by any module scope methods that
are called by methods declared in subclasses of $Own$.

\begin{definition}[{\bf simulation}]\label{def:gsimu}
A simulation is a coupling relation $\SREL$ such that 
\begin{enumerate}
\item
(constructors of $Own$ establish $\SREL$)
For any $\mu,\mu'$, any $\ell,\ell'$ in $locs(Own\subclasses)$ with
$\sigma\,\ell=\ell'$, and any $h,h'$ with $\srel{\sigma}{\heap}{h}{h'}$,
let  
\[ 
\begin{array}{l}
h_1= \ext{h}{\ell}{[\fields(\loctype\,\ell)\mapsto \mathit{defaults}]} \\
h'_1= \ext{h'}{\ell'}{[\fields'(\loctype\,\ell')\mapsto \mathit{defaults}]} \\
h_0 =\means{\self:(\loctype\,\ell)\proves
  \constr(\loctype\,\ell):\CON}\hat{\mu}(h_1,[\self\mapsto\ell]) \\
h'_0 =\means{\self:(\loctype\,\ell')\proves'
  \constr(\loctype\,\ell'):\CON}'\hat{\mu}'(h'_1,[\self\mapsto\ell']) 
\end{array}
\]
Then there is $\sigma_0\supseteq \sigma$ such that $G\;\sigma\;h_0\;h'_0$. 

\item\label{ssimm} (methods of $Own$ preserve $\SREL$) 
Let $\mu\in \nats\to\means{\menv}$ (resp.\ $\mu'\in \nats\to\means{\menv}'$) 
be the approximation chain in the definition of $\means{CT}$ (resp.\ $\means{CT'}'$). 
For every $m$ with $\mtype(m,Own)$ defined
and $\neg\mscope(m,Own)$ or $\protected(m,Own)$, 
the following implications hold for every $i$,
where $\ol{x}= \pars(m,Own)$ and $\ol{T}\TO\T = \mtype(m,Own)$.  
 \begin{enumerate}
 \item\label{ssimA} $ \srel{\menv}{\mu_i}{\mu'_i} \implies 
\srel{(Own,\ol{x},\ol{\T}\TO\T )}{(\means{M}\mu_i)}{(\means{M'}'\mu'_i)}  $ \\[.5ex]
if $m$ has declaration $M$ in $CT(Own)$ and  $M'$ in $CT'(Own)$ 
\item\label{ssimB} $ \srel{\menv}{\mu_i}{\mu'_i} \implies 
\srel{(Own,\ol{x},\ol{\T}\TO\T )}{(\means{M}\mu_i)}{(\restr(\means{M_B}'\mu'_i,Own))}  $\\[.5ex]
if $m$ has declaration $M$ in $CT(Own)$ and is inherited from $B$ in
$CT'(Own)$, with $M_B$  the declaration of $m$ in $B$
\item\label{ssimC} $ \srel{\menv}{\mu_i}{\mu'_i} \implies 
\srel{(Own,\ol{x},\ol{\T}\TO\T )}{(\restr(\means{M_B}\mu_i,Own))}{(\means{M'}'\mu'_i)}  $\\[.5ex]
if $m$ has declaration $M'$ in $CT'(Own)$ and is inherited from $B$ in
$CT(Own)$, with $M_B$  the declaration of $m$ in $B$
\end{enumerate}
\end{enumerate}  
\end{definition}

Instead of Assumption~\ref{ass:first} we need only the following.

\begin{assumption}
$CT$ and $CT'$ are confined class tables for which a (generalized) simulation
$\SREL$ is given.  
\end{assumption}

\begin{theorem}[abstraction]\label{thm:sfat}
$\srell{\menv}{\means{CT}}{\means{CT'}'}$.
\end{theorem}

The proof is essentially the same as the proof of
Theorem~\ref{thm:fat}.  
The definition of $\SREL\:\menv$ requires the relation to be preserved
by those module-scoped methods that are called by subowners, and this
is ensured by Def.~\ref{def:gsimu}(\ref{ssimm}) of simulation.
The lemmas used in the proof are as follows.

\begin{lemma}[{\bf preservation by expressions}]\label{lem:sepres}
For any non-rep class $C\neq Own$ and any constituent expression
$\Gamma\proves e:T$ of a method declared in  $C$, the following
holds:
For all $\sigma$ and all  $(h,\eta)\in\means{\heap\otimes\Gamma}$ and 
$(h',\eta')\in\means{\heap\otimes\Gamma}'$, if 
$\srel{\sigma}{(\heap\otimes\Gamma)}{(h,\eta)}{(h',\eta')}$ then
\[ \srel{\sigma}{(T_\bot)}{(\means{\Gamma\proves
    e:T}(h,\eta))}{(\means{\Gamma\proves' e:T}'(h',\eta'))}   
\epunc .\] 
\end{lemma}
\begin{proof}
  The proof is very similar to the proof of Lemma~\ref{lem:epres} except in
the case of field access. 

For $\Gamma \proves \faccess{e}{f}:T$, the argument is as follows, for any $\sigma$. 
By induction on $e$ we have $\srel{\sigma}{C_\bot}{\ell}{\ell'}$.
In the non-$\bot$ case, $\ell\neq\NIL\neq\ell'$ hence,   
by definition of $\SREL$, $\sigma\,\ell=\ell'$.
By closure of the heaps, $\ell\in \dom\,h$ and $\ell'\in \dom\,h'$.

We consider cases on whether $C<Own$.   
Consider confining partitions $(\Ch * \Oh_1 * \Rh_1 \ldots) = h$ and 
$(\Ch'* \Oh'_1 * \Rh'_1\ldots) = h'$ that have corresponding islands
as in the definition of $\REL\;\heap$.
In the case $C<Own$, we have  $\ell\in locs(Own\subclasses)$ and hence
$\ell$ in some $\dom(\Oh_i)$.  
From $\srel{\sigma}{\heap}{h}{h'}$ we have 
\[ G\;\sigma\;(\Oh_i * hRep_i)\;(\Oh'_i * hRep'_i) \]
and thus $\ell'\in\dom(\Oh'_i)$ by basic coupling  
Def.~\ref{def:ssim}(\ref{ssimea}) and bijectivity of $\sigma$.
Since $C\neq Own$, we know by visibility that $f$ is not in the private
fields $\ol{g}$ of $Own$.
Thus, as $\type(f, \loctype\;\ell)) = T$, we have
$\srel{\sigma}{T}{(h \ell f)}{(h' \ell' f)} $ by
Def.~\ref{def:ssim}(\ref{ssimec}).

In the case  $C\nleq Own$ 
we have $\ell\in\dom(\Ch)$ and hence $\ell'\in\dom(\Ch')$ by
$\sigma\,\ell=\ell'$ and definition $\SREL\;\heap$. 
Hence \[ \srel{\sigma}{(\state{(loctype\:\ell)})}{(h\ell)}{(h'\ell')}\] and thus
$\srel{\sigma}{T}{(h\ell f)}{(h'\ell' f)}$ by definition of
$\SREL\;(\state{(loctype\:\ell)})$.
Note that $loctype\,\ell = loctype\,\ell'$ because $\sigma$ is a typed bijection.
\end{proof}

\begin{lemma}[{\bf preservation by commands}]\label{lem:scpres}
Suppose that $\mu$ and $\mu'$
are confined method environments and $\srel{}{\menv}{\mu}{\mu'}$. 
Then the following holds for any non-rep class $C\neq Own$.
For any constituent command $\Gamma\proves S $ 
in a method declaration in  $CT(C)$, any $\sigma$, and any 
$(h,\eta)\in\means{\heap\otimes\Gamma}$ and 
$(h',\eta')\in\means{\heap\otimes\Gamma}'$, if 
$\etaconf{C}{\eta}{h}$,
$\etaconf{C}{\eta'}{h'}$, 
and $\srel{\sigma}{(\heap\otimes\Gamma)}{(h,\eta)}{(h',\eta')}$ then
there is $\sigma_0\supseteq \sigma$ such that 
\[ 
\srel{\sigma_0}{(\heap\otimes\Gamma)_\bot}{(\means{\Gamma\proves S }\mu(h,\eta))}{
(\means{\Gamma\proves' S }'\mu'(h',\eta'))}
\epunc .\]
\end{lemma}
\begin{proof}
The proof is very similar to the proof of the corresponding Lemma~\ref{lem:cpres} except in
the cases of method call, field update, and most interestingly $\NEW$.  We no longer have
the assumption of parametricity of the allocator, and we must consider
construction of reps in sub-owners.
We also need an analog to Lemma~\ref{lem:cxpres}, saying that constructors 
establishes $\REL$:

\textbf{Claim:}
For all $\sigma$ and all $(h,\ell)\in\means{\heap\otimes C}$ and 
$(h',\ell')\in\means{\heap\otimes C}$,
if $\srel{\sigma}{\heap}{h}{h'}$ and $\srel{\sigma}{C}{\ell}{\ell'}$ then 
there is $\sigma_0\supseteq \sigma$ such that 
$\srel{\sigma_0}{\heap}{h_0}{h'_0}$ where 
\[  
\begin{array}{l}
h_0 = \means{\self:C\proves \constr\:C:\CON}\mu(h,[\self\mapsto\ell]) \\
h'_0 = \means{\self:C\proves' \constr\:C:\CON}\mu'(h',[\self\mapsto\ell']) 
\end{array}
\]
We omit the proof of the claim, which has the same structure as the proof
of Lemma~\ref{lem:cxpres}.  

\medskip
\Case\ $\Gamma \proves\assign{x}{\mcall{e}{m}{\ol{e}}}$.
This goes through as before except for 
the case where $C<Own$.  In that case, the called method may have
module scope and this is why such methods (designated by $\protected$)
are included in the definition of $\SREL\:\menv$.

\medskip
\Case\ $\Gamma \proves\assign{\faccess{e_1}{f}}{e_2} $.  
By Lemma~\ref{lem:sepres} for $e_1$ we have 
$\srel{\sigma}{C}{\ell}{\ell'}$, hence $\sigma\,\ell=\ell'$ definition of $\SREL$.
By Lemma~\ref{lem:sepres} for $e_2$ we have $\srel{\sigma}{U}{d}{d'}$ and hence
$\srel{\sigma}{T}{d}{d'}$  by Lemma~\ref{fact:id:sc}.
To conclude the argument it suffices to show
\begin{trivlist}
\item 
\hfill 
$\srel{\sigma}{\heap}{
  \ext{h}{\ell}{\ext{h\ell}{f}{d}}}{\ext{h'}{\ell'}{\ext{h'\ell'}{f}{d'}} }$
\epunc. 
\hfill $(*)$ 
\end{trivlist}
Consider confining partitions $(\Ch * \Oh_1 * \Rh_1 \ldots) = h$ and 
$(\Ch'* \Oh'_1 * \Rh'_1\ldots) = h'$ that correspond as in the definition of
$\srel{\sigma}{\heap}{h}{h'}$. 
We argue by cases on $C$.

\begin{itemize}
\item $C<Own$: Then $\loctype\,\ell\leq C < Own$.
By $\sigma\,\ell=\ell'$ and $\srel{\sigma}{\heap}{h}{h'}$, there is $i$
such that  $\{\ell\} =\dom(\Oh_i)$ and $\{\ell'\} =\dom(\Oh'_i)$ and 
\[ G\;\sigma\;(\Oh_i * \Rh_i)\;(\Oh'_i * \Rh'_i) 
\epunc .\]
By typing and $C\neq Own$, field $f$ is not in the private fields $\ol{g}$ of $Own$.  So
$(*)$ follows from $\srel{\sigma}{\heap}{h}{h'}$ and $\srel{\sigma}{T}{d}{d'}$.
\item $C\nleq Own$:
As $C$ is non-rep,   
we have $\ell\in\dom\,\Ch$ and $\ell'\in\dom\,\Ch'$.
Moreover,  
$\srel{\sigma}{(\state{(loctype\,\ell)})}{(h\ell)}{(h'\ell')}$ and so by 
$\srel{\sigma}{T}{d}{d'}$ we get
\[ \srel{\sigma}{(\state{(loctype\,\ell)})}{\ext{h \ell}{f}{d}}{\ext{h'
    \ell'}{f}{d'}} \epunc .\]
Hence $(*)$.
\end{itemize}

\Case\ $\Gamma\proves\assign{x}{\new{B}{\ }} $.
By confinement of $CT$, this command is confined and hence the final
states are confined: $\etaconf{C}{\eta_0}{h_0}$ and
$\etaconf{C}{\eta'_0}{h'_0}$.
We have $C\nleq Rep$ and $C\neq Own$.
Let $\ell = \fresh(B,h)$ and $\ell' = \fresh(B,h')$.  
Define $\sigma_1 = \sigma \union \{(\ell,\ell')\}$.  
This makes $\sigma_1$ bijective because $\ell,\ell'$ are fresh and 
$\srel{\sigma}{\heap}{h}{h'}$ implies, by definition, that
$\dom\,\sigma\subseteq\dom\,h$ and $rng\,\sigma\subseteq \dom\,h'$.

By $\srel{\sigma}{\Gamma}{\eta}{\eta'}$ and definition of $\sigma_1$ 
we have
$\srel{\sigma_1}{\Gamma}{\eta_0}{\eta'_0}$. We proceed to show  
$\srel{\sigma_1}{\heap}{h_0}{h'_0}$, by cases on $B$.
\begin{itemize}
\item $B\nleq Own\land B\nleq Rep$:   
We have $\fields\,B=\fields'\,B$ and thus
\[ \srel{\sigma_1}{(\state{B})}{[\fields\,B\mapsto \mathit{defaults}]}{
                         [\fields'\,B\mapsto \mathit{defaults}]} 
\epunc .\]
So, as $B$ is non-rep and $B\neq Own$, we can add $\ell$ to $\Ch$ and
$\ell'$ to $\Ch'$ to get partitions that witness
$\srel{\sigma_1}{\heap}{h_1}{h'_1}$. 
Now the induction hypothesis  yields some $\sigma_0\supseteq
\sigma_1$ such that $\srel{\sigma_0}{\heap}{h_0}{h'_0}$.
We obtain $\srel{\sigma_0}{\Gamma}{\eta_0}{\eta'_0}$ from 
$\srel{\sigma_1}{\Gamma}{\eta_0}{\eta'_0}$ 
because $\sigma_0\supseteq \sigma_1$.

\item $B\leq Own$:   
By basic coupling, Def.~\ref{def:ssim}, we get
$\sigma_0$ with $G\,\sigma_0\,h_2\,h'_2$.  Moreover,
$h_2$ and $h'_2$
are owner islands and the confining partitions for $h,h'$ extend to ones
for $h*h_2$. and $h'*h'_2$ with $\sigma_0$.  Finally, by definition of
$\SREL$ we get $\srel{\sigma_0}{\heap}{h_0}{h'_0}$ as $h_0=h*h_2$ and $h'_0
= h'*h'_2$.

\item $B\leq Rep$:
Here, $C\leq Own$ or $C\leq Rep$, as otherwise the command would not be
confined.
Let $j$ be such that $\eta\,\self\in\dom(\Oh_j*\Rh_j)$.
Add $\ell$ to $\Rh_j$ and $\ell'$ to $\Rh'_j$.  This yields 
$\srel{\sigma_0}{\heap}{h_0}{h'_0}$ with $h_0=h*h_2$ and $h'_0=h'*h'_2$.\qed
\end{itemize}
\end{proof}

\section{Static analysis}\label{sect:static}

This section gives a syntax directed static analysis.  It checks a
property called safety.  Safety is shown to imply confinement.

The input is a well formed class table and designated class
names $Own$ and $Rep$.   With one exception, only rep and owner code
(including subclasses) is  constrained.  The exception is for \NEW: a
client cannot construct a new rep.  For practical application, this
can be ensured in a modular way: $Rep$ and its subclasses 
would simply be declared with module scope.

The analysis is given for the language extended in Sect.~\ref{ssect:mscope}
with module-scoped methods.  For the original language, 
$\mscope(m,C)$ can be taken to be false for all $m$ and $C$. 

\begin{definition}[safe]\label{def:static}
Class table $CT$ is \dt{safe} iff for every $C$ and every $m$ with
$\mtype(m,C)=\ol{\T}\TO\T$ the following hold.
\begin{enumerate}
\item
If $m$ is declared in $C$ by
  $\T\;m(\ol{T}\;\ol{x})\{\S\}$ then
$ \ol{x}:\ol{\T},\self:C,\result:T \aproves S$ where $\aproves$ is the
safety relation defined in the sequel.
\item
$\self:C\aproves \constr\:C$, for all $C$
\item\label{safec} If  $C\leq Own$ and $\neg\mscope(m,C)$ then   $T\ncomp Rep$.
\item
If $m$ is inherited in $Own$ from some $B>Own$ then 
$\ol{T}\ncomp Rep$.
\item\label{safee} No $m$ is inherited in $Rep$ from any $B>Rep$.  
\end{enumerate}
\end{definition}

The safety relation $\aproves$ is defined by the following rules. 
There is no restriction on field declarations per se. A client can have a
$Rep$ type field, but can assign only \NULL\ to it.

\begin{infig}{Safety for expressions}
$ 
\begin{array}{c}
\Gamma \aproves x:\Gamma x
\quad 
\Gamma \aproves \NULL:B
\quad 
\Gamma \aproves \ITT:\UNIT
\quad
\Gamma \aproves \TRUE:\BOOL
\quad 
\Gamma \aproves \FALSE:\BOOL
\\[2.5ex]

\Rule{
\begin{array}{c}
C=(\Gamma\,\self)\quad \Gamma \aproves e:C 
\quad(f:\T)\in \dfields\, C\\
C=Own \land e\neq\self \implies T\ncomp Rep\\
C<Own \implies T\ncomp Rep
\end{array}}
{\Gamma \aproves \faccess{e}{f}:\T}
\\[2.5ex]

\Rule{
\begin{array}{c}
\Gamma \aproves e_1:\T \quad \Gamma\aproves e_2:\T \\
\end{array}}
{\Gamma \aproves \eqtest{e_1}{e_2}:\BOOL}
\qquad 
\Rule{\Gamma \aproves e:D \quad B\leq D}
     {\Gamma \aproves \cast{B}{e}:B}
\qquad 
\Rule{\Gamma \aproves e:D \quad B\leq D}
     {\Gamma \aproves \is{e}{B}:\BOOL}
\end{array}
$
\end{infig}

For expressions, the analysis imposes restrictions on field accesses
and nothing else.
If $\faccess{e}{f}$ appears in the body of an owner method, then 
a $Rep$ can be accessed only via the private fields of $Own$; this
requires $e$ to be $\self$. If $\faccess{e}{f}$ appears in a
sub-owner, then the private fields of 
$Own$ cannot be accessed, hence the result cannot be a $Rep$.

For commands, the rules impose restrictions on $\NEW$, field update,
and method call.
The conditions on field update are analogous to those for field access.
For an object construction $\assign{x}{\new{B}{}}$ in the body of a
client method, it cannot create a new rep. And, if it appears in a
subclass of $Rep$, it cannot create a new owner as this would break
confinement of the heap. 

For method call $\assign{x}{\mcall{e}{m}{\ol{e}}}$, the condition
labelled $(a)$ says 
that if $m$ is a client method called from a subclass of $Own$ or $Rep$,
then $m$ cannot be passed reps as parameters. Conditions $(b)$ and $(c)$
consider method calls from an owner class or its subclasses: $(b)$ says
that if $m$'s type is comparable to $Own$ then reps can be passed as
parameters only if $e$ is $\self$. 
Finally, $(c)$ says that if $m$'s type is comparable to $Rep$ then 
no owner, other than itself, can be passed as parameter ---otherwise
confinement will be violated.

\begin{infig}{Safety for commands}
\(
\begin{array}{c}
\Rule{
  \begin{array}{c}
x\neq \self \quad B\leq\Gamma x\quad B\neq \OBJECT\\
C\nleq Rep \land C\nleq Own \implies B\nleq Rep\\
C\leq Rep \implies B\nleq Own
  \end{array}}
     {\Gamma \aproves \assign{x}{\new{B}{\ }} }
\qquad 
\Rule{
\begin{array}{c}
C=(\Gamma\,\self)\qquad
(f:\T)\in \dfields\, C \\
\Gamma \aproves e_1:C \quad \Gamma \aproves e_2:U \quad U\leq T\\
C=Own \land e_1\neq\self \implies U\ncomp Rep\\
C<Own \implies U\ncomp Rep
\end{array}}
{\Gamma \aproves \assign{\faccess{e_1}{f}}{e_2} }
\\[2.5ex]

\Rule{
\begin{array}{c}
\Gamma \aproves e:D \quad
\mtype(m,D) = \ol{\T}\TO \T \quad T\leq \Gamma\,x \\
\Gamma  \aproves\ol{e}:\ol{U} \quad \ol{U}\leq\ol{T} \quad x\neq \self \\
C=(\Gamma\,\self)\qquad \mscope(m,D) \implies C\leq Own \lor C\leq Rep \\ 
\begin{array}{l}
(a)\quad (C\leq Own \lor C\leq Rep) \land D\not\leq Rep \land D\not\leq Own \implies
\ol{T}\ncomp Rep\\
(b)\quad C\leq Own \implies D\ncomp Own \lor (e=\self) \lor \ol{T}\ncomp Rep\\
(c)\quad C\leq Own \implies D\ncomp Rep \lor (\all{e_i \in\ol{e}}{e_i\neq\self \implies
\T_i \ncomp Own})
\end{array}
\end{array}}
{\Gamma \aproves \assign{x}{\mcall{e}{m}{\ol{e}}} }
\\[2.5ex]

\Rule{
\begin{array}{c}
C=(\Gamma\,\self)\quad 
\mtype(m,\super\,C) = \ol{\T}\TO \T \\
\Gamma \aproves \ol{e}:\ol{U} \quad \ol{U}\leq\ol{T}
\quad x\neq\self
\quad T\leq \Gamma\,x \\
\end{array}}
{\Gamma \aproves \assign{x}{\mcall{\SUPER}{m}{\ol{e}}} }
\\[2.5ex]

\Rule{
\begin{array}{c}
x\neq\self \quad
\Gamma \aproves e:\T \quad \T\leq \Gamma\, x
\end{array}}
{\Gamma \aproves \assign{x}{e} }

\qquad 
\Rule{
\begin{array}{c}
\Gamma \aproves \S_1
\quad
\Gamma \aproves \S_2
\end{array}}
{\Gamma \aproves \seq{\S_1}{\S_2} }
\\[2.5ex]

\Rule{
\begin{array}{c}
\Gamma \aproves e:\BOOL\quad 
\Gamma \aproves \S_1  \quad
\Gamma \aproves \S_2
\end{array}}
{\Gamma \aproves \ifelse{e}{\S_1}{\S_2} }
\qquad 
\Rule{
\begin{array}{c}
\Gamma \aproves e:U \quad U\leq\T\quad 
(\Gamma,x:\T) \aproves \S
\end{array}}
{\Gamma \aproves\var{\T\; x}{e}{S} }
\end{array}
\)
\end{infig}

\begin{theorem}[soundness]\label{thm:sasound}
If $CT$ is safe then it is confined.
\end{theorem}
\begin{proof}
Items (\ref{safec})--(\ref{safee}) in the definition of safety are the
same as items  (\ref{tconfc})--(\ref{tconfd}) in the definition of
confinement for class tables.  For items (\ref{tconfa}) and
(\ref{tconfe}), the confinement of method and constructor bodies follows from safety
thereof, by Lemmas~\ref{lem:aconf}, 
 \ref{lem:xsound}, and 
 \ref{lem:csound} to follow.
\end{proof}

\begin{lemma}[argument values confined]\label{lem:aconf}
Suppose $\Gamma \proves e:D$ and $\Gamma \proves \ol{e}:\ol{U}$ are
confined.
\begin{enumerate}
\item\label{aconfa}
If $\Gamma \aproves \assign{x}{\mcall{e}{m}{\ol{e}}}$ then
$\Gamma \proves \assign{x}{\mcall{e}{m}{\ol{e}}}$ has confined arguments.
\item\label{aconfb}
If $\Gamma \aproves \assign{x}{\mcall{\SUPER}{m}{\ol{e}}}$ then
$\Gamma \proves \assign{x}{\mcall{\SUPER}{m}{\ol{e}}}$ has confined arguments.
\end{enumerate}
\end{lemma}
\begin{proof}
We give the argument for (\ref{aconfa}); the argument for (\ref{aconfb}) is
similar (see Appendix).  

As in Def.~\ref{def:argconf}, let $C=(\Gamma\,\self)$. Assume $\conf{\mu}$
and $\etaconf{C}{\eta}{h}$. Let $\ell=\means{\Gamma \proves e:D}\mu(h,\eta)$,
let $\ol{d}=\means{\Gamma \proves \ol{e}:\ol{U}}\mu(h,\eta)$, and let
$\eta_1 = [\self\mapsto\ell, \ol{x}\mapsto\ol{d}]$. Finally, let $\ell\neq\NIL,
\ell\neq\bot$ and $\ol{d}\neq\bot$.

Because $\Gamma \aproves \assign{x}{\mcall{e}{m}{\ol{e}}}$
holds we can use conditions (a)--(c) in the analysis rule for method call.
Now the proof proceeds by cases on $C$ with subcases on $\loctype\,\ell$. In
each case we show $\etaconf{(\loctype\,\ell)}{\eta_1}{h}$.
\begin{itemize}
\item $C\nleq Rep \land C\nleq Own$: Because $e$ and $\ol{e}$ are confined at $C$, we
have $\ell\not\in locs(Rep\subclasses)$ and $d_i\not\in locs(Rep\subclasses)$ for
all $d_i\in\ol{d}$. Thus $rng\,\eta_1\intersect locs(Rep\subclasses)=\Empty$
proving $\etaconf{(\loctype\,\ell)}{\eta_1}{h}$ by
Def.~\ref{def:envconf}(\ref{envconfa}).
\item $C\leq Own$: Choose a confining partition and let $j$ be such that
$\eta\,\self \in \dom(\Oh_j)$. Because $e$ and $\ol{e}$ are confined at $C$, we
have $\ell\in locs(Rep\subclasses) \implies \ell\in\dom(\Rh_j)$ and
$d_i\in locs(Rep\subclasses) \implies d_i\in\dom(\Rh_j)$ for all $d_i\in\ol{d}$.
Now we go by cases on $\loctype\,\ell$:
\begin{itemize}
\item $\loctype\,\ell \nleq Rep \land \loctype\,\ell \nleq Own$: Because
$\loctype\,\ell \leq D$ we have $D\nleq Rep \land D\nleq Own$. Hence by
condition (a) of the analysis, $\ol{T}\ncomp Rep$. Thus
$rng\,\eta_1\intersect locs(Rep\subclasses)=\Empty$ proving
$\etaconf{(\loctype\,\ell)}{\eta_1}{h}$ by
Def.~\ref{def:envconf}(\ref{envconfa}).
\item $\loctype\,\ell \leq Own$: Hence $\ell\in\dom(\Oh_k)$ for some $k$. Because
$\loctype\,\ell \leq D$ we have, $D\leq Own \lor Own\leq D$. If $e=\self$ then
$\ell=(\eta\,\self)$ and $k=j$. Then $rng\,\eta_1\intersect
locs(Rep\subclasses)= \ol{d}\intersect locs(Rep\subclasses) 
\subseteq \dom(\Rh_j)=\dom(\Rh_k)$. Thus
$\etaconf{(\loctype\,\ell)}{\eta_1}{h}$ by 
Def.~\ref{def:envconf}(\ref{envconfb}). If $e\neq\self$ then by condition (b)
of the analysis, $\ol{T}\ncomp Rep$. Hence $rng\,\eta_1\intersect locs(Rep\subclasses)=
\Empty$ proving $\etaconf{(\loctype\,\ell)}{\eta_1}{h}$ by
Def.~\ref{def:envconf}(\ref{envconfb}).
\item $\loctype\,\ell \leq Rep$: Hence $\ell\in\dom(\Rh_j)$
by confinement of $e$ at $C$. As $\loctype\,\ell\leq D$
we have, $D\leq Rep \lor Rep\leq D$. By Def.~\ref{def:envconf}(\ref{envconfc}),
to show $\etaconf{(\loctype\,\ell)}{\eta_1}{h}$, we must show
$rng\,\eta_1\intersect \locsOR \subseteq \dom(\Oh_j * \Rh_j)$. For any
$d_i\in locs(Own\subclasses)$, because $\loctype\,d_i \leq T_i$, we have
$T_i\leq Own \lor Own \leq T_i$. So by condition (c) of the analysis, $e_i=\self$,
hence $d_i = (\eta\,\self) \in \dom(\Oh_j)$. For any $d_i\in locs(Rep\subclasses)$
we have $d_i\in\dom(\Rh_j)$ by confinement of $\ol{e}$ at $C$.
\end{itemize}
\item $C\leq Rep$: Choose a confining partition and let $j$ be such that
$\eta\,\self \in \dom(\Rh_j)$. Because $e$ and $\ol{e}$ are confined at $C$, we
have $\ell\in \locsOR \implies \ell\in\dom(\Oh_j * \Rh_j)$ and
$d_i\in \locsOR \implies d_i\in\dom(\Oh_j * \Rh_j)$ for all $d_i\in\ol{d}$.
Now we go by cases on $\loctype\,\ell$.
\begin{itemize}
\item $\loctype\,\ell \nleq Rep \land \loctype\,\ell \nleq Own$: Because
$\loctype\,\ell \leq D$ we have $D\nleq Rep \land D\nleq Own$. Hence by
condition (a) of the analysis, $\ol{T}\ncomp Rep$. Thus
$rng\,\eta_1\intersect locs(Rep\subclasses)=\Empty$ proving
$\etaconf{(\loctype\,\ell)}{\eta_1}{h}$ by
Def.~\ref{def:envconf}(\ref{envconfa}).
\item $\loctype\,\ell \leq Own$: Hence $\ell\in\dom(\Oh_j)$.
Now $rng\,\eta_1\intersect locs(Rep\subclasses) \subseteq \dom(\Rh_j)$
as required for $\etaconf{(\loctype\,\ell)}{\eta_1}{h}$ by
Def.~\ref{def:envconf}(\ref{envconfb}).
\item $\loctype\,\ell \leq Rep$: Hence $\ell\in\dom(\Rh_j)$. Now
$rng\,\eta_1\intersect \locsOR \subseteq \dom(\Oh_j * \Rh_j)$ as required for
$\etaconf{(\loctype\,\ell)}{\eta_1}{h}$, by
Def.~\ref{def:envconf}(\ref{envconfc}). \qed
\end{itemize}
\end{itemize}
\end{proof}

\begin{lemma}[soundness for expressions]\label{lem:esound}
If $\Gamma \aproves e:T$ then $\Gamma \proves e:T$ is confined.
\end{lemma}
\begin{proof}
Let $C=(\Gamma\,\self)$. Now we go by induction on $\Gamma \aproves
e:T$.  Assume $\etaconf{C}{\eta}{h}$
and $ d=\means{\Gamma \proves e:T}(h,\eta) \neq \bot$ for each case of
$e$.  

\medskip
\Case\ $\Gamma \aproves \faccess{e}{f}:T$. Then $d = h\ell f$.
We consider cases on $C$.
\begin{itemize}
\item $C\nleq Rep \land C\nleq Own$: We must show
$d\not\in locs(Rep\subclasses)$.
Because $\loctype\,\ell \leq C$, we have $\ell\not\in \locsOR$. So
$\ell$ is in the client part of a confining partition and by
Def.~\ref{def:hconf}(\ref{hconfa}) we have $d\not\in
locs(Rep\subclasses)$.

\item $C\leq Own$: 
Consider a confining partition and $j$ such that $\eta\,\self\in \dom(\Oh_j)$.
We must show $d\in locs(Rep\subclasses) \implies d\in\dom(\Rh_j)$.
Assume $d\in locs(Rep\subclasses)$. 
If $C=Own$, 
we have two subcases: If $e=\self$ we get $\ell=\eta\,\self$, so $i=j$ and
$d\in\dom(\Rh_j)$; if $e\neq\self$ then by the analysis we get $T\ncomp Rep$ so
$d\not\in locs(Rep\subclasses)$, falsifying the antecedent. This concludes the
case $C=Own$. If $C<Own$ then by the analysis $T\ncomp Rep$ so
$d\not\in locs(Rep\subclasses)$, falsifying the antecedent.  

\item $C\leq Rep$: 
Consider a confining partition and $j$ such that $\eta\,\self\in \dom(\Rh_j)$.
We must show $d\in \locsOR\implies d\in\dom(\Oh_j * \Rh_j)$. Since
$\loctype\,\ell\leq C$, we 
have $\ell\in locs(Rep\subclasses)$ and by induction on $e$ we get
$\ell\in\dom(\Rh_j)$ via Def.~\ref{def:econf}(\ref{econfc}). Because
$h$ is confined we get $d\in\dom(\Oh_j * \Rh_j)$ by
Def.~\ref{def:hconf}(\ref{hconfd}). 

\end{itemize}
The remaining cases are similar; see Appendix.
\qed\end{proof}

\begin{lemma}[soundness for constructors]\label{lem:xsound}
Suppose that $\self:C\aproves \constr\:C$ for all $C$ and let $\mu$ be arbitrary.
Then the constructor semantics is confined in the following sense:
For all $(h,\eta)$ with $\etaconf{C}{\eta}{h}$ we have $\conf{h_0}$ and $h\hext h_0$
where $h_0= \means{\self:C\aproves \constr\:C:\CON}\mu(h,\eta) \neq \bot$.
\end{lemma}
\begin{proof}
By well founded induction on $C$ using the order $\well$ in an argument
similar to that for Lemma~\ref{lem:cxpres}.  See Appendix.
\qed\end{proof}

\begin{lemma}[soundness for commands]\label{lem:csound}
If $\Gamma \aproves S$ then $\Gamma \proves S$ is confined.
\end{lemma}

\begin{proof}
Let $C=(\Gamma\,\self)$. Now we go by induction on $\Gamma \aproves S$ and
by cases on C. Assume $\etaconf{C}{\eta}{h}$ and $\conf{\mu}$
and $\means{\Gamma \proves S}\mu(h,\eta) \neq \bot$. Let
$(h_0,\eta_0)=\means{\Gamma \proves S}\mu(h,\eta)$. In each case we must
show $h_0$ is confined and $\etaconf{C}{\eta_0}{h_0}$.

\medskip
\Case\ $\Gamma \aproves \fassign{e_1}{f}{e_2}$.
Here $\eta_0=\eta$ and $h_0= \ext{h}{\ell}{\ext{h\ell}{f}{d}}$.
Because $\Gamma \aproves e_1:C$ and $\Gamma \aproves e_2:U$, by
Lemma~\ref{lem:esound}, $e_1$ and $e_2$ are confined at C. We must
first show that $h_0$ is confined and then show $\etaconf{C}{\eta_0}{h_0}$. By
$\etaconf{C}{\eta}{h}$ we know there is a confining partition
$h = \Ch * \ldots$.
We partition $h_0$ using the given partition for $h$.
That is, the domain for each block, say $\Ch^0$, is the same as the
corresponding block for $h$, say $\Ch$. 
We claim this partition is confining for $h_0$.  
It then follows by  Def.~\ref{def:hext} that $h\hext h_0$.
Then by Lemma~\ref{lem:b}, we get $\etaconf{C}{\eta}{h_0}$, hence
$\etaconf{C}{\eta_0}{h_0}$. It remains to show
the claim for which we need to show the conditions in Def.~\ref{def:hconf}.
We go by cases on C.  

\begin{itemize}
\item $C\nleq Rep \land C\nleq Own$: Only condition~(\ref{hconfa}) in
Def.~\ref{def:hconf} can possibly be violated.
By $\etaconf{C}{\eta}{h}$ we obtain
$rng\,\eta\intersect locs(Rep\subclasses) = \Empty$.
Because $\loctype\,\ell\leq C$ we have $\ell\in\dom(\Ch^0)$. By confinement of
$e_2$, $d\not\in locs(Rep\subclasses)$. Hence $\Ch^0\npoints \Rh^0_j$
for all $j$.
\item $C\leq Own$: Let $\eta\,\self\in\dom(\Oh^0_i)$ for some $i$.
Only conditions~(\ref{hconfb}) and~(\ref{hconfc}) in
Def.~\ref{def:hconf} can possibly be violated. Because
$\loctype\,\ell\leq C$, $\ell\in\dom(\Oh^0_j)$ for some $j$.
Because $e_2$ is confined at $C$ we have, $d\in locs(Rep\subclasses)\implies
d\in\dom(\Rh^0_i)$. We consider the case $C = Own$ and $e=\self$. Then
$\ell=\eta\,\self$ and $i=j$, establishing
condition~(\ref{hconfb}). By typing, $f\in\ol{g}$. Hence 
$\Oh_i\npointsx{\ol{g}}\Rh^0_i$ establishing condition~(\ref{hconfc}). In the
case $e\neq\self$, by the analysis we have $U\ncomp Rep$ thus establishing
conditions~(\ref{hconfb}) and~(\ref{hconfc}).

Now we consider the case $C<Own$. By the analysis we have $U\ncomp Rep$ thus
establishing conditions~(\ref{hconfb}) and~(\ref{hconfc}).
\item $C\leq Rep$: Let $\eta\,\self\in\dom(\Rh^0_i)$ for some $i$.
Only condition~(\ref{hconfd}) in Def.~\ref{def:hconf} can
possibly be violated. Because $\loctype\,\ell\leq C$, $\ell\in locs(Rep\subclasses)$.
By confinement of $e_1$ at $C$, we have $\ell\in \dom(\Rh^0_i)$. And, by confinement
of $e_2$ at $C$, we have $d\in\locsOR \implies d\in\dom(\Oh^0_i * \Rh^0_i)$. This
establishes condition~(\ref{hconfd}).
\end{itemize}

\Case\ $\Gamma \aproves \assign{x}{\mcall{e}{m}{\ol{e}}}$.
Here $h_0=h_1$ and $\eta_0=\ext{\eta}{x}{d_1}$.
Because $\Gamma \aproves e:D$ and $\Gamma \aproves \ol{e}:\ol{U}$, by
Lemma~\ref{lem:esound}, $e$ and $\ol{e}$ are confined at $C$.
By the analysis and Lemma~\ref{lem:aconf} we have
$\etaconf{(\loctype\,\ell)}{\eta_1}{h}$. Then by assumption $\conf{\mu}$
we get $\etaconf{(\loctype\,\ell)}{\eta_1}{h_0}$. Hence $h_0$ is confined.
To show $\etaconf{C}{\eta_0}{h_0}$, we go by cases on $C$:
\begin{itemize}
\item $C\nleq Rep \land C\nleq Own$: By $\conf{\mu}$,
$d_1\not\in locs(Rep\subclasses)$.
Hence $rng\,\eta_0\intersect locs(Rep\subclasses) = \Empty$ by
$\etaconf{C}{\eta}{h}$.
\item $C\leq Own$: Let $\eta\,\self \in\dom(\Oh_j)$ for some $j$ in
the confining partition of $h$. By $\conf{\mu}$,
$d_1\not\in locs(Rep\subclasses)$ and $h\hext h_0$. Because $x\neq\self$,
we have $\eta_0\,\self = \eta\,\self$.
Hence $rng\,\eta_0\intersect locs(Rep\subclasses)
= rng\,\eta\intersect locs(Rep\subclasses)
\subseteq \dom(\Rh_j)$ by $\etaconf{C}{\eta}{h}$. As $h\hext h_0$,
$\dom(\Rh_j)\subseteq\dom(\Rh_{0_j})$. That is,
$rng\,\eta_0\intersect locs(Rep\subclasses) \subseteq \dom(\Rh_{0_j})$.
\item $C\leq Rep$:  Because $\loctype\,\ell\leq C$, let $\ell\in\dom(Rh_j)$
for some $j$ in the confining partition of $h$.
By $\conf{\mu}$, $d_1\in \locsOR \implies d_1\in \dom(\Oh_{0_j} * \Rh_{0_j})$
and $h\hext h_0$.
Hence $rng\,\eta_0\intersect \locsOR \subseteq  \dom(\Oh_{0_j} * \Rh_{0_j})$ by
$\etaconf{C}{\eta}{h}$ and Def.~\ref{def:hext}.
\end{itemize}

\Case\ $\Gamma \aproves \assign{x}{\new{B}{}}$.
First, we claim $\etaconf{B}{\eta_1}{h_1}$ and $h\hext h_1$.
Then by Lemma~\ref{lem:xsound} we get $\etaconf{B}{\eta_1}{h_0}$ and
$h_1\hext h_0$. So $h\hext h_0$ and by Lemma~\ref{lem:b}
$\etaconf{C}{\eta}{h_0}$.
To conclude, we argue that $\etaconf{C}{\ext{\eta}{x}{\ell}}{h_0}$ by 
cases on $C$.
\begin{itemize}
\item $C\nleq Own\land C\nleq Rep$: then $B\nleq Rep$ so $\ell\not\in
  locs(Rep\subclasses)$ by typing and hence
  $\etaconf{C}{\ext{\eta}{x}{\ell}}{h_0}$.
\item $C\leq Own$: 
Let $h = \Ch * (\Oh_1 * \Rh_1) \ldots (\Oh_k * \Rh_k)$ be a confining
partition of $h$, and $j$ such that $\eta\,\self$ be in $\dom(\Oh_j)$.
If $B\nleq Rep$ then $\etaconf{C}{\ext{\eta}{x}{\ell}}{h_0}$ by definition.  
If $B\leq Rep$ then we must show $\ell\in \dom(\Rh^0_j)$ where $h_0$ has  confining
extension $h_0= \Ch^0 * (\Oh^0_1 * \Rh^0_1) \ldots$.  
This is defined just as in the proof of Lemma~\ref{lem:c}, and we choose to
put $\ell$ and the objects it constructs in $\Rh_j$ to obtain $\Rh^0_j$.
\item $C\leq Rep$:
By the static analysis, $B\nleq Own$.  So $\Oh^0_j=\Oh_j$.
Thus
$rng\,\ext{\eta}{x}{\ell}\intersect \locsOR \subseteq
\dom(\Oh_j*\Rh^0_j)$.  
We choose to put $\ell$ and the objects it constructs in $\Rh_j$ to obtain
$\Rh^0_j$, which makes the inclusion hold.
\end{itemize}
It remains to prove the claims $\etaconf{B}{\eta_1}{h_1}$ and $h\hext h_1$. 
In the semantic definition, 
$h_1 =\ext{h}{\ell}{[\fields\,B\mapsto\mathit{defaults}]}$ where 
$\ell = \fresh(B,h)$.  
Define $Bh = [\ell\mapsto [\fields\,B\mapsto\mathit{defaults}]]$ so 
$h_1 = h* Bh$.  
Let $\eta_1 = [\self\mapsto \ell]$.  
Next, we argue that $h\hext h_1$ and $\etaconf{B}{\eta_1}{h_1}$.
Because $h$ is closed, $\ell$ is not in the range of any object state in
$h$. To construct an extending partition it suffices to deal with the
new object, as its addition cannot violate confinement of existing
objects. We define the extension and argue by cases on $B$.
\begin{itemize}
\item $B\nleq Own\land B\nleq Rep$.
For a confining partition of $h_1$ we extend that for $h$ by 
defining $\Ch^0 = \Ch * Bh$ and using the given partition of owner islands.
Because $\mathit{defaults}$ contains no locations, this is a confining
partition and we have $\etaconf{B}{\eta_1}{h_1}$. 

\item $B\leq Own$.  We extend the partition by adding an island 
$\Oh^0_{k+1}*\Rh^0_{k+1}$ with 
$\Oh^0_{k+1} =  Bh$ and $\Rh^0_{k+1}=\Empty$.  This is a confining partition 
because $\mathit{defaults}$ has no locations and we have 
$\etaconf{B}{\eta_1}{h_1}$ because $rng\,\eta_1$ has no reps.   

\item $B\leq Rep$.  Then, by the analysis we have $C\leq Own$ or $C\leq Rep$;
moreover as $x\neq\self$, we have $\eta\;\self\neq\ell$, so
$\eta\;\self\in\dom(\Oh_j * \Rh_j)$ for some $j$. 
Then we can obtain a confining extension by adding 
$Bh$ to $\Rh_j$, as $\mathit{defaults}$ has no locations.
As $rng\,\eta_1= \{\ell\}$, we have $\etaconf{B}{\eta_1}{h_1}$ 
by definition.
\end{itemize}
This concludes the argument for $h\hext h_1$ and
$\etaconf{B}{\eta_1}{h_1}$.

\medskip
The remaining cases are similar and can be found in the appendix.
\qed
\end{proof}


\section{Discussion and related work}\label{sect:disc}

Programmers draw pictures of pointers in heap-based data structures  and often
manage to get things right as far as the presence of pointers goes.  For
example, lists don't get disconnected.  The absence of pointers is harder
to picture and many bugs are due to unexpected aliasing.  Expectations
are raised through use of encapsulation constructs such as private
fields and modules,  but heap structure is not entirely manifested in language 
constructs.  
Simulation relations are often used for reasoning about abstractions 
and here too aliasing presents a challenge: Multiple instances of an
abstraction may reference a shared client object or be shared by  
multiple clients ---but client references to representation objects
can violate encapsulation.
Various notions of ownership confinement have been proposed
for encapsulation of objects.  We have formalized one and shown that
clients are independent from confined representations.  Independence
is formalized by an abstraction theorem that licenses reasoning about
equivalence of class implementations using simulation relations.  Confinement
is formalized by drawing boundaries that signify the absence of
pointers.

\subsection{Related work}
\paragraph*{Representation independence}

The main proof technique for representation independence is so fundamental that it has
appeared in many places, with a variety of names, e.g., simulation, logical relations,
abstraction mappings, relational parametricity (e.g.,
\cite{Plotkin73,Reynolds84,Lynch-Vaandrager,deRdataref}).     
Among the many uses of simulations are program transformations and
justification of logics for reasoning about data abstraction and
modification of encapsulated state.

Representation independence results are known for general transition
systems~\cite{Milner71,Lynch-Vaandrager}, first 
order imperative languages~\cite{refine:refine,deRdataref}, higher order 
functional~\cite{Reynolds84,Mitchell86,Mitchell91,Mitchell96,PowerRobinLR}
and higher order imperative languages~\cite{OHearn:Tennant,sdr},
and sequential object-oriented programs without heap allocation
(\cite{CNfm02} treats a language with class-based visibility and  
\cite{Reddy-classes} treats one with instance-based visibility).
As far as we know, our results are the first for shared references to
mutable state, a ubiquitous feature in object-oriented and imperative programs.
(The lacuna is mentioned in~\cite{Grossman00}.) 

A widely held view seems to be that classical techniques based on denotational
semantics and logical relations are inadequate in the face of the
complex language features of interest.  The combination of 
local state with higher order procedures makes it difficult to prove representation
independence even for Algol, where procedures can be passed as arguments but not
assigned to state variables~\cite{OHearn:Tennant}. Objects exhibit similar features.  

Difficulties with denotational semantics led to considerable advances using small-step
operational semantics~\cite{NewtonOper}.
However,  to get an adequate induction hypothesis for an
abstraction theorem, parametricity needs to be imposed on the latent
effects of procedure abstractions, either as a property to be proved or
as an intrinsic feature of the semantic model
\cite{ReynoldsEssence,OHearn:Tennant}.
These conditions are most easily expressed in terms of a denotational model,
but if procedures can be stored in the heap on which they act, 
difficult domain equations must be
solved.\footnote{Recently~\citeN{LevyCSL} used functor categories  to
  give a denotational model for a higher order language with
  pointers, but the model does not capture relational parametricity and the 
  language has neither object-oriented features nor recursive types.
}
Recursive data types also lead to nontrivial domain equations. 
Even if solutions can be found, they may be quite complex structures
that are difficult to understand and work with.

One of the most relevant works using operational semantics is that
of~\citeN{Grossman00} where 
representation independence is approached using a dynamic notion of ownership
by \dt{principals} as in the security literature.  To prove that
clients are independent from the representation of an 
abstraction provided by a host program, a wrapper construct is 
used to tag code fragments with their owner (e.g., client or ``host''), and to
provide an opaque type for the client's view of the abstraction.   
This is  a promising approach, but the results so far only show
``independence of evaluation'', which is analogous to the special 
case of simulation used for non-interference in analysis of
information flow~\cite{VolpanoSI96,Abadi99}. 
Although \citeN{Grossman00} offer their work as a simpler alternative to domain
theoretic semantics, the technical treatment is somewhat intricate by the time the
language is extended to include references, recursive and polymorphic types. 

Except for parametric polymorphism, we  treat all these features, as
well as others such as subclassing, dynamic binding,  type tests and
casts.  Although Java syntax seems less elegant than, say, lambda
calculus, it has several features that ease the difficulties.
Owing to name-based type equivalence and subtyping, and the binding of methods to objects
via their class, we can use a denotational model with quite simple domains and fixpoint
definitions in the manner of~\citeN{Strachey} (cf.\ Sect.~\ref{sect:mse}).

For applications in security and automated static checking, it is
important to devise robust, comprehensible models that support not
only the idealized languages of research studies but also the full
languages used in practice.  Denotational semantics has conceptual
advantages, at least if the domains are simple enough to have a clear
operational significance.  
However, we admit that our enthusiasm for the efficacy of denotational
techniques has been tempered by the irritation of flushing out bugs in
intricate definitions and induction hypotheses.  

Our abstraction theorem and identity extension lemma can be used
directly to
prove equivalence of programs, where a program is a command in the
context of a class table and designated class $C$.
It would be reasonable to use a notion of 
equivalence based on field visibility: states would be equated if they
are equal after hiding all fields except those visible in $C$.   
But this would beg the question whether hiding imposes encapsulation
that is not intrinsic to the language.   
In this paper we use the finer equivalence on programs: for commands
to be equivalent they must yield outcomes that are identical after
garbage collection. Thus encapsulation is formulated in terms of 
private fields and confined reps but the identity extension lemma is
expressed, in effect, in terms of local variable blocks 
(in the style of, e.g., \citeN{refine:refine}).

Besides the ``client interface'' provided by public methods
and analogous to the interfaces studied in previous work on
representation independence, 
a class also has a ``protected'' interface to its subclasses.  The
combination of protected and public interfaces is complicated, but a
thorough treatment of representation independence for object-oriented
programs must take it into account.   For reasoning about the
protected interface, work on behavioral subclassing has used
simulations to connect a class with its subclass~\cite{LiskovWing,Leavens-Dhara00} but
a formal connection has not been made with the use of simulations to
connect alternative representations.

\paragraph*{Confinement}

Quite a few confinement disciplines have been proposed, by 
\citeN{Hogg}, \citeN{Almeida},\citeN{Vitek00}, \citeN{NoblePotter},
\citeN{FTfJPmuller}, \citeN{BoylandBury}, \citeN{JavaConcur},
\citeN{Aldrich02}, and \citeN{ClarkeDiss} 
(the latter has a more comprehensive recent survey).
Most proposals have significant shortcomings; they
disallow important design patterns  or are not efficiently checkable. 
Although the aim is to achieve encapsulation and thereby support modular
reasoning in one form or another, few proposals have been formally justified in these terms
---none in terms of representation independence.    

Several works justify a syntactic discipline by proving that it
ensures a confinement invariant
\cite{FTfJPmuller,ClarkeDiss,Aldrich02}.
Others go further and show some form of modular reasoning principle,
as we discuss in detail below.
Existing justifications involve disparate techniques and objectives,
so that it is quite hard to assess and compare confinement
disciplines.  One of our contributions is to show how standard
semantic techniques can be used for such assessments.

The fact that type names are semantically relevant lets us use them to formulate in 
semantic terms a condition similar to the ownership confinement
notions of \citeN{Mueller01}, \citeN{NoblePotter} and their
predecessors \cite{Hogg,Almeida}.  
Whereas several papers emphasize reachability via paths, our
formulation of confinement emphasizes partitioning of heap objects and
the one-step points-to relation.  In this we were inspired by the work of
\citeN{ReynoldsPtrs} that shows the efficacy of reasoning about partition
blocks that may have dangling pointers.  

Reasoning on the assumption of confinement is a separate concern 
from enforcement or checking of confinement.
Semantic considerations led us to a flexible, syntax-directed static
analysis, but other analysis techniques such as
model checking or theorem proving for (an approximation of) the
semantic confinement property could be interesting.

It is interesting to note that we get a strong reasoning principle on
the basis of ownership confinement alone, in a form that can be
checked without program annotations.  By contrast, other works use
annotations and combine ownership with uniqueness and effects (e.g.,
read-only) \cite{Clarke02,Aldrich02,Mueller01}. 

Confinement figures heavily in the verification logics of 
\citeN{MuellerPoetzsch-Heffter00} and in some work by
the group of Nelson and Leino~\cite{SRC160toplas,SRC156}
where it is needed for sound reasoning about the ``modifies clause''
framing the scope of effects. 
Subsequent to the present work, \citeN{Clarke02} state results on
reasoning about effects, using a confinement discipline imposed using
code annotations for confinement and effects.   
These works are concerned with delimiting the scope of effects, 
which is an important aspect of modular reasoning, but they do not
address representation independence.  

There has been much work on capturing encapsulation via visibility
(lexical scope), using existential types and subsumption 
(see~\cite{BruceCP99,KimBruceBook,PierceBook} and references
therein). 
None of these works addresses the problem of confinement;  they are
concerned with the complex typing issues for object oriented
languages.  

It is interesting to note that one of the main difficulties
in designing safe and flexible type systems is due to the desire to
eliminate or minimize the use of type 
testing and casting which are seen as loopholes that subvert
type-based encapsulation.
Indeed, parametric polymorphism has been much pursued as a
means to cope with generic patterns that, in current practice, are
usually coded using subsumption, casts, and type \OBJECT\ (a recent
reference is the textbook by~\citeN{KimBruceBook}). 
Although parametric polymorphism has obvious merit,
our results show that casts and type tests are themselves relationally
parametric.  It is behavioral subclassing which is at risk in some
uses of casts and tests.  This does not contradict \cite{Reynolds84}
because our language has a nominal type system \cite{PierceBook}; it
is the name of a type, not its set of values, that is involved with
tests and casts.

Our aim is to deal with the rich languages currently in use, rather
than to advance language design.  It is challenging to formalize the
syntax precisely yet perspicuously.  
Rather than devising our own idiosyncratic formalization, we adapted
that of  \citeN{Featherweight}.  The details differ, as our language 
includes imperative constructs and non-public scoping and their
main concern is type soundness.

\subsection{Future challenges}

The language for which our results are given encompasses many
important features of object oriented languages.  
Two major features
are missing and will require substantial additional work: concurrency
and parametric polymorphism.
The interaction between parametric and subtyping
polymorphism is non-trivial and there are a number of competing type
systems.
Some languages, e.g., C++, have parametric polymorphism but with
significant limitations; for Java, parametric types are a late addition.
We expect to extend our work to them  in the future.

Ownership confinement is appropriate for reasoning about many designs
in practice and we have shown through a series of examples that our 
notion is applicable to widely used designs such as the observer and
factory patterns.   Two important issues are beyond the reach of our
work (and much of the previous work on confinement).
The first is multiple ownership.  A canonical example is a
collection class with iterators.  The reps for the collection are
nodes of a data structure.  The collection object mediates additions
and deletions. To allow enumeration of elements of the collection it
is common to use iterator objects which need access to the nodes of
the data structure.  Static analyses have been given that allow
some form of multiple owners \cite{ClarkeDiss,Mueller01,Aldrich02}. Although our
formalization of islands can be extended easily to encompass multiple
owners, it is not as clear how to extend the notion of simulation in a
useful way.   Our result formalizes the notion that an owner instance
provides an abstraction and this is easily expressed in 
terms of the class construct.  The generalization can perhaps be
expressed by grouping the related owners (e.g., the collection class
and the iterator class) in a module,  but this is left for future work.

The other challenging issue for confinement is ownership transfer.
Consider a queue that owns objects representing tasks to be
performed.  For load balancing, tasks may be moved from one queue to
another.  In this case a task is owned by just one queue at a time and
in a given state the system is confined according to the definition in
this paper.  A sequential program for transferring ownership from one
queue might look as follows:
\textsf{q2.task\assym q1.task;  q1.task\assym \NULL.}   
From a confined initial state this need not lead to a confined final
state: there could be other references to \code{task}.  But it does
lead to a confined final state if \code{q2.task} is initially the only
existing reference to the task.  Unique references have been
extensively studied so let us assume that a static analysis is given for
uniqueness.  Even with uniqueness, our theory fails to apply, for two
reasons.  The first reason is a small one: in the intermediate state two different
owners reference the same task.  This problem is well
known and can be surmounted:  It is easy to add to our language an
atomic command with the effect of the above sequence
\cite{minsky96towards} and to show,
given uniqueness, that it is confined.  For practical purposes one
would use a static analysis to check that \code{q1.task} is a dead
expression \cite{BoylandBury}.  

The second reason our theory does not apply is a technical one.  To
show that a method call is confined, we need that the caller's
environment is confined in the final heap assuming it was confined in
the initial one.  We get this by using a condition stronger than
confinement: from a confined state, a command or method yields a final
heap that extends the initial one in the sense of Def.~\ref{def:hext}.  
All commands of our language yield heaps extended in this sense so all method
meanings have this property. (See the proof of Theorem~\ref{thm:cfix}.)   
But, by definition of extension, $h\hext h_0$ says that reps
that exist in $h$ have the same owners in $h_0$ as in $h$, disallowing
ownership transfer. 


For static analysis there are some more modest issues worthy of
investigation.  The simple conditions of Def.~\ref{def:tconf}
ensure suitable confinement of the class table but they are
unnecessarily strong.  Methods inherited into rep classes are not 
risky if they do not leak $\self$; such ``anonymous methods'' can be statically 
checked as shown by \citeN{Vitek00}  and \citeN{Grothoff} in work on
module-based confinement.\footnote{In fact the cited work is concerned with
  pragmatic aspects of the analysis and does not formalize a semantic
  property ensured by the analysis.}
The conditions of our static analysis may also admit useful variations.   

Having shown that simulation is sound one might proceed to study
completeness.  It is  not the case that our confinement conditions are
necessary in general for simulations to be preserved.
A trivial simulation might depend on no confinement at all.  
Also, a rep could be leaked but not exploited by any client.
One can see confinement as a kind of simulation which happens to be a
rectangular predicate: $h$ relates to $h'$ just if $h$ and $h'$ are
confined, independent of each other. This suggests folding the
confinement condition into the simulation relation, an idea which is
currently under study by Uday Reddy and Hongseok Yang for a
Pascal-like language.\footnote{Their aim is to explicate the semantic
  structure of languages involving heap storage
Their approach should lead to a lucid account on par with
parametricity models for other languages~\cite{Reynolds84,ReynoldsEssence,Reddy-classes}.  
They have defined a parametricity semantics for a Pascal-like
language~\cite{YangReddy-heap} in which heap cells are tuples of
pointers and  integers rather than objects with scoped fields. 
Several challenges remain to be addresssed, if this approach is to
provide a foundation for reasoning about instance-based abstractions
in  Java-like languages using a practical confinement discipline.
For example,  nominal types and class-based visibility (which is not
modelled by naive use of existential types).}
For practical reasoning the benefits of treating
confinement separately are clear: it accords with informal design
practice, is amenable to static checking, and ensures soundness for a
straightforward and modular notion of coupling. 

The more practical question is how to express basic couplings and
prove the simulation property for owner methods.
To formalize the couplings for the observer examples one needs a
formalism for inductive predicates on recursive data structures;
separation logic appears promising for this purpose \cite{ReynoldsPtrsLics}.

As we discussed in conjunction with Example~\ref{ex:ver},
representation independence licenses reasoning about
equivalence of programs that are structurally similar
\cite{BanerjeeHR01,Riecke93}.
This is quite adequate for uses of 
simulations such as static analyses and relating alternative interpretations for
primitives, such as the lazy and eager access control implementations
for Java~\cite{popl02}.   
But for abstraction in program development, typically called data refinement, it is
not uncommon to consider significantly different program structures and this calls
for a full program logic in which something like the abstraction theorem appears as a 
proof rule.   
For first-order imperative languages, several proof systems have been
given for reasoning about two versions of an
abstraction \cite{deRdataref}.  Typically, relations (especially
``abstraction functions'') are used to derive from one version the
specification of the other version, which is then proved correct in a
program logic.  Logics for imperative object-oriented languages are at
an early stage of development
\cite{AbadiLeino,FM99,Poetzsch-HeffterMueller99,HuismanJacobs,Huisman02,ReynoldsPtrsLics}.

\appendix
\section*{APPENDIX}

\section{Additional proofs}

\subsection*{Proof of Lemma~\ref{lem:a}}

By cases on $C$ and $B$.  It suffices to
consider $C<B$ and to deal with confinement of $\eta$ in $h$.  
\begin{itemize}
\item $C\leq Rep$.  Then the hypothesis of the Lemma is falsified because 
$\eta\,\self\in locs(Rep\subclasses)$.
\item $C\nleq Own \land C\nleq Rep$.  Then $B\nleq Own \land B\nleq Rep$, 
so $\conf{C}{\eta}{h} \iff \etaconf{B}{\eta}{h}$ because both $C$ and $B$
are subject to condition (\ref{envconfa}) in Definition~\ref{def:envconf}.
\item $C< B\leq Own$.  Again, both $B$ and $C$ are subject to the same condition,
here (\ref{envconfb}) in  Definition~\ref{def:envconf}. 
\item $C\leq Own<B$.  We have $\etaconf{B}{\eta}{h}\implies
  \etaconf{C}{\eta}{h}$ by implication between the consequents of
  (\ref{envconfa}) and (\ref{envconfb}) in Definition~\ref{def:envconf}.  The converse
  holds owing to hypothesis $rng\,\eta\intersect locs(Rep\subclasses) = \Empty$.
\end{itemize}

\subsection*{Proof of Lemma~\ref{lem:b}}

By cases on $C$.
In the case $C\nleq Own\land C\nleq Rep$, we have 
$\etaconf{C}{\eta}{h}\iff\etaconf{C}{\eta}{h_0}$ because
Definition~\ref{def:envconf}(\ref{envconfa}) of $\conf{C}$ is 
independent of the heap. 
For the cases $C\leq Own$ and $C\leq Rep$, we 
show $\etaconf{C}{\eta}{h_0}$ using  $h\hext h_0$.  
First, by definition of $\hext$ we have $\conf{h_0}$.  To show that 
$\eta$ is confined in $h_0$ for $C$, suppose 
\[ h= \Ch * \Oh_1 * \Rh_1 * \ldots * \Oh_k * \Rh_k \] is a
confining partition of $h$.  Let $j$ be such that
$rng\,\eta\intersect\locsOR \subseteq \dom(\Oh_j*\Rh_j)$.
Suppose, by $h\hext h_0$, that this partition is extended by confining
partition  $h_0 = \Ch^0 * \Oh^0_1 * \Rh^0_1 * \ldots $.
In the case $C\leq Own$, we have 
$rng\,\eta\intersect locs(Rep\subclasses)\subseteq
\dom(\Rh_j)\subseteq \dom(\Rh^0_j)$, using $\etaconf{C}{\eta}{h}$ and the
definition $\hext$. The case $C\leq Rep$ is similar.

\subsection*{Additional cases for Lemma~\ref{lem:c}}

\Case\  $\Gamma\proves\assign{x}{e} $.
Here the heap is unchanged: $h_0 = h$ and the result holds by reflexivity
of $\hext$.

\medskip
\Case\  $\Gamma \proves \assign{x}{\mcall{\SUPER}{m}{\ol{e}}}$. 
The same argument as for method call $e.m$.

\medskip\Case\ $\Gamma \proves \seq{\S_1}{\S_2}$. 
Let $(\eta_1, h_1) = \means{\Gamma \proves \S_1 }\mu(h,\eta)$.
By induction on $S_1$ we have $h\hext h_1$.  
By confinement of $S_1$ we have $\etaconf{C}{\eta_1}{h_1}$.
So we can use induction on $S_2$ to obtain $h_1\hext h_0$ and then $h\hext
h_0$ by transitivity of $\hext$.

\medskip\Case\ $\Gamma\proves \ifelse{e}{\S_1}{\S_2}$.
By induction on $\S_1$ and $\S_2$, using confinement of $\S_1$ and $\S_2$.

\medskip\Case\ $\Gamma \proves\var{\T\; x}{e}{S}$.
Let $\eta_1 = \ext{\eta}{x}{\means{\Gamma\proves e:U}(h,\eta)}$. 
By $\etaconf{C}{\eta}{h}$ and confinement of $e$ we have $\etaconf{C}{\eta_1}{h}$.
Then by induction on $\S$, using confinement of $\S$, we get $h\hext
h_0$. 

\subsection*{Proof of Lemma~\ref{lem:eqdep}}

By induction on depth.
If $C\nleq Own$ then the equality is direct from
Definition~\ref{def:comp}(\ref{simb}).   
If $C\leq Own$ then it is possible that $CT(Own)$ declares $m$ but
$CT'(Own)$ does not (or vice versa).  But in that case, by 
Definition~\ref{def:comp}(\ref{simd}) we have $\mtype(m,C)=\mtype'(m,C)$ so
$m$ must be declared in a superclass, whence   
$depth(m,C)=1+depth(m,\super\, C)=1+depth'(m,\super\, C)=depth'(m,C)$.  

\subsection*{Proof of Lemma~\ref{lem:inh}}

Let $\Gamma_B = (\ol{x}:\ol{T},\self:B)$ and $\Gamma_C =
(\ol{x}:\ol{T},\self:C)$.
To show 
\begin{trivlist}
\item 
\hfill $\rel{(B,\ol{x},\ol{T}\to T)}{(\restr(d,B))}{(\restr(d',B))} $
\hfill $(*)$  
\end{trivlist}
consider $(h,\eta)\in\means{\heap\otimes\Gamma_B}$ and $(h',\eta')\in
\means{\heap\otimes\Gamma_B}'$ 
such that $\etaconf{B}{\eta}{h}$, $\etaconf{B}{\eta'}{h'}$, and  
$\rel{(\heap\otimes\Gamma_B)}{(h,\eta)}{(h',\eta')}$.
By definition of $restr$ we have 
$restr(d,B)(h,\eta) = d(h,\eta)$ and 
$restr(d',B)(h',\eta') = d'(h',\eta')$.
So for $(*)$ it remains to show
\begin{trivlist}
\item 
\hfill $\rel{(\heap\otimes\T)_\bot}{(d
    (h,\eta))}{(d'(h',\eta'))} $ 
\hfill $(\dagger)$ 
\end{trivlist}
By Lemma~\ref{lem:semty}(\ref{stya}) we have
$(h,\eta)\in\means{\heap\otimes\Gamma_C}$ and $(h',\eta')\in
\means{\heap\otimes\Gamma_C}'$.
By hypothesis, $C$ is non-rep so $B$ is also non-rep.  
As $\ol{T}$ is non-rep, we have 
$rng\,\eta\intersect locs(Rep\subclasses) = \Empty$ and 
$rng\,\eta'\intersect locs(Rep'\subclasses) = \Empty$. 
Thus Lemma~\ref{lem:a} is applicable to $\eta,\eta'$ and 
using hypothesis $B<C$ we obtain $\etaconf{C}{\eta}{h}$, and
$\etaconf{C}{\eta'}{h'}$.  
Thus we have established the antecedents needed to use hypothesis
$\rel{(C,\ol{x},\ol{T}\to T)}{d}{d'}$  
to obtain $(\dagger)$.

\subsection*{Proof of Lemma~\ref{lem:epres}}

\Case\ $\Gamma \proves x:T$. 
Then $\rel{T_\bot}{(\eta x)}{(\eta' x)}$ by $\rel{\Gamma}{\eta}{\eta'}$, so
the result follows by semantics of $x:T$.

\medskip\Case\ $\Gamma \proves \NULL:B$. 
Then semantics is $\NIL$ and $\rel{B_\bot}{\NIL}{\NIL}$ by definition of
$\REL$.

\medskip\Case\ $\Gamma \proves \ITT:\UNIT$.
Similar to $\NULL$, as are the cases $\TRUE$ and $\FALSE$.  

\medskip\Case\ $\Gamma \proves \eqtest{e_1}{e_2}:\BOOL$.
Then, using identifiers from the semantic definition as usual, we consider
cases on $d_1$.  If $d_1=\bot$ then $d'_1=\bot=d_1$ by induction on $e_1$
and definition of $\REL\,T$.  Hence, by semantics of $\eqtest{e_1}{e_2}$,
$\means{\Gamma \proves \eqtest{e_1}{e_2}:\BOOL}(h,\eta) =
\means{\Gamma \proves \eqtest{e_1}{e_2}:\BOOL}'(h',\eta')$ 
and thus 
\begin{trivlist}
\item 
\hfill $
\rel{\BOOL_\bot}{(\means{\Gamma \proves \eqtest{e_1}{e_2}:\BOOL}(h,\eta))}{
(\means{\Gamma \proves \eqtest{e_1}{e_2}:\BOOL}'(h',\eta'))}$ \hfill $(*)$ 
\end{trivlist}
The argument is symmetric for $d_2=\bot$.  

If none of $d_1,d'_1,d_2,d'_2$ are $\bot$ then, by induction on $e_1$
we have $\rel{(T_1)_\bot}{d_1}{d'_1}$.  Thus, by Lemma~\ref{fact:id}, $d_1=d'_1$.
Similarly, $d_2=d'_2$.  Hence $d_1=d_2$ iff $d'_1=d'_2$, whence
the result $(*)$ holds by semantics.

\subsection*{Additional cases for Lemma~\ref{lem:cpres}}

\Case\ $\Gamma  \proves\assign{x}{\mcall{\SUPER}{m}{\ol{e}}} $. 

By $\rel{\Gamma}{\eta}{\eta'}$ we have $\rel{C}{\ell}{\ell'}$, hence
$\ell=\ell'$ by Lemma~\ref{fact:id}.  
By $\etaconf{C}{\eta}{h}$ and $\etaconf{C}{\eta'}{h'}$ we have 
$\ell\not\in locs(Rep\subclasses)$ and $\ell\not\in locs(Rep'\subclasses)$.
Let $\eta_1=[\self\mapsto\ell,\ol{x}\mapsto\ol{d}]$
and $\eta'_1=[\self\mapsto\ell,\ol{x}\mapsto\ol{d}']$.  
By confinement of $\assign{x}{\mcall{\SUPER}{m}{\ol{e}}}$ (Definition~\ref{def:cconf}) we
have confined arguments, i.e., 
$\etaconf{(\super\,C)}{\eta_1}{h}$ and $\etaconf{(\super\,C)}{\eta'_1}{h'}$ 

By Lemma~\ref{lem:epres} for $\ol{e}$, and considering the non-$\bot$ case, we have
$\rel{\ol{U}}{\ol{d}}{\ol{d}'}$, whence, by Lemma~\ref{fact:id:c}, 
$\rel{\ol{T}}{\ol{d}}{\ol{d}'}$.  
From $\rel{C}{\ell}{\ell'}$ we get $\rel{(\super\,C)}{\ell}{\ell'}$
by Lemma~\ref{fact:id:c}, 
and thus $\rel{[\ol{x}:\ol{T},this:\super\,C]}{\eta_1}{\eta'_1}$.
From $\rel{\menv}{\mu}{\mu'}$ we get
\[ \rel{(\super\,C,
  \mtype(m,\super\,C))}{(\mu(\super\,C)m)}{(\mu'(\super\,C)m)}
\]
hence, as $h,h',\eta_1,\eta'_1$ are confined and related, 
$\rel{(\heap\otimes\T)}{(h_1,d_1)}{(h'_1,d'_1)}$ where 
$(h_1,d_1) =\mu(\super\,C)m(h,\eta)$ and  
$(h'_1,d'_1) =\mu'(\super\,C)m(h',\eta')$.
Thus $\rel{T}{d_1}{d'_1}$ and $\rel{\heap}{h_1}{h'_1}$.
It remains to show that the updated stores
$\ext{\eta}{x}{d_1}$ and $\ext{\eta'}{x}{d'_1}$ are related. This follows from 
$\rel{T}{d_1}{d'_1}$ and $T\leq \Gamma\,x$ using Lemma~\ref{fact:id:c}.

\medskip\Case\  $\Gamma  \proves \seq{\S_1}{\S_2} $. 

As usual, we consider the non-$\bot$ case.
By induction on $S_1$ we have
$\rel{(\heap\otimes\Gamma)_\bot}{(h_1,\eta_1)}{(h'_1,\eta'_1)}$. 
Moreover, as $S_1$ is a constituent of a method in $CT$ and $CT'$, 
by confinement of $S_1$ we have  $\etaconf{C}{\eta_1}{h_1}$ and $\etaconf{C}{\eta'_1}{h'_1}$,
so we can use induction on $S_2$ to obtain the result.

\medskip\Case\  $\Gamma \proves \ifelse{e}{\S_1}{\S_2} $.
Similar to case of sequence, but also using Lemma~\ref{lem:epres} for $e$.


\medskip\Case\  $\Gamma \proves\var{\T\; x}{e}{S} $.

By Lemma~\ref{lem:epres} for $e$ we have $\rel{U_\bot}{d}{d'}$.  If
$d=\bot$, then $d'=\bot$ and both semantics yield $\bot$.  Otherwise, we
have $\rel{\T}{d}{d'}$ by the corollary to 
Lemma~\ref{fact:id}.  Thus, from $\rel{\Gamma}{\eta}{\eta'}$ we obtain
$\rel{(\Gamma,x:\T)}{\eta_1}{\eta'_1}$ where 
$\eta_1 = \ext{\eta}{x}{d}$ and $\eta'_1=\ext{\eta'}{x}{d'}$ as in the
semantic definition.
In order to use induction on $S$, we need to show 
$\etaconf{C}{\eta_1}{h}$ and $\etaconf{C}{\eta'_1}{h'}$.
From condition (\ref{tconfa}) in Definition~\ref{def:tconf} of confinement
for $CT$, $e$ is confined.
In the case $C\nleq Own$, confinement of $e$ yields $d\not\in locs(Rep\subclasses)$.  
and thus $\etaconf{C}{\eta_1}{h}$.
In the case $C<Own$, confinement of $e$ yields $d\in
locs(Rep\subclasses)\implies d\in\dom(\Rh_j)$ for some partition and $j$
with $\eta\,\self\in\dom(\Oh_j)$.  This is the condition required  
for $\etaconf{C}{\eta_1}{h}$ in this case.
Similarly, we get $\etaconf{C}{\eta'_1}{h'}$.
Now, by induction on $S$ we get that both semantics are $\bot$ or else the
result states from $S$ satisfy 
$\rel{(\heap\otimes\Gamma,x:\T)_\bot}{(h_1,\eta_2)}{(h'_1,\eta'_2)}$.
In the latter case, 
$\rel{(\heap\otimes\Gamma)}{(h_1,(\eta_2\downharpoonright x))}
{(h'_1,(\eta'_2\downharpoonright x))}$ as required.

\subsection*{Proof of Lemma~\ref{lem:aconf}(\ref{aconfb})} 

Again the proof proceeds by cases on $C$. In
each case we show $\etaconf{(\super\,C)}{\eta_1}{h}$, noting that
$\ell=(\eta\,\self)$.
\begin{itemize}
\item $C\nleq Rep \land C\nleq Own$: By confinement of $\eta$ at $C$, we have
$\ell\not\in locs(Rep\subclasses)$. Because $\ol{e}$ is confined at $C$,
we have $d_i\not\in locs(Rep\subclasses)$ for all $d_i\in\ol{d}$. Thus
$rng\,\eta_1\intersect locs(Rep\subclasses)=\Empty$. And, since $C<\super\,C$,
we have $\etaconf{(\super\,C)}{\eta_1}{h}$ by Lemma~\ref{lem:a} and
Definition~\ref{def:envconf}(\ref{envconfa}).
%
\item $C\leq Own$: Choose a confining partition and let $j$ be such that
$\ell=(\eta\,\self) \in \dom(\Oh_j)$. Since $C<\super\,C$ we have $\super\,C \leq
Own$ ($Own < \super\,C$ is impossible by definition of $\super\,C$).
Because $\ol{e}$ is confined at $C$, we have
$d_i\in locs(Rep\subclasses) \implies d_i\in\dom(\Rh_j)$ for all $d_i\in\ol{d}$.
Thus $rng\,\eta_1\intersect locs(Rep\subclasses) \subseteq \dom(\Rh_j)$ proving
$\etaconf{(\super\,C)}{\eta_1}{h}$ by Definition~\ref{def:envconf}(\ref{envconfb}).
\item $C\leq Rep$: Choose a confining partition and let $j$ be such that
$\ell=(\eta\,\self) \in \dom(\Rh_j)$. Since $C<\super\,C$ we have $\super\,C \leq
Rep$ ($Rep < \super\,C$ is impossible by definition of $\super\,C$).
Because $\ol{e}$ is confined at $C$, we have
$d_i\in \locsOR \implies d_i\in\dom(\Oh_j * \Rh_j)$ for all $d_i\in\ol{d}$.
Thus $rng\,\eta_1\intersect \locsOR \subseteq \dom(\Oh_j * \Rh_j)$ proving
$\etaconf{(\super\,C)}{\eta_1}{h}$ by Definition~\ref{def:envconf}(\ref{envconfc}).
\end{itemize}

\subsection*{Additional cases for Lemma~\ref{lem:esound}}

\Case\ $\Gamma \aproves x:\Gamma\,x$. Then $d=\eta\,x$. Confinement of $x$
follows because the conditions for $d$ are exactly the same as the conditions for
$\eta$ and $\eta$ is confined.

\medskip
\Case s\  $\Gamma \aproves \NULL:B$, $\Gamma \aproves \TRUE:\BOOL$,
$\Gamma \aproves \FALSE:\BOOL$, $\Gamma \aproves \ITT:\UNIT$,
$\Gamma \aproves \is{e}{B}:\BOOL$. For $\NULL$ the
result holds since $\NIL\not\in\Loc$ and for $\TRUE, \FALSE, \ITT, \is{e}{B}$
the result holds by Lemma~\ref{lem:nlec}.

\medskip
\Case\ $\Gamma \aproves (B)e:B$. Then $d=\ell$ and the result follows
by induction on $e$ for each subcase of $C$.

\subsection*{Proof of Lemma~\ref{lem:xsound}}

First we show that $h_1$ is confined, where we have the following cases on
$\super\,C$:
\begin{itemize}
\item $\super\,C = \OBJECT$: then $h_1=h$.  So $\conf{h}$ by
hypothesis and $h\hext h_1$ by reflexivity of $\hext$.
\item $\super\,C < \OBJECT$: as $\super\,C\well C$, we can appeal
to induction for $\super\,C$ to obtain $\conf{h_1}$ and $h\hext h_1$.
Now by Lemma~\ref{lem:b} we have $\etaconf{C}{\eta}{h_1}$.  
\end{itemize}
It remains to show $\conf{h_0}$ and $h\hext h_0$.
This is a consequence of a more general

\textbf{Claim:} For the given $C$, suppose $\self:C\proves S$ is a command
with no method calls and $\self:C\aproves S$.  Moreover, suppose that for
any $\new{B}{}$ that occurs in $S$ we have $B\cdep C$. Then $\self:C\proves
S$ is confined. 

Applying the claim to $\constr\:C$, we get  $\conf{h_0}$.  
Then Lemma~\ref{lem:cx} applies, to yield  $h_1\hext h_0$.  So finally
$h\hext h_0$ by transitivity.

The proof of the claim is by structural induction on $S$.  The argument is
the same as the proof of Lemma~\ref{lem:csound}, except 
that in the case of $\NEW$ that proof appeals to 
Lemma~\ref{lem:xsound} whereas here we appeal to the induction hypothesis.
This use of induction is sound because for any $\new{B}{}$ in the
constructor, $B\cdep C$ and hence $B\well C$.

\subsection*{Additional cases for Lemma~\ref{lem:csound}}

\Case\ $\Gamma \aproves \assign{x}{e}$. Here $h_0=h$, hence confinement of $h_0$
follows because $\etaconf{C}{\eta}{h}$. To show $\etaconf{C}{\eta_0}{h_0}$, we
go by cases on $C$. First, as $\Gamma \aproves e:T$, by
Lemma~\ref{lem:esound} we have $e$ is confined. Choose a confining partition
of $h$ and let $j$ be such that $\eta\,\self \in\dom(\Oh_j)$. As $x\neq\self$
we have $\eta\,\self = \eta_0\,\self$.
\begin{itemize}
\item $C\nleq Rep \land C\nleq Own$: We must show
$rng\,\eta_0\intersect locs(Rep\subclasses) = \Empty$, which follows because
$d\not\in locs(Rep\subclasses)$ and because
$rng\,\eta\intersect locs(Rep\subclasses) = \Empty$ by $\etaconf{C}{\eta}{h}$.
\item $C\leq Own$: We must show
$rng\,\eta_0\intersect locs(Rep\subclasses) \subseteq \dom(\Rh_j)$, which
follows because $\{d\}\intersect locs(Rep\subclasses) \subseteq \dom(\Rh_j)$
by confinement of $e$ at $C$ and because
$rng\,\eta\intersect locs(Rep\subclasses) \subseteq \dom(\Rh_j)$ by
$\etaconf{C}{\eta}{h}$.
\item $C\leq Rep$: We must show
$rng\,\eta_0\intersect \locsOR \subseteq \dom(\Oh_j * \Rh_j)$, which
follows because $\{d\}\intersect \locsOR \subseteq \dom(\Oh_j * \Rh_j)$
by confinement of $e$ at $C$ and because
$rng\,\eta\intersect \locsOR \subseteq \dom(\Oh_j * \Rh_j)$ by
$\etaconf{C}{\eta}{h}$.
\end{itemize}

\Case\ $\Gamma \aproves \assign{x}{\mcall{\SUPER}{m}{\ol{e}}}$.
Here $h_0=h_1$ and $\eta_0=\ext{\eta}{x}{d_1}$.
Because $\Gamma \aproves \ol{e}:\ol{U}$, by
Lemma~\ref{lem:esound}, $\ol{e}$ is confined at $C$.
By Lemma~\ref{lem:aconf} we have $\etaconf{(\super\,C)}{\eta_1}{h}$.
Then by assumption $\conf{\mu}$
we get $\etaconf{(\super\,C)}{\eta_1}{h_0}$. Hence $h_0$ is confined.
To show $\etaconf{C}{\eta_0}{h_0}$, we go by cases on $C$. Recall that
$\ell=\eta\,\self$, and, as $x\neq\self$, $\ell=\eta_0\,\self$.
\begin{itemize}
\item $C\nleq Rep \land C\nleq Own$: As $C<\super\,C$ we have
$\super\,C\nleq Rep \land \super\,C\nleq Own$. By $\conf{\mu}$,
$d_1\not\in locs(Rep\subclasses)$.
Hence $rng\,\eta_0\intersect locs(Rep\subclasses) = \Empty$ by
$\etaconf{C}{\eta}{h}$.
\item $C\leq Own$: Let $\eta\,\self \in\dom(\Oh_j)$ for some $j$ in
the confining partition of $h$. As $C<\super\,C$ we have
either $\super\,C\leq Own$ ($Own<\super\,C$ is impossible by definition of
$\super$). By $\conf{\mu}$,
$d_1\not\in locs(Rep\subclasses)$ and $h\hext h_0$.
Hence $rng\,\eta_0\intersect locs(Rep\subclasses)
= rng\,\eta\intersect locs(Rep\subclasses)
\subseteq \dom(\Rh_j)$ by $\etaconf{C}{\eta}{h}$. As $h\hext h_0$,
$\dom(\Rh_j)\subseteq\dom(\Rh_{0_j})$. That is,
$rng\,\eta_0\intersect locs(Rep\subclasses) \subseteq \dom(\Rh_{0_j})$.
\item $C\leq Rep$:  Because $\loctype\,\ell\leq C$, let $\ell\in\dom(Rh_j)$
for some $j$ in the confining partition of $h$. As $C<\super\,C$ we have
either $\super\,C\leq Rep$ ($Rep<\super\,C$ is impossible by definition of
$\super$).
By $\conf{\mu}$, $d_1\in \locsOR \implies d_1\in \dom(\Oh_{0_j} * \Rh_{0_j})$
and $h\hext h_0$.
Hence $rng\,\eta_0\intersect \locsOR \subseteq  \dom(\Oh_{0_j} * \Rh_{0_j})$ by
$\etaconf{C}{\eta}{h}$ and Definition~\ref{def:hext}.
\end{itemize}

\medskip
\Case\ $\Gamma \aproves \seq{\S_1}{\S_2}$. By induction on $\S_1$,
$h_1$ is confined and $\etaconf{C}{\eta_1}{h_1}$. Moreover, if $\S_1$ is
a method call, it has confined argument values. Now by induction on $\S_2$,
$h_2$ is confined and $\etaconf{C}{\eta_2}{h_2}$. And, if $\S_2$ is
a method call, it has confined argument values. Hence all method calls in
$\seq{\S_1}{\S_2}$ have confined argument values.

\medskip
\Case\ $\Gamma \aproves \ifelse{e}{\S_1}{\S_2}$. By Lemma~\ref{lem:nlec},
$e$ is confined at $C$. If $b=\semtrue$, result follows by induction on
$\S_1$ and if $b=\semfalse$, result follows by induction on $\S_2$.

\medskip
\Case\ $\Gamma \aproves\var{\T\; x}{e}{S}$. Because $\Gamma\aproves e:U$
we have by Lemma~\ref{lem:esound} that $e$ is confined at $C$.
And, because $x\neq\self$ and $\etaconf{C}{\eta}{h}$, we get
$\etaconf{C}{\eta_1}{h}$. Since $\Gamma, x:\T \aproves S$, by induction on $S$
we have $\etaconf{C}{\eta_2}{h_1}$ and all method calls in $S$ have confined
argument values. Hence $h_1$ is confined and
$\etaconf{C}{\eta_2\downharpoonright\,x}{h_1}$ and all method calls in
$\var{\T\; x}{e}{S}$ have confined argument values.

\begin{acks}
Our work benefitted from discussions with a number of people as well
as from helpful feedback from the POPL referees.  
For particularly useful technical help and encouragement we thank 
Torben Amtoft,
Steve Bloom, 
Paulo Borba,
Sophia Drossopoulou,
Nevin Heintze, 
Doug Lea, 
Peter M{\"u}ller, 
Peter O'Hearn, 
Uday Reddy, 
David Schmidt, and 
Hongseok Yang.
\end{acks}
  
\bibliographystyle{acmtrans}

\begin{received}
Received Month Year;
revised Month Year; accepted Month Year
\end{received}

\end{document}
